\documentclass[sigconf,nonacm]{acmart}

\usepackage{boldline} 

\AtBeginDocument{%
  \providecommand\BibTeX{{%
    \normalfont B\kern-0.5em{\scshape i\kern-0.25em b}\kern-0.8em\TeX}}}

\begin{document}

\pagestyle{plain}

\title{Leveraging Large Language Models in Conversational Recommender Systems}

\author{Luke Friedman*, Sameer Ahuja, David Allen, Zhenning Tan, Hakim Sidahmed, Changbo Long, Jun Xie, Gabriel Schubiner, Ajay Patel, Harsh Lara, Brian Chu, Zexi Chen and Manoj Tiwari }
\thanks{*Corresponding author: lbfried@google.com}
\affiliation{Google Research}

\begin{abstract}
A Conversational Recommender System (CRS) offers increased transparency and control to users by enabling them to engage with the system through a real-time multi-turn dialogue.  Recently, Large Language Models (LLMs)  have exhibited an unprecedented ability to converse naturally and incorporate world knowledge and common-sense reasoning into language understanding, unlocking the potential of this paradigm.  However, effectively leveraging LLMs within a CRS introduces new technical challenges, including properly understanding and controlling a complex conversation and retrieving from external sources of information.  These issues are exacerbated by a large, evolving item corpus and a lack of conversational data for training.  In this paper, we provide a roadmap for building an end-to-end large-scale CRS using LLMs.  In particular, we propose new implementations for user preference understanding, flexible dialogue management and explainable recommendations as part of an integrated architecture powered by LLMs.  For improved personalization, we describe how an LLM can consume interpretable natural language user profiles and use them to modulate session-level context. To overcome conversational data limitations in the absence of an existing production CRS, we propose techniques for building a controllable LLM-based user simulator to generate synthetic conversations. As a proof of concept we introduce RecLLM, a large-scale CRS for YouTube videos built on LaMDA, and demonstrate its fluency and diverse functionality through some illustrative example conversations.
\end{abstract}

\settopmatter{printacmref=false}
\maketitle

\section{Introduction}

\begin{figure*}[h]
  \centering
  \includegraphics[width=0.8\textwidth,scale=0.3]{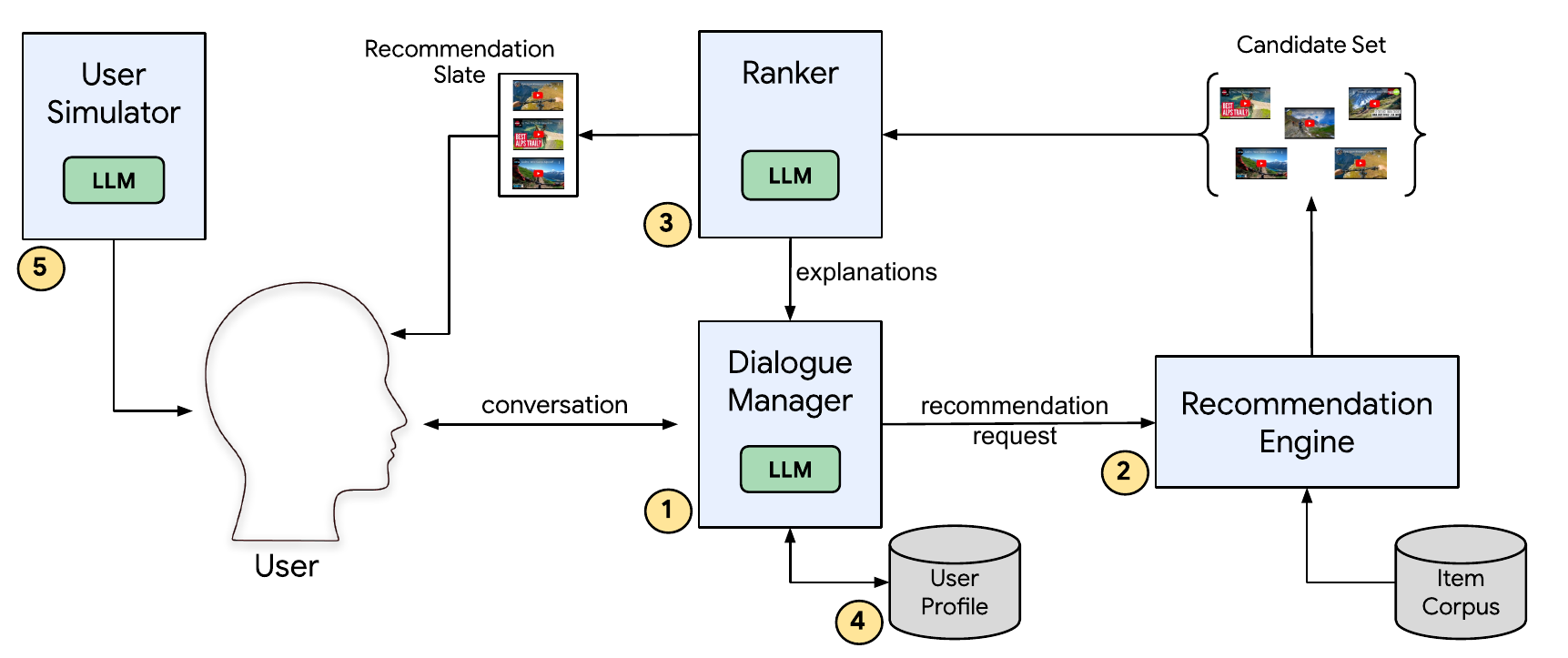}
  \caption{Overview of key contributions from RecLLM.  (1) A dialogue management module uses an LLM to converse with the user, track context and make system calls such as submitting a request to a recommendation engine all as a unified language modeling task. (2) Various solutions are presented for tractable retrieval over a large item corpus within an LLM-based CRS. (3) A ranker module uses an LLM to match preferences extracted from the context of the conversation to item metadata and generate a slate of recommendations that is displayed to the user. The LLM also jointly generates explanations for its decisions that can be surfaced to the user. (4) Interpretable natural language user profiles are consumed by system LLMs to modulate session-level context and increase personalization. (5) A controllable LLM-based user simulator can be plugged into the CRS to generate synthetic conversations for tuning system modules.}
\label{recllm_system}
\end{figure*}

Recommender systems are one of the most prominent success stories of machine learning in industry, delivering personalized content to billions of users over a wide range of domains such as Search, Videos, News, and Shopping. Machine learning algorithms have transformed the way these systems are built within industry; in particular, over the last decade deep learning based systems that thrive in the large data regime have capitalized on the abundance of user interaction data available through products to learn sophisticated statistical correlations and better optimize for key engagement metrics \cite{yt_system}.  However, despite the success of ML in this setting, this increasing reliance on implicit interaction signals like clicks as a proxy for user preference has its downsides as well. It is well documented that many modern large-scale recommender systems encounter problems like surfacing clickbait, propagating societal biases and polarization of the user base \cite{mansoury2020feedback, position_bias, garimella2017long}.  Recommender systems based on point-and-click interfaces also afford the user only a limited channel to communicate with the system and little opportunity to engage in any type of interactive exploration.

A Conversational Recommender System (CRS) gives users more control over their recommendations through the ability to engage in a real-time multi-turn dialogue \cite{jannach2022conversational, gao2021advances}. This enables the system to actively query the user instead of relying solely on prior behavior to infer preferences, and in response the user can provide feedback and refine suggestions over a series of turns.  This new paradigm is a generalization of both recommender and classical search systems, in which typically users direct the system through a single-shot query. Now the system must explore cooperatively with the user and, to maintain a natural flow, sometimes even veer off into modes tangential to the core recommendation task such as undirected chit-chat and question answering (see e.g. \cite{cai2020predicting}). Oftentimes the CRS must also be multimodal, for instance displaying recommendation slates within a visual UI while simultaneously carrying on a conversation through a natural language interface (see e.g. \cite{zhang2018towards}).

Although conversational recommender systems have existed in some form for decades \cite{linden1997interactive, thompson2004personalized}, the recent explosion of Large Language Models (LLMs) \cite{brown2020language, thoppilan2022lamda, chowdhery2022palm} unlocks new opportunities. LLMs have made a huge leap in the ability of machines to converse in a human-like way and can power the natural language interface between the user and the system. LLMs have also shown an unprecedented ability to draw on general world knowledge and utilize some level of common sense reasoning \cite{huang2022towards}, which can be exploited in various ways within a CRS.  For instance, we can try to use an LLM to directly reason about how well an item matches the context of a conversation within a ranking module and generate an intuitive natural language explanation as a byproduct.  Other possible use cases for LLMs behind the scenes include dialogue management, incorporating natural language user profiles for better personalization and building realistic user simulators to generate synthetic data at scale for evaluation and tuning of system components.

Tantalizing as LLMs are as a tool for a CRS, new technical challenges must be overcome to leverage them effectively.  For instance, LLMs are prone to hallucinations and grounding them remains a largely unsolved problem \cite{ji2022survey}.  Also, one of the appeals of LLMs is their sense of naturalness and unpredictability, but when operating in a task-oriented setting this means that controlling an LLM can be more difficult than with a template based system.  Particularly challenging in the recommendation setting is how to interface between the LLM and the underlying recommendation engine.  One approach is to have the LLM \textit{be} the recommendation engine in addition to its role as a dialogue agent (see e.g. \cite{kang2019recommendation}). However, for large-scale recommender applications the item corpus can contain millions or billions of always-changing items, making it challenging for an LLM to memorize the corpus within its parameters.  Alternatively the LLM must somehow connect to an external recommendation engine or database, passing on relevant preference information.  While studied recently \cite{byrne2020tickettalk, rastogi2020towards}, this approach is yet to be solved in the general large-scale recommendation setting. 

In this paper we provide a roadmap for leveraging LLMs in a variety of ways to build a controllable and explainable large-scale CRS.  Key contributions of the proposal are:
\begin{itemize}
\item A dialogue management module that reframes natural language generation, preference understanding, context tracking, and calls to a recommendation engine as a unified language modeling task performed by a single LLM.  
\item
A general conceptual framework for performing retrieval with an LLM over a huge corpus of items. Various solutions are presented depending on efficiency requirements and what data and external APIs are available.  
\item
A joint ranking / explanation module that uses an LLM to extract user preferences from an ongoing conversation and match them to textual artifacts synthesized from item metadata.  As a byproduct of intermediate chain-of-thought reasoning \cite{wei2022chain}, the LLM generates natural language justifications for each item shown to the user, increasing the transparency of the system. 
\item
Incorporation of persistent, interpretable natural language user profiles as additional input to system LLMs, which supplements session-level context and improves the personalized experience. 
\item 
Techniques for building controllable LLM-based user simulators that can be used to generate synthetic conversations for tuning system modules. 
\end{itemize}

As a proof of concept we introduce RecLLM, an LLM-based CRS for YouTube videos powered by LaMDA \cite{thoppilan2022lamda}, and share some example conversations showing the fluency and diverse functionality of the system. Our goal is to make a compelling argument for the promise and viability of LLM-based conversational recommender systems and to take a first step towards realizing this vision in practice.

\section{Problem Scope}
\label{problem_scope_section}

\begin{figure}[h]
  \centering
  \includegraphics[width=1.00\linewidth]{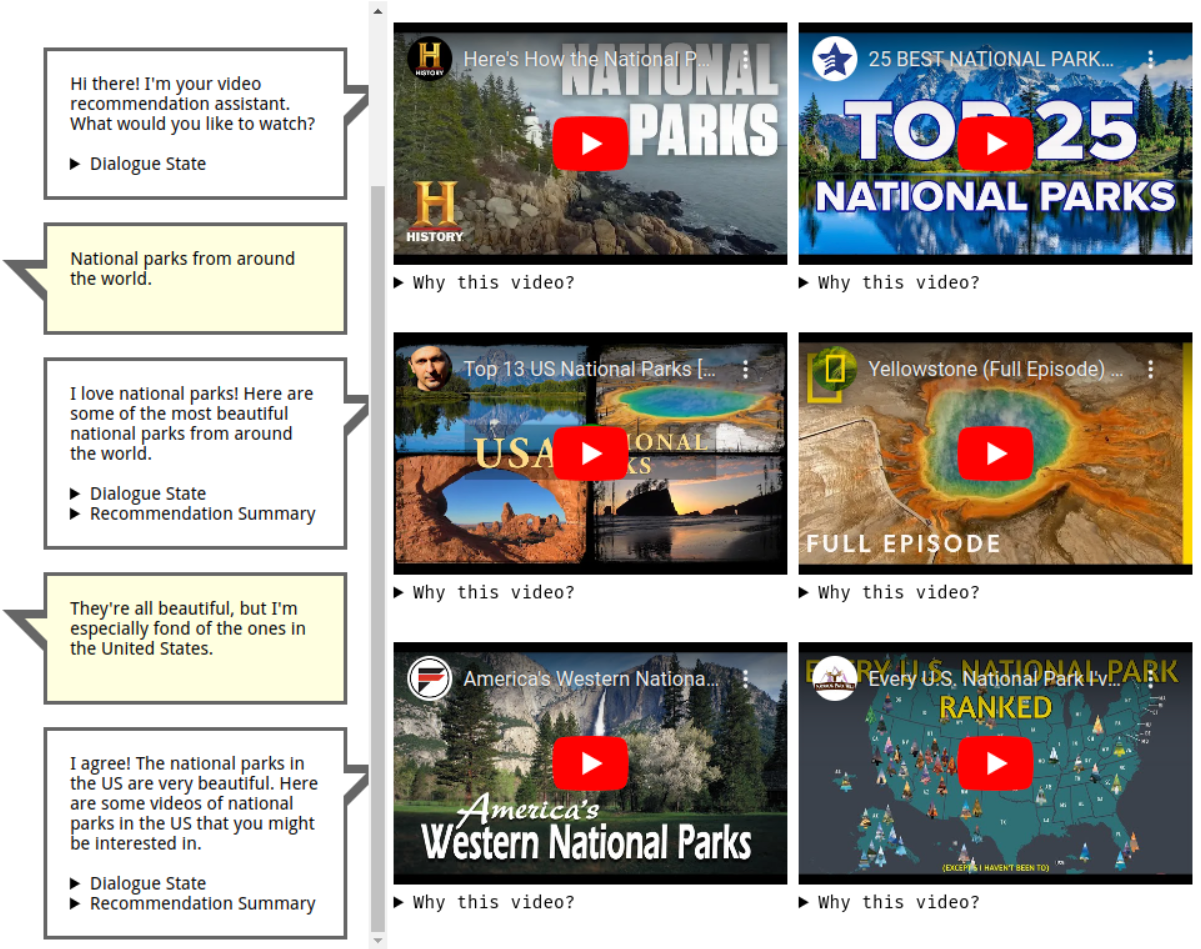}
  \caption{Screenshot of an LLM-based user simulator talking with RecLLM.
}
\label{system_screen_figure}
\end{figure}

\begin{figure*}[h]
  \centering
  \includegraphics[width=\textwidth]{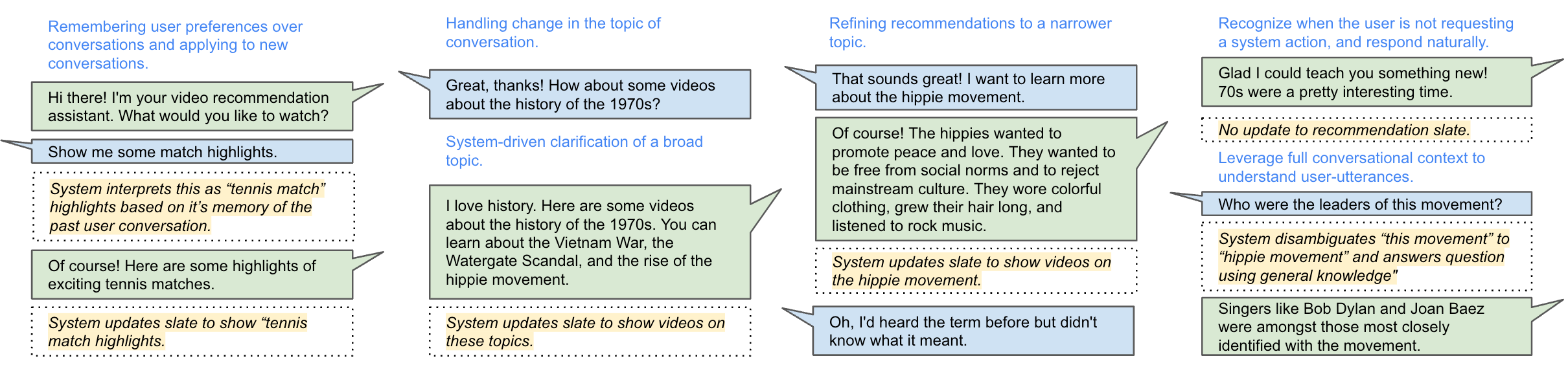}
  \caption{RecLLM possesses many conversational capabilities such as the ability to retain context throughout a session, handle topic shifts and reference items from recommendation slates.
}
\label{core_use_cases_figure}
\end{figure*}

In RecLLM, we represent the CRS in a multi-modal setup comprised of two components: A slate of recommendations and an ongoing conversation between the user and the conversational agent (see Figure \ref{system_screen_figure}). The user outputs a natural language message on their turn, and the agent responds with a natural language message, optionally updating the slate of recommendations based on the conversation. By separating dialogue from the recommendation slate we hope to more accurately reflect how a large-scale CRS would eventually look in a production setting. 

Traditionally, users have interacted with recommender systems via user interface interactions such as viewing the recommended items, or marking recommendations as good or bad via interface widgets \cite{dietz2019designing, ricci2007acquiring}. Although currently in RecLLM we exclude these types of interactions we do not intend to replace them; our eventual goal is to augment them with the more expressive channel of natural language that allows users to better express nuance about their interests.

In terms of the item corpus, RecLLM recommends from the corpus of all public YouTube videos. We make this choice due to two characteristics that increase the applicability of the system to other real-world problems: One, unlike corpora of items that occur frequently in the LLM's training data (e.g., movies and popular music), an LLM cannot feasibly be used to directly recommend YouTube videos and must interface with the corpus. Secondly, it's a large-scale corpus, requiring a scalable approach to recommendations. A natural consequence of building such a system from scratch is that there are no logs of users interacting with this system to jumpstart training of the model(s).  Although RecLLM focuses on YouTube videos, our intention in this paper is to outline a general approach that can be easily extended to many other domains.

While evaluating with initial testers, we found that users expect a CRS that pairs slate recommendations with natural language conversation to possess a wide range of conversational capabilities, such as retaining context, handling topic shifts and referencing slate items. RecLLM focuses on leveraging techniques that can scale over a broad number of these use-cases. In Figure \ref{core_use_cases_figure} a few of the core conversational capabilities currently supported are demonstrated via a mock conversation.

Finally, there are several problems that need to be addressed for conversational agents to become mainstream. These include safety of dialogue, debiasing, consistent personality of agents, etc. In this work we do not attempt to tackle these problems directly, rather focusing on problems that are unique to the setting of conversational recommenders.

\section{System Overview}
\label{system_overview_section}

In this section we take a closer look at key system components of RecLLM (see Figure \ref{recllm_system}).  In particular we focus on dialogue management, retrieval, ranking and explanations, and incorporation of natural language user profiles.

\subsection{Dialogue Management}
\label{dialogue_management_section}

\begin{figure}[h]
  \centering
  \includegraphics[width=0.9\linewidth]{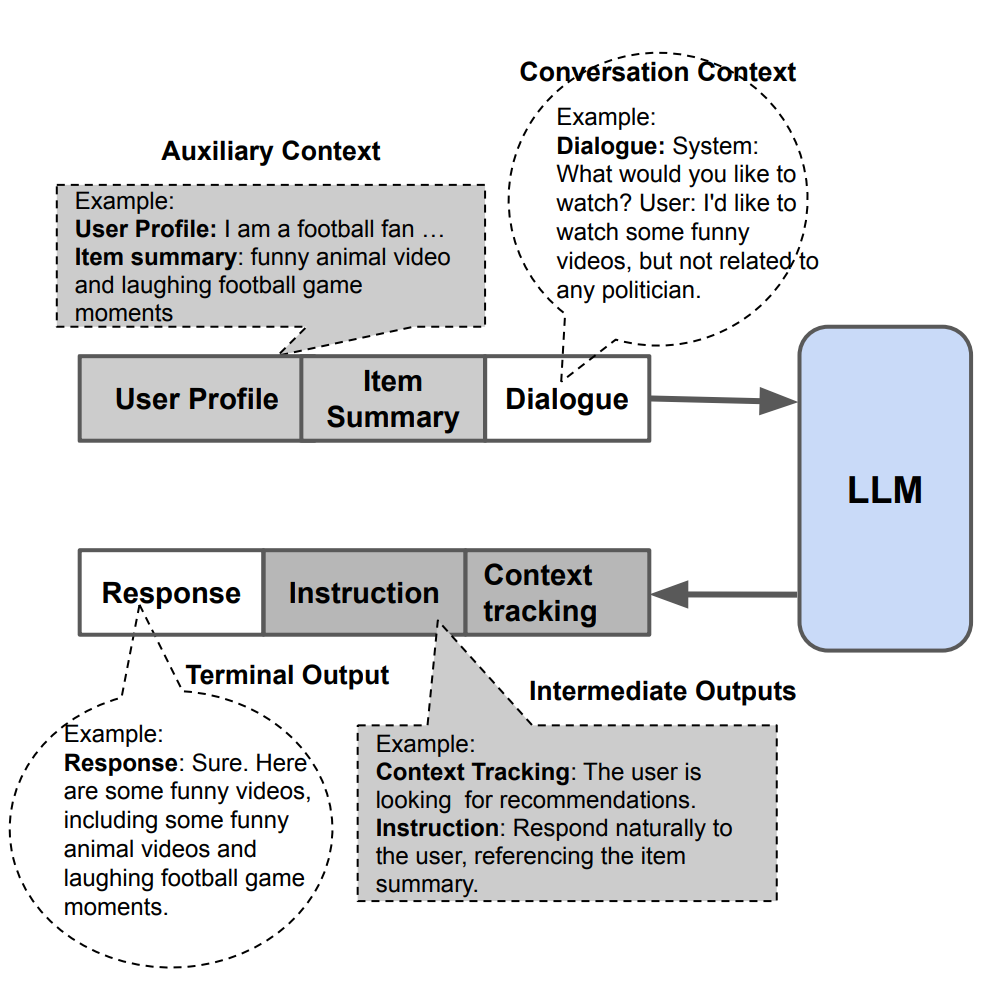}
  \caption{A unified LLM dialogue management module. An LLM takes as input the full session context and outputs a sequence of messages ending in a terminal output that triggers a system action, such as a response to the user.
}
\end{figure}

Dialogue management is the central module of a CRS, acting as the interface between the user and the rest of the system.  It is responsible for guiding the user through a multi-turn exploration of the recommendation corpus and generating sensible, useful, and grounded responses at each turn. In the process it must either implicitly or explicitly perform context tracking to extract useful representations of user preferences and intents. This information can be used to inform the dialogue policy and also as the basis for outputting API calls to initiate system actions (e.g. by sending a search query to a recommendation engine backend, see Section \ref{retrieval_section}). From an end-to-end point of view, given context information (dialogue history, a user profile, item summaries, etc.), the goal of the dialogue manager is to generate system actions to take, as well as an appropriate system utterance.

There are extra challenges and requirements to dialogue management in the context of conversational recommenders:

\begin{itemize}
\item\textbf{Control}: In contrast to open-ended dialogue, a CRS dialogue manager must actively work with the user to explore the recommendation corpus.  This entails a mixed-initiative setup where the system must respond to user requests and also at times actively steer the conversation in a specific direction.  For instance, preference elicitation---in which the system must figure out when and how to best query the user in order to extract maximal information about their preferences---is an entire subfield of CRS dialogue management \cite{ren2021learning, sun2018conversational, chen2019towards, zhang2018towards}.
\item\textbf{Ambiguity}: Compared to task-oriented dialogue there is no clear cut measure of success for a CRS dialogue manager. Although the system should try to ensure that the conversation does not get too far off track the core recommendation task, the goal is not necessarily to minimize the number of turns that it takes the user to find an acceptable item, but rather to provide an overall satisfactory exploratory experience (see, for instance \cite{schnabel2018short}). This means that there is rarely a single objectively "correct" thing for a conversational system to say at any given time, nor an easily defined metric for whether the dialogue manager is doing a good job.  
\item\textbf{Grounding}: One of the main challenges of a CRS dialogue manager is to faithfully ground its responses to the user in the recommendation corpus.  After returning a slate of recommendations, the system should be able to refer to the items in a relevant and factually correct way.  Other sources of external information, such as long term preferences coming from a user profile, may also be injected and the dialogue manager should be able to incorporate them appropriately in the ongoing conversation.
\end{itemize}

Traditionally CRSs take a modular approach to dialogue management, where a hardcoded policy graph maps dialogue states (e.g. intent) to different system actions, such as whether to get a recommendation, ask a question, or chat casually.  Natural language understanding models extract preferences and determine the dialogue states, and separate natural language generation models generate system responses.  Alternatively, in some recent CRSs language models are tuned end-to-end directly to imitate dialogue collected from crowdsource workers, discarding any notion of dialogue states or internal structure.   

In RecLLM we employ a single unified LLM to execute dialogue management purely in terms of language modeling. At each turn the LLM takes as input the prior conversation context along with additional information like textual representations of recommendation slates and user profiles that are potentially injected from external sources. Like the end-to-end approach mentioned above, one of the distinguishing features of this architecture is that there no longer exists a hardcoded policy graph with fixed dialogue states.  Instead, on a given system turn the LLM generates a sequence of natural language outputs that encapsulate all context tracking, intermediate reasoning, natural language generation, and API calls to the rest of the system.  It is hardcoded that certain string patterns in outputs from the dialogue manager trigger system actions. For instance an output "Response: <\emph{message}>" will cause \emph{message} to be shown as a user facing response, and "Request: <\emph{query}>" will cause \emph{query} to be sent to the recommendation engine backend to retrieve a slate of recommendations.  Other outputs of the LLM can function as chain-of-reasoning steps, instructions to itself to follow, or dialogue state tracking inferences.  Unlike the system calls, there are no ingrained rules about the functionality of these intermediate outputs, and conventions about their  use must be taught to the LLM either through in-context few-shot learning or tuning.  

The advantage of this architecture over the modular approach is its simplicity and flexibility. In the modular approach, any new functionality such as the addition of a new user intent or dialogue state has to be engineered into the system, which is a serious impediment to scalability.  The unified LLM architecture shifts the emphasis from engineering-driven to data-driven quality iteration.  To fix a problem or introduce new capabilities, instead of engineering a new component a designer must now create examples that enable the LLM to learn the desired behavior.  This also creates the potential for the dialogue manager to learn new policy states and useful dialogue state tracking artifacts through the generalization abilities of the LLM.  

The main challenge to the unified LLM approach is how to effectively control the dialogue manager and guide it towards a reasonable dialogue policy without explicitly constraining it via hard rules. In our initial implementation we tune our unified LLM on a moderate number of manually generated examples.  In this way we are able to establish some direction about the type of behavior and internal states we would like to see while still relying heavily on the ability of LLMs pretrained on dialogue data to converse naturally with only minimal supervision.  Although we are able to build a functional dialogue manager this way, with only a limited amount of training examples it is difficult to teach the dialogue manager a sophisticated policy tailored to the conversational recommender domain. In Section \ref{tuning_modules_section} we discuss ideas for overcoming this limitation by tuning our dialogue manager and recommendation modules with larger amounts of synthetically generated data.

\subsection{Recommendations and Refinement}
\label{recommendations_section}

Once triggered by the dialogue management module, it is the responsibility of the recommendation module to return a slate of high quality, relevant, and diverse recommendations that will be shown to the user. This can either be an initial recommendation slate or a refinement of an earlier slate from the session based on feedback from the user.  A traditional recommender system chooses items by inferring preferences of the user from some type of user profile or dense representation built from historical data, possibly taking into account other contextual factors (e.g. the location or time of day).  In a search system, the user can supplement these implicit signals with explicit intents, usually through a simple static query.  A primary challenge of a CRS is that now the user can express these explicit intents over the course of a full multi-turn conversation, which the recommendation module must understand and connect to the item corpus. Many traditional recommender systems employ a two stage pipeline, first retrieving candidate items and then ranking them \cite{eksombatchai2018pixie, ma2020off}.  RecLLM follows this strategy, with the added twist that the ranker also jointly generates natural language explanations for why each item is being selected.  

\subsubsection{Retrieval}
\label{retrieval_section}

\begin{figure}[h]
  \centering
  \includegraphics[width=\linewidth]{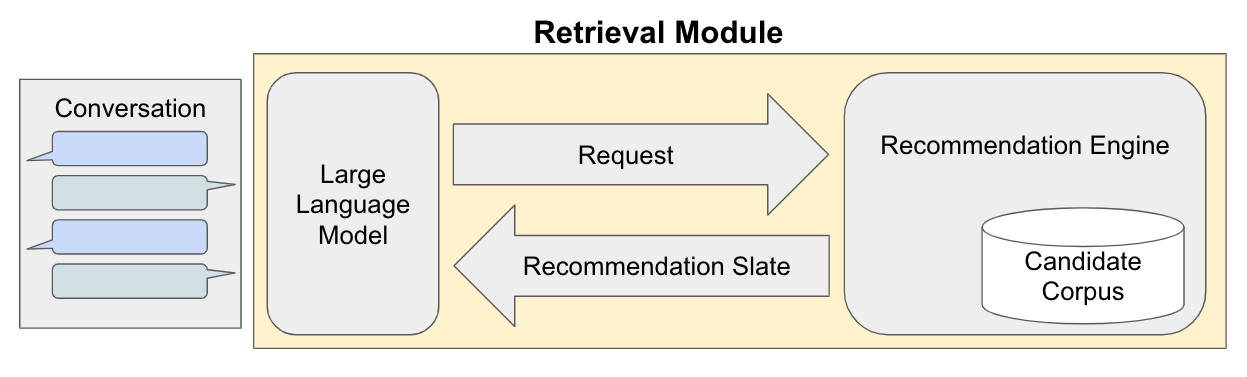}
  \caption{Overview of large-scale retrieval in an LLM-based CRS.}
  \label{concept_bridge_figure}
\end{figure}

The purpose of the retrieval phase is to take the full corpus, which for some domains such as videos or urls may contain hundreds of millions of items, and based on the context select a small number of candidate items (e.g. 100) that will be fed to a downstream ranker.  A key challenge of retrieval is to make this process tractable, as it is not computationally feasible to process each item independently at inference time.  In Figure \ref{concept_bridge_figure} we illustrate a general conceptual framework for retrieval in our problem setting.  An LLM processes the session context and generates a request, either implicitly through a model activation layer or explicitly through its language output interface. A recommendation engine then uses a tractable search algorithm to retrieve candidates from the item corpus. In Table \ref{tab:concept_bridge_representations} we give a few illustrative examples of possible retrieval algorithms that fit into this framework, which we describe in more detail below.

\begin{table}[h]
    \centering
    \resizebox{0.48\textwidth}{!}{
    \begin{tabular}{| c | p{0.15\textwidth} | p{0.15\textwidth} |}
        \hline
        \textbf{Approach} & \textbf{Request Type} & \textbf{Tractable Search Algorithm} \\ \hlineB{4}
        Generalized Dual Encoder Model & Internal LLM embeddings & KNN or ScaNN \cite{DBLP:journals/corr/abs-1908-10396} \\ \hline
        Direct LLM Search & Title or id & Fuzzy lookup \\ \hline
        Concept Based Search & List of concepts & Concept Activation Vector \cite{https://doi.org/10.48550/arxiv.1711.11279} \\ \hline
        Search API Lookup & Search query & Search API \\ \hline
    \end{tabular}}
    \caption{Various possible solutions to large-scale retrieval in a CRS.}
    \label{tab:concept_bridge_representations}
\end{table}

\paragraph{Generalized Dual Encoder Model}

A popular solution to retrieval in traditional deep learning based recommenders is to use a dual encoder model consisting of two neural net towers, one to encode the context and one to encode the items (see e.g \cite{yi2019sampling} and Figure \ref{tuning_figure}a).  Item embeddings can be generated offline using the item tower and stored in an efficient data structure.  An approximate nearest neighbor lookup can then use the generated context embedding to perform a sub-linear time retrieval of item embeddings at inference time \cite{DBLP:journals/corr/abs-1808-04776}. We can extend this approach for conversational recommenders by using an LLM as a context encoder that processes the full ongoing conversation between the user and system along with any other additional context information.  In this case the request sent to the recommendation engine is an embedding, which can be generated by extracting and then projecting a suitable activation layer from the model.

One downside to this approach of pulling embeddings from the internals of an LLM is that it severely hampers our ability to learn a retrieval model in a sample efficient way. Dual encoder models trained from scratch require large amounts of training data to constrain the context tower embeddings to occupy the same subspace as the item tower embeddings.  Sometimes it is possible to use pretrained embeddings on the item side (for instance by taking them from an existing production search or recommender system), but still the context embeddings must be tuned to align with the item embeddings to get good results. LLMs operate via a text-in / text-out interface and much of their power comes from the transfer learning afforded by knowledge gained through extensive pretraining.  By leaving the level of language abstraction we are sacrificing much of this ability to generalize from a small amount of data. 

\paragraph{Direct LLM Search}

In this method the LLM directly outputs ids or titles of items to recommend as text.  The tractable search algorithm is an exact or fuzzy match against items in the corpus and the recommendation engine plays no role beyond this simple matching. The LLM must learn to output these ids/titles through some combination of its pretraining and a corpus-specific fine tuning phase (see e.g \cite{tay2022transformer}).  Given the assumption that our system must be able to return slates from a fixed item corpus, this is the closest thing to having an LLM-based chatbot function directly as a CRS.  The downside to this approach is that because only negligible work is being offloaded to the recommendation engine, the LLM must memorize information about the entire item corpus within its model parameters.  For a large corpus this can be prohibitively expensive in terms of the model size and training data needed, and also makes it difficult to refresh the item corpus without retraining the LLM.

\paragraph{Concept Based Search}

In this method the LLM outputs a list of concepts, which are then embedded and aggregated by the recommendation engine into a single context embedding.  This is used to lookup items through approximate k-nearest neighbor search similar to the generalized dual encoder method.  A technique like Concept Activation Vectors \cite{https://doi.org/10.48550/arxiv.1711.11279} can be used to perform this transformation from concepts to embeddings in the item space. The appeal of this approach is that extracting relevant concepts from a conversation is a natural task that can be taught to an LLM through in-context learning or tuning with a small number of examples.  Also, because only item embeddings are needed (the concept embeddings are derived from these) if pretrained item embeddings can be borrowed from an existing source then no additional tuning of embeddings is required. However, one limitation is that lists of concepts are often a coarse representation of a conversation and similar to continuous bag-of-words methods \cite{mikolov2013efficient} are lossy with respect to word order and other nuances of language, which can negatively affect retrieval quality.  

\paragraph{Search API Lookup}

In this method, the LLM directly outputs a search query, which gets fed into a black-box search API to retrieve items.  Unlike Concept Based Search, which is generic as long as item embeddings can be trained or reused, Search API Lookup is only applicable when such a search API already exists for the domain in question.  However, when available, this type of API is often backed by a sophisticated search stack and can yield higher quality results. Analogous to Concept Based Search, in Search API Lookup the LLM can be taught to output relevant search queries using a small number of examples (see Section \ref{dialogue_management_section}), but the quality of retrieval is limited by the extent to which a search query can properly represent the full context of a conversation.

In Section \ref{tuning_modules_section} we build upon these methods by discussing options for tuning a retrieval model using large-scale synthetic data.

\subsubsection{Ranking / Explanations}
\label{ranking_section}

After candidate items have been retrieved, a ranker decides which of them will be included in the recommendation slate and in what order.  Unlike the retrieval module, the ranking module does not need to perform tractable search over a large corpus and is therefore less constrained in the types of computation that are possible.  In a traditional recommender system, this usually manifests in the ranker crossing context and item features (instead of processing them in separate towers as is done in a dual encoder) and potentially using custom ranking losses during training that directly compare candidate items \cite{cao2007learning}. In the case of RecLLM, we take advantage of this extra room for computation to use an LLM that reasons sequentially about how well an item matches the context and generates a rationalization for its decision as a byproduct.

\begin{figure}[h]
  \centering
  \includegraphics[width=\linewidth]{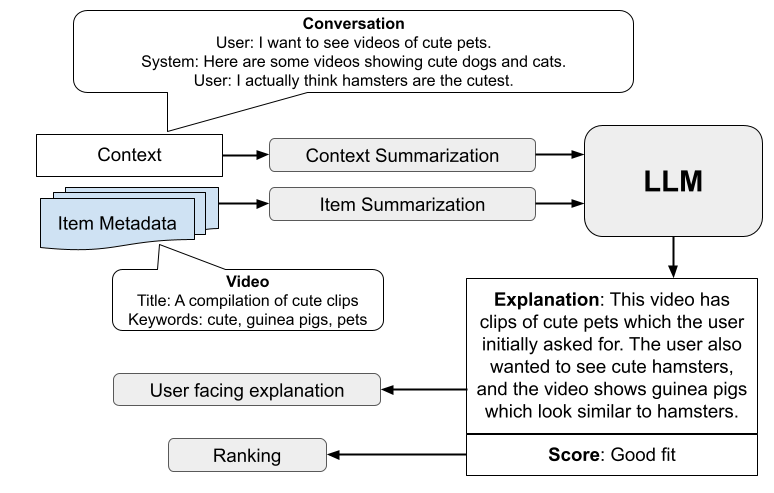}
  \caption{A joint LLM ranking / explanation module. The conversation is used as context for the user’s preferences and the video metadata is used as context for the item. The LLM takes in summaries of the item side and context side to produce a score for the item and an explanation for the score.
  \label{ranker_explainer_figure}
}
\end{figure}

Figure \ref{ranker_explainer_figure} gives a schematic for the LLM ranker.  For each candidate item, the LLM jointly generates a score and a natural language explanation for the score\footnote{There are many proposed solutions for enabling text in / text out LLMs to solve regression problems (i.e. output a score) \cite{liu2021enct5}; within RecLLM we use the simple approach of bucketing the range of possible scores and having the LLM output a semantically meaningful phrase (e.g. "excellent fit") corresponding to a bucket id.}.  These scores implicitly induce a ranking of the items.  The first step is to create a text summarization of the item that fits into the context window of the LLM based on metadata associated with the item.  In the case of a YouTube video recommender, this metadata consists of information such as the title, knowledge graph entities associated with the video, developer description of the video, transcript of the video, and user comments. In the future we would also expect a large multimodal model to directly process the raw video instead of relying only on textual artifacts. This item summarization can be done offline and is necessary in the case where the metadata is high volume (e.g. if we have thousands of user comments). We can view this summarization as a special case of the multi-document summarization problem \cite{liu2019hierarchical}; it is also related to a main challenge of the user profile module (see Section \ref{user_memory_section}), which must summarize large amounts of prior user data into a text format that can be passed into an LLM (or alternatively augment the LLM with the ability to access this information efficiently at inference time). There also can be a similar preprocessing step for summarizing the context information, although this must be done at inference time since unlike for items we cannot enumerate all possible contexts and process them offline.  

Given these item and context summaries as input, the LLM ranker then scores the item using chain-of-thought reasoning, which has been shown to improve the performance of LLMs on these types of classification / regression tasks \cite{wei2022chain}.  The intermediate chain-of-thought reasoning steps generated by the LLM function as explanations for why certain items are eventually included or left out of the recommendation slate.  These explanations can be viewed internally for debugging purposes and also shown to the user, either by including them as input to the dialogue manager that produces utterances within the conversational interface or by postprocessing and including them within pop-up boxes in the visual UI where the recommendation slates are displayed.

\subsection{User Profile}
\label{user_memory_section}

One of the key advantages to a CRS is the ability of the user to articulate their preferences over the course of a session, so that the system can assist them without necessarily needing any prior background information.  Despite this, the personalized experience can be improved if the system has built up a profile of the user beforehand so that there is a mutual starting base to build the conversation on top of. For instance, if a user dislikes jazz music and has shared this previously, they should not have to reiterate this point every new session when searching for music videos.

In traditional deep learning based recommender systems, non-verbal interaction signals such as clicks or ratings are often used to train embedding representations of a user that can be fed into a neural net.  In RecLLM we instead represent users with natural language profiles (see e.g. \cite{radlinski2022natural}), which can be consumed by an LLM.  These are more transparent compared to embeddings and specific pieces of information can usually be attributed to an original source, which aids in explainability.  Also, users can manually edit these natural language profiles, which gives them greater control to monitor and update their preferences.  In RecLLM we build user profiles based on a user's repeated interaction with the system over multiple sessions, although it would be possible to incorporate other data sources as well.  

An important open research question is how to structure a user profile in terms of natural language. Currently in RecLLM we represent a user by a set of salient facts we have extracted from prior sessions (e.g. "I do not like listening to jazz while in the car") similar to \cite{zhang2018personalizing}, although many other more sophisticated schemes are possible.  Another extreme possibility is to avoid any lossiness by  defining a user profile degenerately as the raw conversational history of all sessions the user has had with the system in the past. In this case we would need to implement an efficient mechanism for an LLM to retrieve relevant facts from this raw history at inference time.

There are three main components to the User Profile module, which we now describe.

\paragraph{Memory Extraction}

The purpose of the memory extraction component is to identify when a particular utterance contains a meaningful and enduring fact about the user that can be extracted and added to the user profile.  In RecLLM, this is currently implemented by an LLM using in-context few-shot learning as part of the dialogue management module.  

\paragraph{Triggering and Retrieval}

The triggering and retrieval component decides at what instances during a session it is likely beneficial to query the user profile for supplementary information and to then retrieve the most relevant facts related to the current context.  Currently at each turn RecLLM retrieves a single fact from the user profile by embedding the last user utterance and doing a cosine distance comparison between this embedding and precomputed embeddings of each fact in the user profile.  Triggering is implemented post hoc by thresholding on this minimal cosine distance. Better performance is likely possible by using a separate LLM classifier for triggering, retrieving multiple facts from the user profile, and basing retrieval on the entire conversation context of the session as opposed to just the last utterance.  

\paragraph{System Integration}

Once the user profile information is retrieved, it must be integrated into the rest of the system so that it can influence behavior such as the system's dialogue and API calls to the recommendation engine.  How to properly integrate facts coming from a user profile is a difficult open question, as it is highly context dependent how they should modulate short term preferences expressed by the user in the current session.  For instance, the system may know that the user is allergic to seafood, but if the user explicitly says they want to see some videos about fish recipes to pass along to a friend it's important that the system overrides this preference from the user profile and gives the user what they are asking for.  In RecLLM we use a simple strategy of injecting facts from the user profile into the text input of the dialogue manager (see Section \ref{dialogue_management_section}).  By doing so we allow LLMs powering the dialogue manager to make nuanced decisions about how to utilize this auxiliary information in the context of the ongoing session without having to engineer any hard rules into the system. 

\begin{figure}[h]
  \centering
  \includegraphics[width=1.0\linewidth]{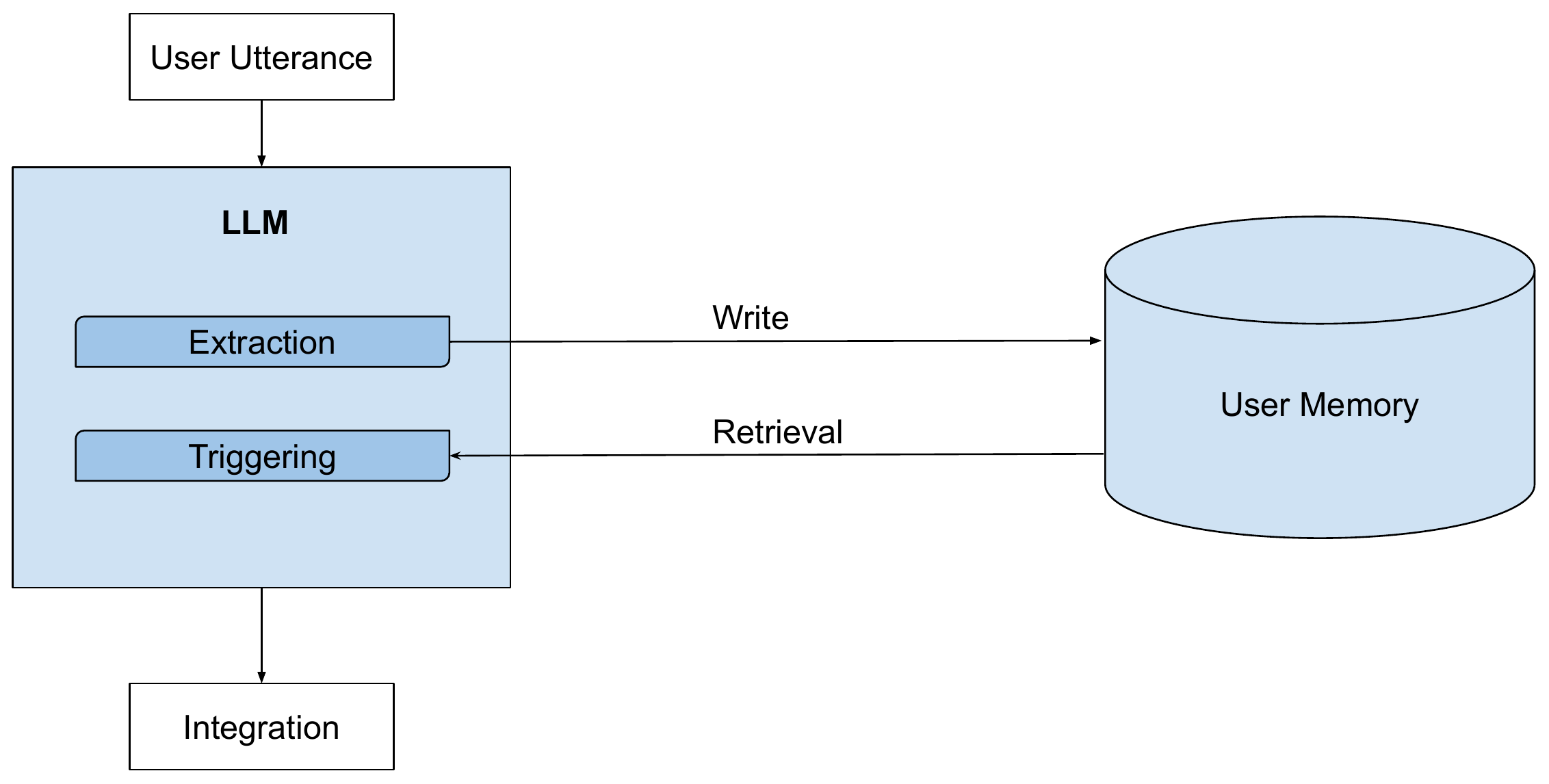}
  \caption{Overview of the architecture incorporating the User Profile module}
\end{figure}

\section{Simulation and Large-scale Tuning}
\label{simulator_and_tuning_section}

A major impediment to building a high-quality industrial CRS is a lack of data available for training and evaluation.  Typically, large-scale recommender systems are trained on user interaction data mined from the logs of existing products; however, conversational recommenders are a nascent technology and for the most part products using this paradigm do not exist yet.  An initial high quality system must be built to make such a product viable, after which a bootstrapping cycle can begin in which real data is generated from the system and then increasingly better versions of the system are trained using that data.  RecLLM deals with the data sparsity problem by exploiting the transfer learning ability of large language models using in-context few-shot learning or fine-tuning on a small number of manually generated examples.  However, we hypothesize that ultimately there is a ceiling to the quality that can be achieved through these approaches, given the long-tail of different scenarios that can arise within a mixed-initiative CRS. In this section we discuss the use of LLM-powered user simulators to generate realistic data at scale and techniques for tuning system components using larger amounts of data.

\subsection{User Simulation}
\label{simulator_section}

\begin{figure}[h]
  \centering
  \includegraphics[width=0.6\linewidth]{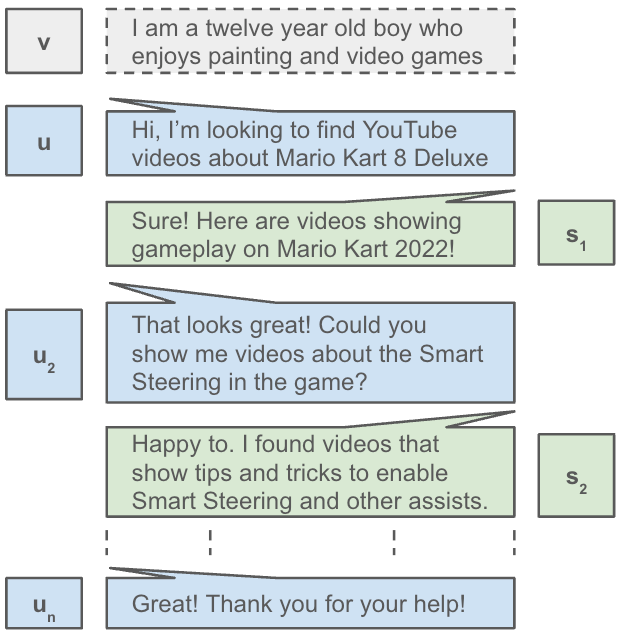}
  \caption{An example of session based control: A single variable (a user profile) is used to condition the user simulator.}
\end{figure}

\begin{figure}[h]
  \centering
  \includegraphics[width=0.8\linewidth]{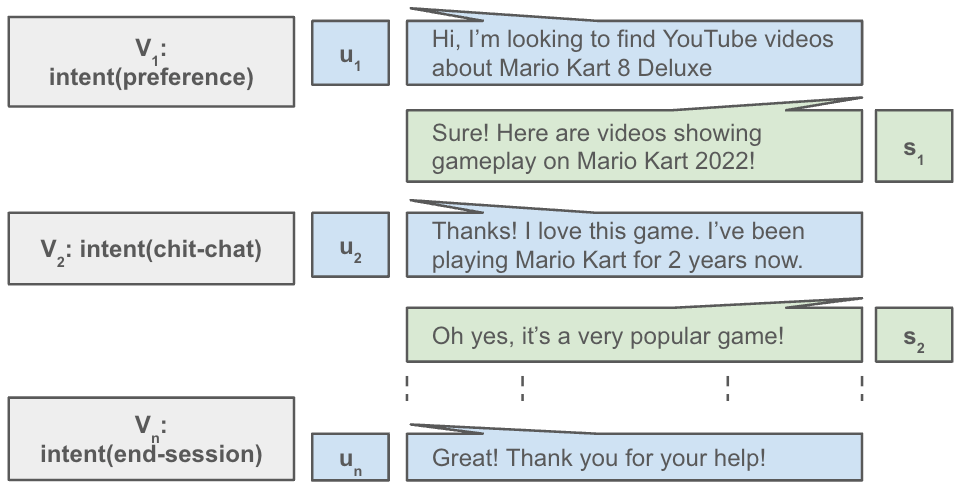}
  \caption{An example of turn level control: A series of variables (user intents) are used to condition the user simulator at each turn.}
\end{figure}

According to the conversational recommender setup considered in this paper (see Section \ref{problem_scope_section}), a session consists of a sequence $S = \{ s_1, u_1, s_2, u_2, ... ,s_n, u_n \}$, where each $u_i$ is a natural language utterance by the user and each $s_i$ is a combination of a natural language utterance and possibly a slate of recommendations by the CRS.  Therefore, a user simulator is defined by a function $f(S') = U_i$, where $S' = \{ s_1, u_1, s_2, u_2, ... ,s_i \}$ is a partial session and $U_i$ is a distribution over possible user utterances $u_i$ continuing the session.  Given a fixed CRS and such a user simulator $f$, we can generate a new sample session by having the CRS and $f$ interact for a given number of turns (i.e. the CRS generates each $s_i$ and $f$ generates each $u_i$).  

The ideal property we would like our user simulator to have when synthetically generating data for evaluation or training is \textbf{realism}: Conversations between the user simulator and CRS should be nearly indistinguishable from conversations between a representative group of real users and the CRS.  Let $\mathbf{R}$ be a set of sessions generated by having real users interact with a particular CRS, and $\mathbf{Q}$ be a set of simulated sessions sampled from the CRS and a user simulator $f$ according to the procedure outlined above.  We offer three possible ways to measure the realism of $f$:
\begin{itemize}
\item Have crowdsource workers attempt to distinguish between simulated sessions coming from $\mathbf{Q}$ and real sessions coming from $\mathbf{R}$.
\item Train a discriminator model \cite{goodfellow2020generative} on the same differentiation task.
\item Let $g(S) \rightarrow [1, k]$ be a function that classifies a session into $k$ categories and let $G = { \{ g_i \} }$ be an ensemble of such classifiers. One way to define such an ensemble is by adapting dialogue state tracking artifacts used within the dialogue management module of a CRS (see Section \ref{dialogue_management_section}).  For instance, we can have a classifier that labels the user intent at a specific turn, or the topics that are covered within a session, or the primary sentiment of a session.  Once defined, we can measure how close the distributions $\mathbf{Q}$ and $\mathbf{R}$ are by matching statistics according to the classifier ensemble $G$.
\end{itemize}

A necessary condition of realism is \textbf{diversity}: Simulated sessions from $\mathbf{Q}$ should have sufficient variation to invoke all the different functionality of a CRS users will encounter in practice when using the system. It may be that in certain situations measuring realism directly is difficult, for instance if collecting a representative set of real user sessions is infeasible.  In this case we can at least attempt to measure the diversity of the user simulator, for instance by defining a notion of entropy of $\mathbf{Q}$ with respect to the classifier ensemble $G$.  

\paragraph{Controlled Simulation}

Our starting point for building a user simulator is the observation that an unconstrained LLM built for dialogue such as LaMDA \cite{thoppilan2022lamda} can interact with a CRS in a similar way to real users.  The LLM takes as input the full history of the ongoing conversation and outputs the next user utterance, analogous to how a CRS dialogue manager can use an LLM to generate system utterances. However, we would like to exhibit greater control over the simulator to increase its realism.  In controlled simulation, we condition the user simulator on additional latent (to the CRS) variables that allow us to guide its behavior in a certain direction. We explore with two different variations:

\begin{itemize}
\item \textbf{Session-level control}: A single variable $v$ is defined at the beginning of the session and is used to condition the user simulator throughout the session.  For instance, we could define $v$ as a user profile such as the ones discussed in Section \ref{user_memory_section}.
\item \textbf{Turn-level control}: A distinct variable $v_i$ is defined at each turn of the session and is used to condition the simulator for that turn.  For instance, we could define each $v_i$ to be a user intent for the simulator to adopt at that turn.
\end{itemize}

In the case of an LLM user simulator, one way to execute the control is to translate the variable into text that can be included as part of the simulator's input along with the rest of the conversation. For instance, for the user profile example we could append the statement "I am a twelve year old boy who enjoys painting and video games" to the beginning of the conversation to induce the LLM to imitate this personality.  To increase realism, one possible strategy is to define session-level or turn-level variables in terms of the classifiers making up one of the ensembles $G$ discussed above and then to sample the variables according to the empirical distribution of the collection of real user sessions $\mathbf{R}$.  Another possibility is to ground the conditioning in trajectories coming from real data from a related product.  For instance, we could look at query sequences submitted by users in a non-conversational search application and sample turn-level variables as trajectories of topics that match these query sequences.

\paragraph{Generating Synthetic Training Data}

To use a user simulator to generate data for supervised training of one of the CRS system modules an additional property is needed: ground truth labels that the system can learn from.  As a toy example, suppose we are trying to learn a sentiment classifier as part of a traditional dialogue state tracking module.  For this we need to generate a set of examples ${S_i, l_i}$, where $S_i$ is a session $s_1, u_1, s_2, u_2, ... s_n, u_n$ and $l_i$ is a ground truth label for the primary user sentiment within $S_i$ coming from a set of possible labels $L$, e.g  \{angry, satisfied, confused, ...\}.  We can use controlled user simulation to solve this problem, by defining a session level variable $v$ over this set of labels $L$.  First we sample a variable $v$ from $L$ (e.g. "angry") and then condition the simulator based on this label, for instance in a priming implementation by appending the message "You are an angry user" to the beginning of the input of the simulator.  If we are able to solve this LLM control problem effectively then we can attach a label $l_i = $"angry"  to the session $S_i$ and trust that with high probability it will be accurate. 

A more ambitious use case is generating data for training the retrieval and ranking modules discussed in Sections \ref{retrieval_section} and \ref{ranking_section}.  For this we can define a session level variable $v$ as a tuple $(x, j)$, where $x$ is an item from the corpus and $j$ is an integer turn index.  Once we sample a $v = (x, j)$, we condition the simulator to generate a session $S = \{ v, s_1, u_1, s_2, u_2, ..., s_j, u_j, ... \} $ such that after $j$ turns the item $x$ is a good match for the context $S$ (i.e. the user would be satisfied if on turn $s_{j+1}$ the system included $x$ within a recommendation slate).  This session can then be used as an input example for training a recommendation module, where the item $x$ is a positive instance and other items from the corpus can be sampled as negatives. This is a far more complex conditioning problem, and a simple zero-shot priming instruction (e.g. "Generate a session such that after $j$ turns item $x$ is a good match for the context") will not work. How to solve this control problem effectively, either through more sophisticated turn level priming or by tuning the user simulator LLM, is an ongoing research effort.

\subsection{Tuning System Modules}
\label{tuning_modules_section}

For the remainder of this section we focus on tuning LLMs within our system using large amounts of synthetically generated data.  For concreteness we examine three modules discussed earlier in the paper: Retrieval (Section \ref{retrieval_section}), Ranking / Explanation (Section \ref{ranking_section}), and Dialogue Management (Section \ref{dialogue_management_section}).

\paragraph{Retrieval}

In Section \ref{simulator_section} we outlined a strategy for generating synthetic training data for tuning a recommendation module.  For retrieval we assume our training examples are tuples of the form $(S', x_{pos}, \{x_{neg}\})$, where $S'$ is a partial session $s_1, u_1, s_2, u_2, ... s_i, u_i$, $x_{pos}$ is an item that is a good match for the context $S'$ (in the sense defined previously) and $\{x_{neg}\}$ is a set of negative items generated by some negative sampling procedure. Given this data, we can tune a Generalized Dual Encoder Model (see Section \ref{retrieval_section}), in which the initial context representation and item representations are each encoded by an LLM.  Regardless of whether we choose to tune only the adapter layers of the two tower model or the LLM params as well, the loss is fully differentiable and normal supervised learning with gradient descent suffices.  

\begin{figure*}[h]
  \centering
  \includegraphics[width=\textwidth]{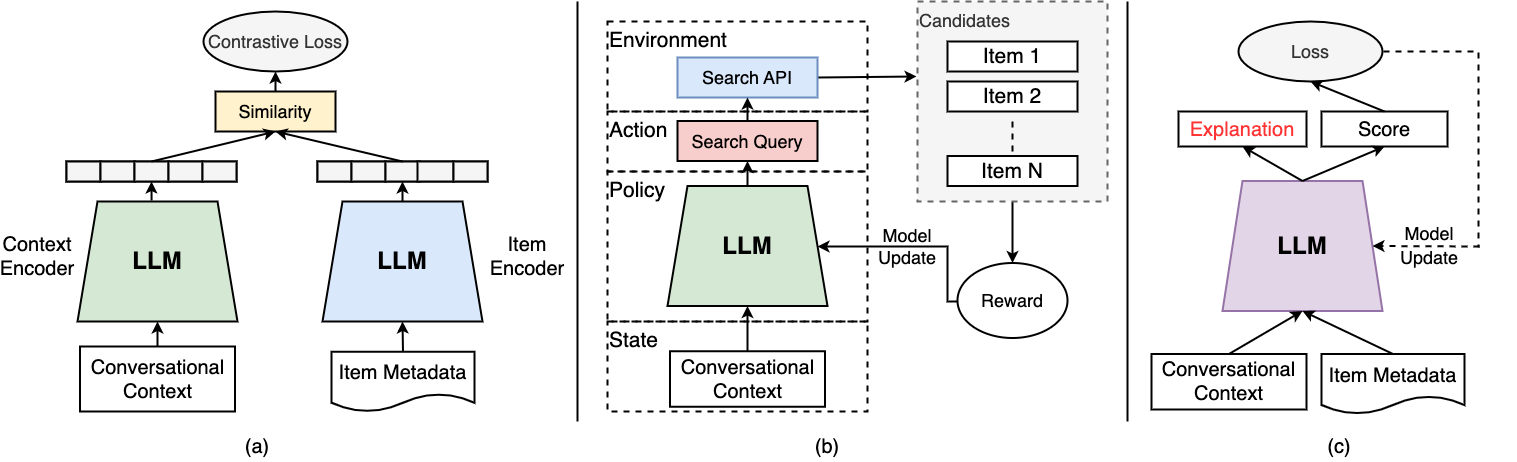}
  \caption{Tuning Recommendation Modules: (a) Tuning a General Dual Encoder retrieval model. (b) Tuning a Search API Lookup retrieval model, framed as a contextual bandits problem. (c) Tuning a joint ranking / explanation model.  The only learning signal comes from ground truth scores, but through self-consistency / bootstrapping tricks it is possible to indirectly tune the explanations as well.}
\label{tuning_figure}
\end{figure*}

In Search API Lookup (see Section \ref{retrieval_section}), an LLM processes the session history and outputs a search query, which then gets passed into a black-box search algorithm.  When this architecture is used, the loss is no longer differentiable and ordinary supervised learning is not possible. Instead, we can reframe the setup as a contextual bandit problem \cite{bouneffouf2020survey}, where the LLM is a policy, the labels are rewards signals, and the black box search algorithm is treated as the environment (see Figure \ref{tuning_figure}b).  If the LLM encoder is shared with other modules we have the choice of tuning protected parameters of the LLM that influence only this task of outputting a search query, or instead tuning shared parameters of the LLM that also influence the behavior of these other modules.

\paragraph{Ranking}

For this use case we assume our training examples are tuples of the form $(S', Y)$, where $S'$ is a partial session $s_1, u_1, s_2, u_2, ..., s_i$ such that $s_i$ contains a recommendation slate and $Y$ is a list of relevancy scores for the items in that slate.  In Section \ref{ranking_section} we present an LLM based ranking module that jointly generates a score for each item and an explanation for that score.  Using this data, we can tune the ranking LLM to predict the ground truth labels as a regression problem.  Using only this relevancy data we cannot directly tune the LLM to generate better explanations, although this is still possible using bootstrapping methods that depend only on labels for the end task (in this case the scoring task) \cite{zelikman2022star, huang2022large}.

\paragraph{Dialogue Management}

In Section \ref{dialogue_management_section} we present a dialogue management module based on a unified LLM that at each turn generates a sequence of natural language outputs, the last of which triggers a system action such as a response to the user or an API call to the recommendation engine.  Our initial implementation involves tuning the LLM on a moderate (e.g. O(1000)) number of example sequences meant to demonstrate desired behavior.  Here we propose a Reinforcement Learning from Human Feedback (see e.g. \cite{ouyang2022training}) strategy for building on this medium-scale tuning:
\begin{enumerate}
\item Generate a set of simulated sessions $\mathbf{Q}$ using a user simulator as outlined in Section \ref{simulator_section}
\item
Have crowdsource workers evaluate our unified LLM by rating per turn responses within $\mathbf{Q}$ in terms of fluency, interestingness, groundedness etc, as well as giving session level ratings based on overall how effective the system was at helping the user explore the recommendations corpus
\item
Train reward models  on this rating data (likely also using LLMs with chain-of-thought reasoning).
\item
Further tune the unified LLM on simulated sessions through reinforcement learning to optimize for proxy rewards generated by these reward models
\end{enumerate}

\section{Related Work}
\label{related_work_section}

In this section we briefly survey prior research related to the main topics covered in this paper; for a more comprehensive treatment covering CRSs and other conversational information seeking applications, see for instance \cite{jannach2021survey, gao2021advances, zamani2022conversational, gao2022neural}

\paragraph{Large Language Models}
Large Language Models based on the transformer architecture \cite{vaswani2017attention} have revolutionized the field of artificial intelligence in recent years, producing state of the art results across a variety of natural language understanding and dialogue tasks \cite{brown2020language, raffel2020exploring, thoppilan2022lamda, chowdhery2022palm}.  In particular these models excel in the zero or few-shot learning setting, where through appropriately engineered prompts they can be adapted to novel tasks without modifying the model parameters \cite{reynolds2021prompt}.  When more training data is available, parameter efficient tuning methods such as prompt tuning can achieve even better performance while still enabling a single LLM to handle multiple sub tasks. \cite{lester2021power, vu2021spot}.  LLMs have also demonstrated an exciting ability to execute multi-step reasoning using chain of thought prompting \cite{wei2022chain}, a faculty that appears to emerge only at certain model scales \cite{wei2022emergent}.  A number of recent papers have employed self-consistency and boosting techniques to amplify this reasoning potential further \cite{wang2022self, zelikman2022star, li2022advance}. One challenge explored in this paper is how to effectively harness these new skills within the conversational recommender space, where scarcity of available training data places a high premium on these types of sample efficient learning methods.  

\paragraph{CRS Datasets}
Evaluation of CRSs is difficult in part due to the generative and open-ended nature of the mixed-initiative dialogue (see for example \cite{jannach2021survey} for a more detailed discussion). Many popular public CRS datasets are based around relatively small domains such as movies and are conversation-only, i.e. recommendations are offered by the system directly within the dialogue without a notion of recommendation slates or other visual elements (see e.g. \cite{li2018towards, zhou2020towards, hayati2020inspired,kang2019recommendation}). These datasets rely on crowdsource workers to provide examples of good system utterances and recommendations that are treated as ground truth. A few other CRS datasets are adapted from non-conversational datasets involving large-scale domains such as e-commerce or Yelp reviews \cite{zhang2018towards, lei2020estimation, zhou2020towards}.  However in this case the conversations are generated either synthetically or through substitution from other sources, and are overly rigid compared to actual human dialogue.  In this paper we focus on recommendations over a large-scale corpus with recommendation slates distinct from the dialogue, a setup that doesn't fit cleanly into any existing offline benchmarks. As future work we are planning to release human evaluations and a public dataset to quantitatively evaluate design alternatives within RecLLM.

\paragraph{Dialogue Management}
Early CRSs did not rely on natural language but instead on simple forms of preference elicitation by the system and "critiquing" by the user \cite{linden1997interactive, burke1997findme, mccarthy2010experience}. When conversational recommenders with natural language interfaces first emerged, they were highly rule-based and limited in their ability to handle mixed-initiative interactions \cite{thompson2004personalized, belkin1995cases}.  Later, CRSs with model based language generation and understanding appeared, although they tended to still be narrowly focused on issues such as when to ask a question of the user versus showing recommendations, and what questions to ask \cite{zhang2020conversational, christakopoulou2018q, sun2018conversational, zhang2018towards}.  Other works have explored learning more flexible dialogue management modules end-to-end, usually by fine-tuning language models on dialogues collected from crowdsource workers \cite{li2018towards, chen2019towards, kang2019recommendation, penha2020does}, although a recent study has indicated that more progress is needed to make these systems practically useful \cite{jannach2020end}.  In some cases the end-to-end approach has been extended to jointly train a separate item recommendation module along with the dialogue \cite{lei2020estimation, lei2020interactive, deng2021unified, recindial}.  The unified LLM dialog management architecture from Section \ref{dialogue_management_section} builds on this prior work by:
\begin{itemize} 
\item Integrating with a recommendation module that can handle a large scale corpus.
\item Learning internal natural language representations such as dialogue state tracking artifacts and self-instructions along with the final dialogue utterances.
\item Incorporating natural language inputs such as user profiles and textual representations of recommendation slates from external sources.
\end{itemize}

\paragraph{Recommendations / Explanations}
In \cite{zamani2022retrieval} the authors define a conceptual framework for Retrieval Enhanced Machine Learning; our framework defined in Section \ref{retrieval_section} is similar in nature but is simplified and focused on capturing existing approaches to retrieval in the recommendation domain.  An overall theme of this paper is how to properly integrate LLMs with external resources, particularly recommendation engines and user profiles, in order to build a better CRS. Some prior research \cite{thoppilan2022lamda, nakano2021webgpt, byrne2020tickettalk, rastogi2020towards} explores with tuning conversational systems through human demonstrations to make calls to an external search API, but not for recommendations over a large corpus.  More generally, it is a fundamental research area in machine learning to augment deep learning models with external memory \cite{weston2014memory, graves2014neural, Wu2022MemorizingT}, and it has been demonstrated that giving LLMs the ability to retrieve from external corpora can improve performance on tasks like question answering and reduce hallucinations \cite{lewis2020retrieval, borgeaud2021improving, shuster2021retrieval}.

Explainability has been a longstanding concern in recommender systems \cite{zhang2020explainable} and a number of works have previously explored jointly generating explanations and recommendations in more traditional recommender systems \cite{zhang2014explicit, chen2018sequential, chen2019dynamic, wang2018reinforcement, mcinerney2018explore, chen2019towards, gao2019explainable}.  Recently, LLMs have been used to explain classifiers and also boost their performance \cite{narang2020wt5, rajani2019explain, lampinen2022can}.  LLMs have also been used for document ranking \cite{nogueira2020document, ju2021text, pradeep2021expando}; however, we are not aware of previous attempts to apply them to ranking problems in the CRS setting or over large-scale corpora where items are represented by heterogeneous metadata, as we do within RecLLM. What type of explanations a recommender system should share is a difficult question (see e.g. \cite{gedikli2014should, park2019design}); in RecLLM we currently have the system give post hoc natural language justifications for item slates, although this still leaves open the question of how to verify their correctness.

\paragraph{User Profile}
A number of recent works explore extracting transparent natural language user profiles in order to personalize open-ended chat bots \cite{zhang2018personalizing, xu2021beyond, mazare2018training}, and recommender systems \cite{balog2019transparent, radlinski2022natural, torbati2021you}.  Our proposal from Section \ref{user_memory_section} is perhaps most closely related to BlenderBot \cite{shuster2022blenderbot}, which also breaks the problem down into separate extraction, triggering, retrieval and generation phases.

\paragraph{Simulation / Large scale training}
Various user simulators have been built for training and evaluating recommender systems, often to support experimentation with reinforcement learning algorithms \cite{ie2019recsim, zhang2020evaluating, rohde2018recogym, zou2020pseudo}.  Recently there has also been a surge in research using LLMs to generate synthetic data for training dialogue systems and text classifiers \cite{dai2022dialog, mehri2022lad, puri2020training, yoo2021gpt3mix}.  Particularly relevant is Unsupervised Data Generation \cite{wang2021towards}, in which an LLM takes a description of a desired label and then generates an input that fits the label.  This input / label pair then becomes a synthetic example that can be used for training.  Controlled simulation from section \ref{simulator_section} employs a similar principle where we condition on a latent variable to generate a simulated session and then use the latent variable as a label for tuning.  However, we are attempting to generate entire conversations (partially generated by a system outside the simulator's control) and more sophisticated techniques than basic few-shot prompting are likely required.

In \cite{xiong2020approximate, hofstatter2021efficiently, ni2021large} a pretrained language model is tuned to process documents as part of a dual encoder retrieval model, and in \cite{henderson2019convert} this is extended to full conversations as in the Generalized Dual Encoder proposal from Section \ref{tuning_modules_section}.  When the ground truth labels do not enable a fully differentiable loss function (such as in Search API Lookup), \cite{stiennon2020learning, ouyang2022training} show it is still effective to tune LLMs for language generation tasks using techniques derived from reinforcement learning.  Other works \cite{chow2022mixture, snell2022context} also use reinforcement learning to tune LLMs for open ended or task based dialogue using reward signals inferred from the conversations (e.g. through sentiment analysis or a notion of task completion). The proposal for tuning a dialogue manager LLM in Section \ref{tuning_modules_section} is an example of Reinforcement Learning from Human Feedback \cite{bai2022training, glaese2022improving, ouyang2022training}, a technique that is often used for teaching LLMs to follow instructions and align better with human values.

\section{RecLLM Prototype}
\label{examples_section}
We have built an initial RecLLM prototype based on the outline shared within this paper.  Retrieval is currently implemented via Search API Lookup (see Section \ref{retrieval_section}) using in-context few-shot learning and a public YouTube search API.  LaMDA \cite{thoppilan2022lamda} is currently used as the underlying LLM powering dialogue management, recommendations and explanations, user profile integration and user simulation within the system. In Appendix \ref{recllm_samples} we share sample sessions from RecLLM demonstrating some of its core competencies.

\section{Ethical Considerations}
\label{ethical_considerations_section}
It is our belief that by leveraging large language models within CRSs we can mitigate some challenging ethical problems that have been noted in many recommender systems.  RecLLM has the following desirable properties:
\begin{itemize} 
\item A recommendation module that reasons over the attributes of items and is less reliant on learning from interaction data such as clicks that are noisy and can promote unintentional biases.
\item The ability to give natural language justifications for why certain recommendations are being shown, which the user can then validate.
\item Opportunity for the user to control their recommendations in a nuanced way through language.
\item Transparent personalization through human interpretable and editable user profiles.
\end{itemize}

On the other hand, our proposed system relies heavily on large language models and therefore inherits all of their well-known problems centered around societal biases learned through pretraining, hallucinations, and expensive use of resources \cite{weidinger2021ethical}.  Various controls are included to constrain the LLMs to the conversational recommender task, but these are unlikely to fully wash away their inherent issues.  Significant further progress needs to be made in areas like debiasing, grounding in factuality and efficient serving before we can safely deploy this type of system in a production setting.

\section{Conclusions and Future Work}
\label{conclusions_and_future_work_section}

In this paper we examine the system architecture of a conversational recommender system and identify areas where large language models can unlock new capabilities, along with the technical challenges that emerge through their use.  In particular we reimagine how LLMs can transform dialogue management, retrieval, ranking and user profiles to improve system quality, give the user greater control and increase transparency throughout the system.  We focus on how to build a large-scale end-to-end CRS without assuming access to logs data coming from an existing product, by utilizing the generalization abilities of LLMs and generating synthetic training data using LLM-powered user simulators.  As a proof of concept we introduce RecLLM and share example conversations highlighting its diverse functionality.  Our hope is that this roadmap can accelerate progress towards a world where controllable and explainable CRSs allow users to explore content within a healthier recommender system ecosystem.

Some important items for future work include:
\begin{itemize}
\item We are planning the release of human evaluations and a public dataset based on our system to quantitatively evaluate design alternatives for RecLLM and help the community better study CRSs in the multimodal, large-scale setting.
\item In this paper we assume a simplified setting where users interact with the system only through conversation.  We would like to generalize our system to handle more realistic scenarios where users give feedback through other channels as well such as clicking on items or like buttons.  We would also like to consider more complicated recommender system UIs containing hierarchical structures such as item shelves as opposed to just flat slates.
\item We have proposed ideas for large-scale tuning of the main system modules based on synthetically generated data, but currently RecLLM relies exclusively on in-context few-shot learning or tuning on small amounts of data collected through crowdsourcing. Successfully proving out these ideas will be critical to properly handle huge item corpora and the full space of possible conversations.
\item We would like to support new use cases that naturally arise in a mixed-initiative conversational recommender dialogue, such as question answering over corpus items.
\end{itemize}

\section{Acknowledgements}
We would like to thank Filip Radlinski, Karan Singhal, Abhinav Rastogi, Raghav Gupta and Yinlam Chow for useful feedback on drafts of this paper.

\bibliographystyle{ACM-Reference-Format}
\bibliography{mybib}


\begin{thebibliography}{115}


\ifx \showCODEN    \undefined \def \showCODEN     #1{\unskip}     \fi
\ifx \showDOI      \undefined \def \showDOI       #1{#1}\fi
\ifx \showISBNx    \undefined \def \showISBNx     #1{\unskip}     \fi
\ifx \showISBNxiii \undefined \def \showISBNxiii  #1{\unskip}     \fi
\ifx \showISSN     \undefined \def \showISSN      #1{\unskip}     \fi
\ifx \showLCCN     \undefined \def \showLCCN      #1{\unskip}     \fi
\ifx \shownote     \undefined \def \shownote      #1{#1}          \fi
\ifx \showarticletitle \undefined \def \showarticletitle #1{#1}   \fi
\ifx \showURL      \undefined \def \showURL       {\relax}        \fi
\providecommand\bibfield[2]{#2}
\providecommand\bibinfo[2]{#2}
\providecommand\natexlab[1]{#1}
\providecommand\showeprint[2][]{arXiv:#2}

\bibitem[\protect\citeauthoryear{Bai, Jones, Ndousse, Askell, Chen, DasSarma,
  Drain, Fort, Ganguli, Henighan, et~al\mbox{.}}{Bai et~al\mbox{.}}{2022}]%
        {bai2022training}
\bibfield{author}{\bibinfo{person}{Yuntao Bai}, \bibinfo{person}{Andy Jones},
  \bibinfo{person}{Kamal Ndousse}, \bibinfo{person}{Amanda Askell},
  \bibinfo{person}{Anna Chen}, \bibinfo{person}{Nova DasSarma},
  \bibinfo{person}{Dawn Drain}, \bibinfo{person}{Stanislav Fort},
  \bibinfo{person}{Deep Ganguli}, \bibinfo{person}{Tom Henighan},
  {et~al\mbox{.}}} \bibinfo{year}{2022}\natexlab{}.
\newblock \showarticletitle{Training a helpful and harmless assistant with
  reinforcement learning from human feedback}.
\newblock \bibinfo{journal}{\emph{arXiv preprint arXiv:2204.05862}}
  (\bibinfo{year}{2022}).
\newblock


\bibitem[\protect\citeauthoryear{Balog, Radlinski, and Arakelyan}{Balog
  et~al\mbox{.}}{2019}]%
        {balog2019transparent}
\bibfield{author}{\bibinfo{person}{Krisztian Balog}, \bibinfo{person}{Filip
  Radlinski}, {and} \bibinfo{person}{Shushan Arakelyan}.}
  \bibinfo{year}{2019}\natexlab{}.
\newblock \showarticletitle{Transparent, scrutable and explainable user models
  for personalized recommendation}. In \bibinfo{booktitle}{\emph{Proceedings of
  the 42nd international acm sigir conference on research and development in
  information retrieval}}. \bibinfo{pages}{265--274}.
\newblock


\bibitem[\protect\citeauthoryear{Belkin, Cool, Stein, and Thiel}{Belkin
  et~al\mbox{.}}{1995}]%
        {belkin1995cases}
\bibfield{author}{\bibinfo{person}{Nicholas~J Belkin}, \bibinfo{person}{Colleen
  Cool}, \bibinfo{person}{Adelheit Stein}, {and} \bibinfo{person}{Ulrich
  Thiel}.} \bibinfo{year}{1995}\natexlab{}.
\newblock \showarticletitle{Cases, scripts, and information-seeking strategies:
  On the design of interactive information retrieval systems}.
\newblock \bibinfo{journal}{\emph{Expert systems with applications}}
  \bibinfo{volume}{9}, \bibinfo{number}{3} (\bibinfo{year}{1995}),
  \bibinfo{pages}{379--395}.
\newblock


\bibitem[\protect\citeauthoryear{Borgeaud, Mensch, Hoffmann, Cai, Rutherford,
  Millican, Driessche, Lespiau, Damoc, Clark, et~al\mbox{.}}{Borgeaud
  et~al\mbox{.}}{2021}]%
        {borgeaud2021improving}
\bibfield{author}{\bibinfo{person}{Sebastian Borgeaud}, \bibinfo{person}{Arthur
  Mensch}, \bibinfo{person}{Jordan Hoffmann}, \bibinfo{person}{Trevor Cai},
  \bibinfo{person}{Eliza Rutherford}, \bibinfo{person}{Katie Millican},
  \bibinfo{person}{George van~den Driessche}, \bibinfo{person}{Jean-Baptiste
  Lespiau}, \bibinfo{person}{Bogdan Damoc}, \bibinfo{person}{Aidan Clark},
  {et~al\mbox{.}}} \bibinfo{year}{2021}\natexlab{}.
\newblock \showarticletitle{Improving language models by retrieving from
  trillions of tokens}.
\newblock \bibinfo{journal}{\emph{arXiv preprint arXiv:2112.04426}}
  (\bibinfo{year}{2021}).
\newblock


\bibitem[\protect\citeauthoryear{Bouneffouf, Rish, and Aggarwal}{Bouneffouf
  et~al\mbox{.}}{2020}]%
        {bouneffouf2020survey}
\bibfield{author}{\bibinfo{person}{Djallel Bouneffouf}, \bibinfo{person}{Irina
  Rish}, {and} \bibinfo{person}{Charu Aggarwal}.}
  \bibinfo{year}{2020}\natexlab{}.
\newblock \showarticletitle{Survey on applications of multi-armed and
  contextual bandits}. In \bibinfo{booktitle}{\emph{2020 IEEE Congress on
  Evolutionary Computation (CEC)}}. IEEE, \bibinfo{pages}{1--8}.
\newblock


\bibitem[\protect\citeauthoryear{Brown, Mann, Ryder, Subbiah, Kaplan, Dhariwal,
  Neelakantan, Shyam, Sastry, Askell, et~al\mbox{.}}{Brown
  et~al\mbox{.}}{2020}]%
        {brown2020language}
\bibfield{author}{\bibinfo{person}{Tom Brown}, \bibinfo{person}{Benjamin Mann},
  \bibinfo{person}{Nick Ryder}, \bibinfo{person}{Melanie Subbiah},
  \bibinfo{person}{Jared~D Kaplan}, \bibinfo{person}{Prafulla Dhariwal},
  \bibinfo{person}{Arvind Neelakantan}, \bibinfo{person}{Pranav Shyam},
  \bibinfo{person}{Girish Sastry}, \bibinfo{person}{Amanda Askell},
  {et~al\mbox{.}}} \bibinfo{year}{2020}\natexlab{}.
\newblock \showarticletitle{Language models are few-shot learners}.
\newblock \bibinfo{journal}{\emph{Advances in neural information processing
  systems}}  \bibinfo{volume}{33} (\bibinfo{year}{2020}),
  \bibinfo{pages}{1877--1901}.
\newblock


\bibitem[\protect\citeauthoryear{Burke, Hammond, and Yound}{Burke
  et~al\mbox{.}}{1997}]%
        {burke1997findme}
\bibfield{author}{\bibinfo{person}{Robin~D Burke}, \bibinfo{person}{Kristian~J
  Hammond}, {and} \bibinfo{person}{BC Yound}.} \bibinfo{year}{1997}\natexlab{}.
\newblock \showarticletitle{The FindMe approach to assisted browsing}.
\newblock \bibinfo{journal}{\emph{IEEE Expert}} \bibinfo{volume}{12},
  \bibinfo{number}{4} (\bibinfo{year}{1997}), \bibinfo{pages}{32--40}.
\newblock


\bibitem[\protect\citeauthoryear{Byrne, Krishnamoorthi, Ganesh, and Kale}{Byrne
  et~al\mbox{.}}{2020}]%
        {byrne2020tickettalk}
\bibfield{author}{\bibinfo{person}{Bill Byrne}, \bibinfo{person}{Karthik
  Krishnamoorthi}, \bibinfo{person}{Saravanan Ganesh}, {and}
  \bibinfo{person}{Mihir~Sanjay Kale}.} \bibinfo{year}{2020}\natexlab{}.
\newblock \showarticletitle{TicketTalk: Toward human-level performance with
  end-to-end, transaction-based dialog systems}.
\newblock \bibinfo{journal}{\emph{arXiv preprint arXiv:2012.12458}}
  (\bibinfo{year}{2020}).
\newblock


\bibitem[\protect\citeauthoryear{Cai and Chen}{Cai and Chen}{2020}]%
        {cai2020predicting}
\bibfield{author}{\bibinfo{person}{Wanling Cai} {and} \bibinfo{person}{Li
  Chen}.} \bibinfo{year}{2020}\natexlab{}.
\newblock \showarticletitle{Predicting user intents and satisfaction with
  dialogue-based conversational recommendations}. In
  \bibinfo{booktitle}{\emph{Proceedings of the 28th ACM Conference on User
  Modeling, Adaptation and Personalization}}. \bibinfo{pages}{33--42}.
\newblock


\bibitem[\protect\citeauthoryear{Cao, Qin, Liu, Tsai, and Li}{Cao
  et~al\mbox{.}}{2007}]%
        {cao2007learning}
\bibfield{author}{\bibinfo{person}{Zhe Cao}, \bibinfo{person}{Tao Qin},
  \bibinfo{person}{Tie-Yan Liu}, \bibinfo{person}{Ming-Feng Tsai}, {and}
  \bibinfo{person}{Hang Li}.} \bibinfo{year}{2007}\natexlab{}.
\newblock \showarticletitle{Learning to rank: from pairwise approach to
  listwise approach}. In \bibinfo{booktitle}{\emph{Proceedings of the 24th
  international conference on Machine learning}}. \bibinfo{pages}{129--136}.
\newblock


\bibitem[\protect\citeauthoryear{Chen, Lin, Zhang, Ding, Cen, Yang, and
  Tang}{Chen et~al\mbox{.}}{2019a}]%
        {chen2019towards}
\bibfield{author}{\bibinfo{person}{Qibin Chen}, \bibinfo{person}{Junyang Lin},
  \bibinfo{person}{Yichang Zhang}, \bibinfo{person}{Ming Ding},
  \bibinfo{person}{Yukuo Cen}, \bibinfo{person}{Hongxia Yang}, {and}
  \bibinfo{person}{Jie Tang}.} \bibinfo{year}{2019}\natexlab{a}.
\newblock \showarticletitle{Towards knowledge-based recommender dialog system}.
\newblock \bibinfo{journal}{\emph{arXiv preprint arXiv:1908.05391}}
  (\bibinfo{year}{2019}).
\newblock


\bibitem[\protect\citeauthoryear{Chen, Xu, Zhang, Tang, Cao, Qin, and Zha}{Chen
  et~al\mbox{.}}{2018}]%
        {chen2018sequential}
\bibfield{author}{\bibinfo{person}{Xu Chen}, \bibinfo{person}{Hongteng Xu},
  \bibinfo{person}{Yongfeng Zhang}, \bibinfo{person}{Jiaxi Tang},
  \bibinfo{person}{Yixin Cao}, \bibinfo{person}{Zheng Qin}, {and}
  \bibinfo{person}{Hongyuan Zha}.} \bibinfo{year}{2018}\natexlab{}.
\newblock \showarticletitle{Sequential recommendation with user memory
  networks}. In \bibinfo{booktitle}{\emph{Proceedings of the eleventh ACM
  international conference on web search and data mining}}.
  \bibinfo{pages}{108--116}.
\newblock


\bibitem[\protect\citeauthoryear{Chen, Zhang, and Qin}{Chen
  et~al\mbox{.}}{2019b}]%
        {chen2019dynamic}
\bibfield{author}{\bibinfo{person}{Xu Chen}, \bibinfo{person}{Yongfeng Zhang},
  {and} \bibinfo{person}{Zheng Qin}.} \bibinfo{year}{2019}\natexlab{b}.
\newblock \showarticletitle{Dynamic explainable recommendation based on neural
  attentive models}. In \bibinfo{booktitle}{\emph{Proceedings of the AAAI
  Conference on Artificial Intelligence}}, Vol.~\bibinfo{volume}{33}.
  \bibinfo{pages}{53--60}.
\newblock


\bibitem[\protect\citeauthoryear{Chow, Tulepbergenov, Nachum, Ryu, Ghavamzadeh,
  and Boutilier}{Chow et~al\mbox{.}}{2022}]%
        {chow2022mixture}
\bibfield{author}{\bibinfo{person}{Yinlam Chow}, \bibinfo{person}{Aza
  Tulepbergenov}, \bibinfo{person}{Ofir Nachum}, \bibinfo{person}{MoonKyung
  Ryu}, \bibinfo{person}{Mohammad Ghavamzadeh}, {and} \bibinfo{person}{Craig
  Boutilier}.} \bibinfo{year}{2022}\natexlab{}.
\newblock \showarticletitle{A Mixture-of-Expert Approach to RL-based Dialogue
  Management}.
\newblock \bibinfo{journal}{\emph{arXiv preprint arXiv:2206.00059}}
  (\bibinfo{year}{2022}).
\newblock


\bibitem[\protect\citeauthoryear{Chowdhery, Narang, Devlin, Bosma, Mishra,
  Roberts, Barham, Chung, Sutton, Gehrmann, et~al\mbox{.}}{Chowdhery
  et~al\mbox{.}}{2022}]%
        {chowdhery2022palm}
\bibfield{author}{\bibinfo{person}{Aakanksha Chowdhery},
  \bibinfo{person}{Sharan Narang}, \bibinfo{person}{Jacob Devlin},
  \bibinfo{person}{Maarten Bosma}, \bibinfo{person}{Gaurav Mishra},
  \bibinfo{person}{Adam Roberts}, \bibinfo{person}{Paul Barham},
  \bibinfo{person}{Hyung~Won Chung}, \bibinfo{person}{Charles Sutton},
  \bibinfo{person}{Sebastian Gehrmann}, {et~al\mbox{.}}}
  \bibinfo{year}{2022}\natexlab{}.
\newblock \showarticletitle{Palm: Scaling language modeling with pathways}.
\newblock \bibinfo{journal}{\emph{arXiv preprint arXiv:2204.02311}}
  (\bibinfo{year}{2022}).
\newblock


\bibitem[\protect\citeauthoryear{Christakopoulou, Beutel, Li, Jain, and
  Chi}{Christakopoulou et~al\mbox{.}}{2018}]%
        {christakopoulou2018q}
\bibfield{author}{\bibinfo{person}{Konstantina Christakopoulou},
  \bibinfo{person}{Alex Beutel}, \bibinfo{person}{Rui Li},
  \bibinfo{person}{Sagar Jain}, {and} \bibinfo{person}{Ed~H Chi}.}
  \bibinfo{year}{2018}\natexlab{}.
\newblock \showarticletitle{Q\&R: A two-stage approach toward interactive
  recommendation}. In \bibinfo{booktitle}{\emph{Proceedings of the 24th ACM
  SIGKDD International Conference on Knowledge Discovery \& Data Mining}}.
  \bibinfo{pages}{139--148}.
\newblock


\bibitem[\protect\citeauthoryear{Covington, Adams, and Sargin}{Covington
  et~al\mbox{.}}{2016}]%
        {yt_system}
\bibfield{author}{\bibinfo{person}{Paul Covington}, \bibinfo{person}{Jay
  Adams}, {and} \bibinfo{person}{Emre Sargin}.}
  \bibinfo{year}{2016}\natexlab{}.
\newblock \showarticletitle{Deep Neural Networks for YouTube Recommendations}.
  In \bibinfo{booktitle}{\emph{Proceedings of the 10th ACM Conference on
  Recommender Systems}}. \bibinfo{address}{New York, NY, USA}.
\newblock


\bibitem[\protect\citeauthoryear{Dai, Chaganty, Zhao, Amini, Rashid, Green, and
  Guu}{Dai et~al\mbox{.}}{2022}]%
        {dai2022dialog}
\bibfield{author}{\bibinfo{person}{Zhuyun Dai}, \bibinfo{person}{Arun~Tejasvi
  Chaganty}, \bibinfo{person}{Vincent~Y Zhao}, \bibinfo{person}{Aida Amini},
  \bibinfo{person}{Qazi~Mamunur Rashid}, \bibinfo{person}{Mike Green}, {and}
  \bibinfo{person}{Kelvin Guu}.} \bibinfo{year}{2022}\natexlab{}.
\newblock \showarticletitle{Dialog inpainting: Turning documents into dialogs}.
  In \bibinfo{booktitle}{\emph{International Conference on Machine Learning}}.
  PMLR, \bibinfo{pages}{4558--4586}.
\newblock


\bibitem[\protect\citeauthoryear{Deng, Li, Sun, Ding, and Lam}{Deng
  et~al\mbox{.}}{2021}]%
        {deng2021unified}
\bibfield{author}{\bibinfo{person}{Yang Deng}, \bibinfo{person}{Yaliang Li},
  \bibinfo{person}{Fei Sun}, \bibinfo{person}{Bolin Ding}, {and}
  \bibinfo{person}{Wai Lam}.} \bibinfo{year}{2021}\natexlab{}.
\newblock \showarticletitle{Unified conversational recommendation policy
  learning via graph-based reinforcement learning}. In
  \bibinfo{booktitle}{\emph{Proceedings of the 44th International ACM SIGIR
  Conference on Research and Development in Information Retrieval}}.
  \bibinfo{pages}{1431--1441}.
\newblock


\bibitem[\protect\citeauthoryear{Dietz, Myftija, and W{\"o}rndl}{Dietz
  et~al\mbox{.}}{2019}]%
        {dietz2019designing}
\bibfield{author}{\bibinfo{person}{Linus~W Dietz}, \bibinfo{person}{Saadi
  Myftija}, {and} \bibinfo{person}{Wolfgang W{\"o}rndl}.}
  \bibinfo{year}{2019}\natexlab{}.
\newblock \showarticletitle{Designing a conversational travel recommender
  system based on data-driven destination characterization}. In
  \bibinfo{booktitle}{\emph{ACM RecSys workshop on recommenders in tourism}}.
  \bibinfo{pages}{17--21}.
\newblock


\bibitem[\protect\citeauthoryear{Eksombatchai, Jindal, Liu, Liu, Sharma,
  Sugnet, Ulrich, and Leskovec}{Eksombatchai et~al\mbox{.}}{2018}]%
        {eksombatchai2018pixie}
\bibfield{author}{\bibinfo{person}{Chantat Eksombatchai},
  \bibinfo{person}{Pranav Jindal}, \bibinfo{person}{Jerry~Zitao Liu},
  \bibinfo{person}{Yuchen Liu}, \bibinfo{person}{Rahul Sharma},
  \bibinfo{person}{Charles Sugnet}, \bibinfo{person}{Mark Ulrich}, {and}
  \bibinfo{person}{Jure Leskovec}.} \bibinfo{year}{2018}\natexlab{}.
\newblock \showarticletitle{Pixie: A system for recommending 3+ billion items
  to 200+ million users in real-time}. In \bibinfo{booktitle}{\emph{Proceedings
  of the 2018 world wide web conference}}. \bibinfo{pages}{1775--1784}.
\newblock


\bibitem[\protect\citeauthoryear{Gao, Lei, He, de~Rijke, and Chua}{Gao
  et~al\mbox{.}}{2021}]%
        {gao2021advances}
\bibfield{author}{\bibinfo{person}{Chongming Gao}, \bibinfo{person}{Wenqiang
  Lei}, \bibinfo{person}{Xiangnan He}, \bibinfo{person}{Maarten de Rijke},
  {and} \bibinfo{person}{Tat-Seng Chua}.} \bibinfo{year}{2021}\natexlab{}.
\newblock \showarticletitle{Advances and challenges in conversational
  recommender systems: A survey}.
\newblock \bibinfo{journal}{\emph{AI Open}}  \bibinfo{volume}{2}
  (\bibinfo{year}{2021}), \bibinfo{pages}{100--126}.
\newblock


\bibitem[\protect\citeauthoryear{Gao, Wang, Wang, and Xie}{Gao
  et~al\mbox{.}}{2019}]%
        {gao2019explainable}
\bibfield{author}{\bibinfo{person}{Jingyue Gao}, \bibinfo{person}{Xiting Wang},
  \bibinfo{person}{Yasha Wang}, {and} \bibinfo{person}{Xing Xie}.}
  \bibinfo{year}{2019}\natexlab{}.
\newblock \showarticletitle{Explainable recommendation through attentive
  multi-view learning}. In \bibinfo{booktitle}{\emph{Proceedings of the AAAI
  Conference on Artificial Intelligence}}, Vol.~\bibinfo{volume}{33}.
  \bibinfo{pages}{3622--3629}.
\newblock


\bibitem[\protect\citeauthoryear{Gao, Xiong, Bennett, and Craswell}{Gao
  et~al\mbox{.}}{2022}]%
        {gao2022neural}
\bibfield{author}{\bibinfo{person}{Jianfeng Gao}, \bibinfo{person}{Chenyan
  Xiong}, \bibinfo{person}{Paul Bennett}, {and} \bibinfo{person}{Nick
  Craswell}.} \bibinfo{year}{2022}\natexlab{}.
\newblock \showarticletitle{Neural approaches to conversational information
  retrieval}.
\newblock \bibinfo{journal}{\emph{arXiv preprint arXiv:2201.05176}}
  (\bibinfo{year}{2022}).
\newblock


\bibitem[\protect\citeauthoryear{Garimella and Weber}{Garimella and
  Weber}{2017}]%
        {garimella2017long}
\bibfield{author}{\bibinfo{person}{Venkata Rama~Kiran Garimella} {and}
  \bibinfo{person}{Ingmar Weber}.} \bibinfo{year}{2017}\natexlab{}.
\newblock \showarticletitle{A long-term analysis of polarization on Twitter}.
  In \bibinfo{booktitle}{\emph{Eleventh international AAAI conference on web
  and social media}}.
\newblock


\bibitem[\protect\citeauthoryear{Gedikli, Jannach, and Ge}{Gedikli
  et~al\mbox{.}}{2014}]%
        {gedikli2014should}
\bibfield{author}{\bibinfo{person}{Fatih Gedikli}, \bibinfo{person}{Dietmar
  Jannach}, {and} \bibinfo{person}{Mouzhi Ge}.}
  \bibinfo{year}{2014}\natexlab{}.
\newblock \showarticletitle{How should I explain? A comparison of different
  explanation types for recommender systems}.
\newblock \bibinfo{journal}{\emph{International Journal of Human-Computer
  Studies}} \bibinfo{volume}{72}, \bibinfo{number}{4} (\bibinfo{year}{2014}),
  \bibinfo{pages}{367--382}.
\newblock


\bibitem[\protect\citeauthoryear{Glaese, McAleese, Tr{\k{e}}bacz, Aslanides,
  Firoiu, Ewalds, Rauh, Weidinger, Chadwick, Thacker, et~al\mbox{.}}{Glaese
  et~al\mbox{.}}{2022}]%
        {glaese2022improving}
\bibfield{author}{\bibinfo{person}{Amelia Glaese}, \bibinfo{person}{Nat
  McAleese}, \bibinfo{person}{Maja Tr{\k{e}}bacz}, \bibinfo{person}{John
  Aslanides}, \bibinfo{person}{Vlad Firoiu}, \bibinfo{person}{Timo Ewalds},
  \bibinfo{person}{Maribeth Rauh}, \bibinfo{person}{Laura Weidinger},
  \bibinfo{person}{Martin Chadwick}, \bibinfo{person}{Phoebe Thacker},
  {et~al\mbox{.}}} \bibinfo{year}{2022}\natexlab{}.
\newblock \showarticletitle{Improving alignment of dialogue agents via targeted
  human judgements}.
\newblock \bibinfo{journal}{\emph{arXiv preprint arXiv:2209.14375}}
  (\bibinfo{year}{2022}).
\newblock


\bibitem[\protect\citeauthoryear{Goodfellow, Pouget-Abadie, Mirza, Xu,
  Warde-Farley, Ozair, Courville, and Bengio}{Goodfellow et~al\mbox{.}}{2020}]%
        {goodfellow2020generative}
\bibfield{author}{\bibinfo{person}{Ian Goodfellow}, \bibinfo{person}{Jean
  Pouget-Abadie}, \bibinfo{person}{Mehdi Mirza}, \bibinfo{person}{Bing Xu},
  \bibinfo{person}{David Warde-Farley}, \bibinfo{person}{Sherjil Ozair},
  \bibinfo{person}{Aaron Courville}, {and} \bibinfo{person}{Yoshua Bengio}.}
  \bibinfo{year}{2020}\natexlab{}.
\newblock \showarticletitle{Generative adversarial networks}.
\newblock \bibinfo{journal}{\emph{Commun. ACM}} \bibinfo{volume}{63},
  \bibinfo{number}{11} (\bibinfo{year}{2020}), \bibinfo{pages}{139--144}.
\newblock


\bibitem[\protect\citeauthoryear{Graves, Wayne, and Danihelka}{Graves
  et~al\mbox{.}}{2014}]%
        {graves2014neural}
\bibfield{author}{\bibinfo{person}{Alex Graves}, \bibinfo{person}{Greg Wayne},
  {and} \bibinfo{person}{Ivo Danihelka}.} \bibinfo{year}{2014}\natexlab{}.
\newblock \showarticletitle{Neural turing machines}.
\newblock \bibinfo{journal}{\emph{arXiv preprint arXiv:1410.5401}}
  (\bibinfo{year}{2014}).
\newblock


\bibitem[\protect\citeauthoryear{Guo, Geng, Simcha, Chern, Kumar, and Wu}{Guo
  et~al\mbox{.}}{2019}]%
        {DBLP:journals/corr/abs-1908-10396}
\bibfield{author}{\bibinfo{person}{Ruiqi Guo}, \bibinfo{person}{Quan Geng},
  \bibinfo{person}{David Simcha}, \bibinfo{person}{Felix Chern},
  \bibinfo{person}{Sanjiv Kumar}, {and} \bibinfo{person}{Xiang Wu}.}
  \bibinfo{year}{2019}\natexlab{}.
\newblock \showarticletitle{New Loss Functions for Fast Maximum Inner Product
  Search}.
\newblock \bibinfo{journal}{\emph{CoRR}}  \bibinfo{volume}{abs/1908.10396}
  (\bibinfo{year}{2019}).
\newblock
\showeprint[arXiv]{1908.10396}
\urldef\tempurl%
\url{http://arxiv.org/abs/1908.10396}
\showURL{%
\tempurl}


\bibitem[\protect\citeauthoryear{Hayati, Kang, Zhu, Shi, and Yu}{Hayati
  et~al\mbox{.}}{2020}]%
        {hayati2020inspired}
\bibfield{author}{\bibinfo{person}{Shirley~Anugrah Hayati},
  \bibinfo{person}{Dongyeop Kang}, \bibinfo{person}{Qingxiaoyang Zhu},
  \bibinfo{person}{Weiyan Shi}, {and} \bibinfo{person}{Zhou Yu}.}
  \bibinfo{year}{2020}\natexlab{}.
\newblock \showarticletitle{INSPIRED: Toward sociable recommendation dialog
  systems}.
\newblock \bibinfo{journal}{\emph{arXiv preprint arXiv:2009.14306}}
  (\bibinfo{year}{2020}).
\newblock


\bibitem[\protect\citeauthoryear{Henderson, Casanueva, Mrk{\v{s}}i{\'c}, Su,
  Wen, and Vuli{\'c}}{Henderson et~al\mbox{.}}{2019}]%
        {henderson2019convert}
\bibfield{author}{\bibinfo{person}{Matthew Henderson},
  \bibinfo{person}{I{\~n}igo Casanueva}, \bibinfo{person}{Nikola
  Mrk{\v{s}}i{\'c}}, \bibinfo{person}{Pei-Hao Su}, \bibinfo{person}{Tsung-Hsien
  Wen}, {and} \bibinfo{person}{Ivan Vuli{\'c}}.}
  \bibinfo{year}{2019}\natexlab{}.
\newblock \showarticletitle{ConveRT: Efficient and accurate conversational
  representations from transformers}.
\newblock \bibinfo{journal}{\emph{arXiv preprint arXiv:1911.03688}}
  (\bibinfo{year}{2019}).
\newblock


\bibitem[\protect\citeauthoryear{Hofst{\"a}tter, Lin, Yang, Lin, and
  Hanbury}{Hofst{\"a}tter et~al\mbox{.}}{2021}]%
        {hofstatter2021efficiently}
\bibfield{author}{\bibinfo{person}{Sebastian Hofst{\"a}tter},
  \bibinfo{person}{Sheng-Chieh Lin}, \bibinfo{person}{Jheng-Hong Yang},
  \bibinfo{person}{Jimmy Lin}, {and} \bibinfo{person}{Allan Hanbury}.}
  \bibinfo{year}{2021}\natexlab{}.
\newblock \showarticletitle{Efficiently teaching an effective dense retriever
  with balanced topic aware sampling}. In \bibinfo{booktitle}{\emph{Proceedings
  of the 44th International ACM SIGIR Conference on Research and Development in
  Information Retrieval}}. \bibinfo{pages}{113--122}.
\newblock


\bibitem[\protect\citeauthoryear{Huang and Chang}{Huang and Chang}{2022}]%
        {huang2022towards}
\bibfield{author}{\bibinfo{person}{Jie Huang} {and} \bibinfo{person}{Kevin
  Chen-Chuan Chang}.} \bibinfo{year}{2022}\natexlab{}.
\newblock \showarticletitle{Towards Reasoning in Large Language Models: A
  Survey}.
\newblock \bibinfo{journal}{\emph{arXiv preprint arXiv:2212.10403}}
  (\bibinfo{year}{2022}).
\newblock


\bibitem[\protect\citeauthoryear{Huang, Gu, Hou, Wu, Wang, Yu, and Han}{Huang
  et~al\mbox{.}}{2022}]%
        {huang2022large}
\bibfield{author}{\bibinfo{person}{Jiaxin Huang},
  \bibinfo{person}{Shixiang~Shane Gu}, \bibinfo{person}{Le Hou},
  \bibinfo{person}{Yuexin Wu}, \bibinfo{person}{Xuezhi Wang},
  \bibinfo{person}{Hongkun Yu}, {and} \bibinfo{person}{Jiawei Han}.}
  \bibinfo{year}{2022}\natexlab{}.
\newblock \showarticletitle{Large language models can self-improve}.
\newblock \bibinfo{journal}{\emph{arXiv preprint arXiv:2210.11610}}
  (\bibinfo{year}{2022}).
\newblock


\bibitem[\protect\citeauthoryear{Ie, Hsu, Mladenov, Jain, Narvekar, Wang, Wu,
  and Boutilier}{Ie et~al\mbox{.}}{2019}]%
        {ie2019recsim}
\bibfield{author}{\bibinfo{person}{Eugene Ie}, \bibinfo{person}{Chih-wei Hsu},
  \bibinfo{person}{Martin Mladenov}, \bibinfo{person}{Vihan Jain},
  \bibinfo{person}{Sanmit Narvekar}, \bibinfo{person}{Jing Wang},
  \bibinfo{person}{Rui Wu}, {and} \bibinfo{person}{Craig Boutilier}.}
  \bibinfo{year}{2019}\natexlab{}.
\newblock \showarticletitle{Recsim: A configurable simulation platform for
  recommender systems}.
\newblock \bibinfo{journal}{\emph{arXiv preprint arXiv:1909.04847}}
  (\bibinfo{year}{2019}).
\newblock


\bibitem[\protect\citeauthoryear{Jannach and Chen}{Jannach and Chen}{2022}]%
        {jannach2022conversational}
\bibfield{author}{\bibinfo{person}{Dietmar Jannach} {and} \bibinfo{person}{Li
  Chen}.} \bibinfo{year}{2022}\natexlab{}.
\newblock \showarticletitle{Conversational Recommendation: A Grand AI
  Challenge}.
\newblock \bibinfo{journal}{\emph{arXiv preprint arXiv:2203.09126}}
  (\bibinfo{year}{2022}).
\newblock


\bibitem[\protect\citeauthoryear{Jannach and Manzoor}{Jannach and
  Manzoor}{2020}]%
        {jannach2020end}
\bibfield{author}{\bibinfo{person}{Dietmar Jannach} {and}
  \bibinfo{person}{Ahtsham Manzoor}.} \bibinfo{year}{2020}\natexlab{}.
\newblock \showarticletitle{End-to-End Learning for Conversational
  Recommendation: A Long Way to Go?}. In \bibinfo{booktitle}{\emph{IntRS@
  RecSys}}. \bibinfo{pages}{72--76}.
\newblock


\bibitem[\protect\citeauthoryear{Jannach, Manzoor, Cai, and Chen}{Jannach
  et~al\mbox{.}}{2021}]%
        {jannach2021survey}
\bibfield{author}{\bibinfo{person}{Dietmar Jannach}, \bibinfo{person}{Ahtsham
  Manzoor}, \bibinfo{person}{Wanling Cai}, {and} \bibinfo{person}{Li Chen}.}
  \bibinfo{year}{2021}\natexlab{}.
\newblock \showarticletitle{A survey on conversational recommender systems}.
\newblock \bibinfo{journal}{\emph{ACM Computing Surveys (CSUR)}}
  \bibinfo{volume}{54}, \bibinfo{number}{5} (\bibinfo{year}{2021}),
  \bibinfo{pages}{1--36}.
\newblock


\bibitem[\protect\citeauthoryear{Ji, Lee, Frieske, Yu, Su, Xu, Ishii, Bang,
  Madotto, and Fung}{Ji et~al\mbox{.}}{2022}]%
        {ji2022survey}
\bibfield{author}{\bibinfo{person}{Ziwei Ji}, \bibinfo{person}{Nayeon Lee},
  \bibinfo{person}{Rita Frieske}, \bibinfo{person}{Tiezheng Yu},
  \bibinfo{person}{Dan Su}, \bibinfo{person}{Yan Xu}, \bibinfo{person}{Etsuko
  Ishii}, \bibinfo{person}{Yejin Bang}, \bibinfo{person}{Andrea Madotto}, {and}
  \bibinfo{person}{Pascale Fung}.} \bibinfo{year}{2022}\natexlab{}.
\newblock \showarticletitle{Survey of hallucination in natural language
  generation}.
\newblock \bibinfo{journal}{\emph{Comput. Surveys}} (\bibinfo{year}{2022}).
\newblock


\bibitem[\protect\citeauthoryear{Ju, Yang, and Wang}{Ju et~al\mbox{.}}{2021}]%
        {ju2021text}
\bibfield{author}{\bibinfo{person}{Jia-Huei Ju}, \bibinfo{person}{Jheng-Hong
  Yang}, {and} \bibinfo{person}{Chuan-Ju Wang}.}
  \bibinfo{year}{2021}\natexlab{}.
\newblock \showarticletitle{Text-to-text Multi-view Learning for Passage
  Re-ranking}. In \bibinfo{booktitle}{\emph{Proceedings of the 44th
  International ACM SIGIR Conference on Research and Development in Information
  Retrieval}}. \bibinfo{pages}{1803--1807}.
\newblock


\bibitem[\protect\citeauthoryear{Kang, Balakrishnan, Shah, Crook, Boureau, and
  Weston}{Kang et~al\mbox{.}}{2019}]%
        {kang2019recommendation}
\bibfield{author}{\bibinfo{person}{Dongyeop Kang}, \bibinfo{person}{Anusha
  Balakrishnan}, \bibinfo{person}{Pararth Shah}, \bibinfo{person}{Paul Crook},
  \bibinfo{person}{Y-Lan Boureau}, {and} \bibinfo{person}{Jason Weston}.}
  \bibinfo{year}{2019}\natexlab{}.
\newblock \showarticletitle{Recommendation as a communication game:
  Self-supervised bot-play for goal-oriented dialogue}.
\newblock \bibinfo{journal}{\emph{arXiv preprint arXiv:1909.03922}}
  (\bibinfo{year}{2019}).
\newblock


\bibitem[\protect\citeauthoryear{Kim, Wattenberg, Gilmer, Cai, Wexler, Viegas,
  and Sayres}{Kim et~al\mbox{.}}{2017}]%
        {https://doi.org/10.48550/arxiv.1711.11279}
\bibfield{author}{\bibinfo{person}{Been Kim}, \bibinfo{person}{Martin
  Wattenberg}, \bibinfo{person}{Justin Gilmer}, \bibinfo{person}{Carrie Cai},
  \bibinfo{person}{James Wexler}, \bibinfo{person}{Fernanda Viegas}, {and}
  \bibinfo{person}{Rory Sayres}.} \bibinfo{year}{2017}\natexlab{}.
\newblock \showarticletitle{Interpretability Beyond Feature Attribution:
  Quantitative Testing with Concept Activation Vectors (TCAV)}.
\newblock  (\bibinfo{year}{2017}).
\newblock
\urldef\tempurl%
\url{https://doi.org/10.48550/ARXIV.1711.11279}
\showDOI{\tempurl}


\bibitem[\protect\citeauthoryear{Lampinen, Dasgupta, Chan, Matthewson, Tessler,
  Creswell, McClelland, Wang, and Hill}{Lampinen et~al\mbox{.}}{2022}]%
        {lampinen2022can}
\bibfield{author}{\bibinfo{person}{Andrew~K Lampinen}, \bibinfo{person}{Ishita
  Dasgupta}, \bibinfo{person}{Stephanie~CY Chan}, \bibinfo{person}{Kory
  Matthewson}, \bibinfo{person}{Michael~Henry Tessler},
  \bibinfo{person}{Antonia Creswell}, \bibinfo{person}{James~L McClelland},
  \bibinfo{person}{Jane~X Wang}, {and} \bibinfo{person}{Felix Hill}.}
  \bibinfo{year}{2022}\natexlab{}.
\newblock \showarticletitle{Can language models learn from explanations in
  context?}
\newblock \bibinfo{journal}{\emph{arXiv preprint arXiv:2204.02329}}
  (\bibinfo{year}{2022}).
\newblock


\bibitem[\protect\citeauthoryear{Lei, He, Miao, Wu, Hong, Kan, and Chua}{Lei
  et~al\mbox{.}}{2020a}]%
        {lei2020estimation}
\bibfield{author}{\bibinfo{person}{Wenqiang Lei}, \bibinfo{person}{Xiangnan
  He}, \bibinfo{person}{Yisong Miao}, \bibinfo{person}{Qingyun Wu},
  \bibinfo{person}{Richang Hong}, \bibinfo{person}{Min-Yen Kan}, {and}
  \bibinfo{person}{Tat-Seng Chua}.} \bibinfo{year}{2020}\natexlab{a}.
\newblock \showarticletitle{Estimation-action-reflection: Towards deep
  interaction between conversational and recommender systems}. In
  \bibinfo{booktitle}{\emph{Proceedings of the 13th International Conference on
  Web Search and Data Mining}}. \bibinfo{pages}{304--312}.
\newblock


\bibitem[\protect\citeauthoryear{Lei, Zhang, He, Miao, Wang, Chen, and
  Chua}{Lei et~al\mbox{.}}{2020b}]%
        {lei2020interactive}
\bibfield{author}{\bibinfo{person}{Wenqiang Lei}, \bibinfo{person}{Gangyi
  Zhang}, \bibinfo{person}{Xiangnan He}, \bibinfo{person}{Yisong Miao},
  \bibinfo{person}{Xiang Wang}, \bibinfo{person}{Liang Chen}, {and}
  \bibinfo{person}{Tat-Seng Chua}.} \bibinfo{year}{2020}\natexlab{b}.
\newblock \showarticletitle{Interactive path reasoning on graph for
  conversational recommendation}. In \bibinfo{booktitle}{\emph{Proceedings of
  the 26th acm sigkdd international conference on knowledge discovery \& data
  mining}}. \bibinfo{pages}{2073--2083}.
\newblock


\bibitem[\protect\citeauthoryear{Lester, Al-Rfou, and Constant}{Lester
  et~al\mbox{.}}{2021}]%
        {lester2021power}
\bibfield{author}{\bibinfo{person}{Brian Lester}, \bibinfo{person}{Rami
  Al-Rfou}, {and} \bibinfo{person}{Noah Constant}.}
  \bibinfo{year}{2021}\natexlab{}.
\newblock \showarticletitle{The power of scale for parameter-efficient prompt
  tuning}.
\newblock \bibinfo{journal}{\emph{arXiv preprint arXiv:2104.08691}}
  (\bibinfo{year}{2021}).
\newblock


\bibitem[\protect\citeauthoryear{Lewis, Perez, Piktus, Petroni, Karpukhin,
  Goyal, K{\"u}ttler, Lewis, Yih, Rockt{\"a}schel, et~al\mbox{.}}{Lewis
  et~al\mbox{.}}{2020}]%
        {lewis2020retrieval}
\bibfield{author}{\bibinfo{person}{Patrick Lewis}, \bibinfo{person}{Ethan
  Perez}, \bibinfo{person}{Aleksandra Piktus}, \bibinfo{person}{Fabio Petroni},
  \bibinfo{person}{Vladimir Karpukhin}, \bibinfo{person}{Naman Goyal},
  \bibinfo{person}{Heinrich K{\"u}ttler}, \bibinfo{person}{Mike Lewis},
  \bibinfo{person}{Wen-tau Yih}, \bibinfo{person}{Tim Rockt{\"a}schel},
  {et~al\mbox{.}}} \bibinfo{year}{2020}\natexlab{}.
\newblock \showarticletitle{Retrieval-augmented generation for
  knowledge-intensive nlp tasks}.
\newblock \bibinfo{journal}{\emph{Advances in Neural Information Processing
  Systems}}  \bibinfo{volume}{33} (\bibinfo{year}{2020}),
  \bibinfo{pages}{9459--9474}.
\newblock


\bibitem[\protect\citeauthoryear{Li, Ebrahimi~Kahou, Schulz, Michalski,
  Charlin, and Pal}{Li et~al\mbox{.}}{2018}]%
        {li2018towards}
\bibfield{author}{\bibinfo{person}{Raymond Li}, \bibinfo{person}{Samira
  Ebrahimi~Kahou}, \bibinfo{person}{Hannes Schulz}, \bibinfo{person}{Vincent
  Michalski}, \bibinfo{person}{Laurent Charlin}, {and} \bibinfo{person}{Chris
  Pal}.} \bibinfo{year}{2018}\natexlab{}.
\newblock \showarticletitle{Towards deep conversational recommendations}.
\newblock \bibinfo{journal}{\emph{Advances in neural information processing
  systems}}  \bibinfo{volume}{31} (\bibinfo{year}{2018}).
\newblock


\bibitem[\protect\citeauthoryear{Li, Lin, Zhang, Fu, Chen, Lou, and Chen}{Li
  et~al\mbox{.}}{2022}]%
        {li2022advance}
\bibfield{author}{\bibinfo{person}{Yifei Li}, \bibinfo{person}{Zeqi Lin},
  \bibinfo{person}{Shizhuo Zhang}, \bibinfo{person}{Qiang Fu},
  \bibinfo{person}{Bei Chen}, \bibinfo{person}{Jian-Guang Lou}, {and}
  \bibinfo{person}{Weizhu Chen}.} \bibinfo{year}{2022}\natexlab{}.
\newblock \showarticletitle{On the Advance of Making Language Models Better
  Reasoners}.
\newblock \bibinfo{journal}{\emph{arXiv preprint arXiv:2206.02336}}
  (\bibinfo{year}{2022}).
\newblock


\bibitem[\protect\citeauthoryear{Linden, Hanks, and Lesh}{Linden
  et~al\mbox{.}}{1997}]%
        {linden1997interactive}
\bibfield{author}{\bibinfo{person}{Greg Linden}, \bibinfo{person}{Steve Hanks},
  {and} \bibinfo{person}{Neal Lesh}.} \bibinfo{year}{1997}\natexlab{}.
\newblock \showarticletitle{Interactive assessment of user preference models:
  The automated travel assistant}. In \bibinfo{booktitle}{\emph{User
  Modeling}}. Springer, \bibinfo{pages}{67--78}.
\newblock


\bibitem[\protect\citeauthoryear{Liu, Shakeri, Yu, and Li}{Liu
  et~al\mbox{.}}{2021}]%
        {liu2021enct5}
\bibfield{author}{\bibinfo{person}{Frederick Liu}, \bibinfo{person}{Siamak
  Shakeri}, \bibinfo{person}{Hongkun Yu}, {and} \bibinfo{person}{Jing Li}.}
  \bibinfo{year}{2021}\natexlab{}.
\newblock \showarticletitle{EncT5: Fine-tuning T5 Encoder for
  Non-autoregressive Tasks}.
\newblock \bibinfo{journal}{\emph{arXiv preprint arXiv:2110.08426}}
  (\bibinfo{year}{2021}).
\newblock


\bibitem[\protect\citeauthoryear{Liu and Lapata}{Liu and Lapata}{2019}]%
        {liu2019hierarchical}
\bibfield{author}{\bibinfo{person}{Yang Liu} {and} \bibinfo{person}{Mirella
  Lapata}.} \bibinfo{year}{2019}\natexlab{}.
\newblock \showarticletitle{Hierarchical transformers for multi-document
  summarization}.
\newblock \bibinfo{journal}{\emph{arXiv preprint arXiv:1905.13164}}
  (\bibinfo{year}{2019}).
\newblock


\bibitem[\protect\citeauthoryear{Ma, Zhao, Yi, Yang, Chen, Tang, Hong, and
  Chi}{Ma et~al\mbox{.}}{2020}]%
        {ma2020off}
\bibfield{author}{\bibinfo{person}{Jiaqi Ma}, \bibinfo{person}{Zhe Zhao},
  \bibinfo{person}{Xinyang Yi}, \bibinfo{person}{Ji Yang},
  \bibinfo{person}{Minmin Chen}, \bibinfo{person}{Jiaxi Tang},
  \bibinfo{person}{Lichan Hong}, {and} \bibinfo{person}{Ed~H Chi}.}
  \bibinfo{year}{2020}\natexlab{}.
\newblock \showarticletitle{Off-policy learning in two-stage recommender
  systems}. In \bibinfo{booktitle}{\emph{Proceedings of The Web Conference
  2020}}. \bibinfo{pages}{463--473}.
\newblock


\bibitem[\protect\citeauthoryear{Mansoury, Abdollahpouri, Pechenizkiy,
  Mobasher, and Burke}{Mansoury et~al\mbox{.}}{2020}]%
        {mansoury2020feedback}
\bibfield{author}{\bibinfo{person}{Masoud Mansoury}, \bibinfo{person}{Himan
  Abdollahpouri}, \bibinfo{person}{Mykola Pechenizkiy},
  \bibinfo{person}{Bamshad Mobasher}, {and} \bibinfo{person}{Robin Burke}.}
  \bibinfo{year}{2020}\natexlab{}.
\newblock \showarticletitle{Feedback loop and bias amplification in recommender
  systems}. In \bibinfo{booktitle}{\emph{Proceedings of the 29th ACM
  international conference on information \& knowledge management}}.
  \bibinfo{pages}{2145--2148}.
\newblock


\bibitem[\protect\citeauthoryear{Mazar{\'e}, Humeau, Raison, and
  Bordes}{Mazar{\'e} et~al\mbox{.}}{2018}]%
        {mazare2018training}
\bibfield{author}{\bibinfo{person}{Pierre-Emmanuel Mazar{\'e}},
  \bibinfo{person}{Samuel Humeau}, \bibinfo{person}{Martin Raison}, {and}
  \bibinfo{person}{Antoine Bordes}.} \bibinfo{year}{2018}\natexlab{}.
\newblock \showarticletitle{Training millions of personalized dialogue agents}.
\newblock \bibinfo{journal}{\emph{arXiv preprint arXiv:1809.01984}}
  (\bibinfo{year}{2018}).
\newblock


\bibitem[\protect\citeauthoryear{McCarthy, Salem, and Smyth}{McCarthy
  et~al\mbox{.}}{2010}]%
        {mccarthy2010experience}
\bibfield{author}{\bibinfo{person}{Kevin McCarthy}, \bibinfo{person}{Yasser
  Salem}, {and} \bibinfo{person}{Barry Smyth}.}
  \bibinfo{year}{2010}\natexlab{}.
\newblock \showarticletitle{Experience-based critiquing: Reusing critiquing
  experiences to improve conversational recommendation}. In
  \bibinfo{booktitle}{\emph{International Conference on Case-Based Reasoning}}.
  Springer, \bibinfo{pages}{480--494}.
\newblock


\bibitem[\protect\citeauthoryear{McInerney, Lacker, Hansen, Higley, Bouchard,
  Gruson, and Mehrotra}{McInerney et~al\mbox{.}}{2018}]%
        {mcinerney2018explore}
\bibfield{author}{\bibinfo{person}{James McInerney}, \bibinfo{person}{Benjamin
  Lacker}, \bibinfo{person}{Samantha Hansen}, \bibinfo{person}{Karl Higley},
  \bibinfo{person}{Hugues Bouchard}, \bibinfo{person}{Alois Gruson}, {and}
  \bibinfo{person}{Rishabh Mehrotra}.} \bibinfo{year}{2018}\natexlab{}.
\newblock \showarticletitle{Explore, exploit, and explain: personalizing
  explainable recommendations with bandits}. In
  \bibinfo{booktitle}{\emph{Proceedings of the 12th ACM conference on
  recommender systems}}. \bibinfo{pages}{31--39}.
\newblock


\bibitem[\protect\citeauthoryear{Mehri, Altun, and Eskenazi}{Mehri
  et~al\mbox{.}}{2022}]%
        {mehri2022lad}
\bibfield{author}{\bibinfo{person}{Shikib Mehri}, \bibinfo{person}{Yasemin
  Altun}, {and} \bibinfo{person}{Maxine Eskenazi}.}
  \bibinfo{year}{2022}\natexlab{}.
\newblock \showarticletitle{LAD: Language Models as Data for Zero-Shot Dialog}.
\newblock \bibinfo{journal}{\emph{arXiv preprint arXiv:2207.14393}}
  (\bibinfo{year}{2022}).
\newblock


\bibitem[\protect\citeauthoryear{Mikolov, Chen, Corrado, and Dean}{Mikolov
  et~al\mbox{.}}{2013}]%
        {mikolov2013efficient}
\bibfield{author}{\bibinfo{person}{Tomas Mikolov}, \bibinfo{person}{Kai Chen},
  \bibinfo{person}{Greg Corrado}, {and} \bibinfo{person}{Jeffrey Dean}.}
  \bibinfo{year}{2013}\natexlab{}.
\newblock \showarticletitle{Efficient estimation of word representations in
  vector space}.
\newblock \bibinfo{journal}{\emph{arXiv preprint arXiv:1301.3781}}
  (\bibinfo{year}{2013}).
\newblock


\bibitem[\protect\citeauthoryear{Nakano, Hilton, Balaji, Wu, Ouyang, Kim,
  Hesse, Jain, Kosaraju, Saunders, et~al\mbox{.}}{Nakano et~al\mbox{.}}{2021}]%
        {nakano2021webgpt}
\bibfield{author}{\bibinfo{person}{Reiichiro Nakano}, \bibinfo{person}{Jacob
  Hilton}, \bibinfo{person}{Suchir Balaji}, \bibinfo{person}{Jeff Wu},
  \bibinfo{person}{Long Ouyang}, \bibinfo{person}{Christina Kim},
  \bibinfo{person}{Christopher Hesse}, \bibinfo{person}{Shantanu Jain},
  \bibinfo{person}{Vineet Kosaraju}, \bibinfo{person}{William Saunders},
  {et~al\mbox{.}}} \bibinfo{year}{2021}\natexlab{}.
\newblock \showarticletitle{WebGPT: Browser-assisted question-answering with
  human feedback}.
\newblock \bibinfo{journal}{\emph{arXiv preprint arXiv:2112.09332}}
  (\bibinfo{year}{2021}).
\newblock


\bibitem[\protect\citeauthoryear{Narang, Raffel, Lee, Roberts, Fiedel, and
  Malkan}{Narang et~al\mbox{.}}{2020}]%
        {narang2020wt5}
\bibfield{author}{\bibinfo{person}{Sharan Narang}, \bibinfo{person}{Colin
  Raffel}, \bibinfo{person}{Katherine Lee}, \bibinfo{person}{Adam Roberts},
  \bibinfo{person}{Noah Fiedel}, {and} \bibinfo{person}{Karishma Malkan}.}
  \bibinfo{year}{2020}\natexlab{}.
\newblock \showarticletitle{Wt5?! training text-to-text models to explain their
  predictions}.
\newblock \bibinfo{journal}{\emph{arXiv preprint arXiv:2004.14546}}
  (\bibinfo{year}{2020}).
\newblock


\bibitem[\protect\citeauthoryear{Ni, Qu, Lu, Dai, {\'A}brego, Ma, Zhao, Luan,
  Hall, Chang, et~al\mbox{.}}{Ni et~al\mbox{.}}{2021}]%
        {ni2021large}
\bibfield{author}{\bibinfo{person}{Jianmo Ni}, \bibinfo{person}{Chen Qu},
  \bibinfo{person}{Jing Lu}, \bibinfo{person}{Zhuyun Dai},
  \bibinfo{person}{Gustavo~Hern{\'a}ndez {\'A}brego}, \bibinfo{person}{Ji Ma},
  \bibinfo{person}{Vincent~Y Zhao}, \bibinfo{person}{Yi Luan},
  \bibinfo{person}{Keith~B Hall}, \bibinfo{person}{Ming-Wei Chang},
  {et~al\mbox{.}}} \bibinfo{year}{2021}\natexlab{}.
\newblock \showarticletitle{Large dual encoders are generalizable retrievers}.
\newblock \bibinfo{journal}{\emph{arXiv preprint arXiv:2112.07899}}
  (\bibinfo{year}{2021}).
\newblock


\bibitem[\protect\citeauthoryear{Nogueira, Jiang, and Lin}{Nogueira
  et~al\mbox{.}}{2020}]%
        {nogueira2020document}
\bibfield{author}{\bibinfo{person}{Rodrigo Nogueira}, \bibinfo{person}{Zhiying
  Jiang}, {and} \bibinfo{person}{Jimmy Lin}.} \bibinfo{year}{2020}\natexlab{}.
\newblock \showarticletitle{Document ranking with a pretrained
  sequence-to-sequence model}.
\newblock \bibinfo{journal}{\emph{arXiv preprint arXiv:2003.06713}}
  (\bibinfo{year}{2020}).
\newblock


\bibitem[\protect\citeauthoryear{Ouyang, Wu, Jiang, Almeida, Wainwright,
  Mishkin, Zhang, Agarwal, Slama, Ray, et~al\mbox{.}}{Ouyang
  et~al\mbox{.}}{2022}]%
        {ouyang2022training}
\bibfield{author}{\bibinfo{person}{Long Ouyang}, \bibinfo{person}{Jeff Wu},
  \bibinfo{person}{Xu Jiang}, \bibinfo{person}{Diogo Almeida},
  \bibinfo{person}{Carroll~L Wainwright}, \bibinfo{person}{Pamela Mishkin},
  \bibinfo{person}{Chong Zhang}, \bibinfo{person}{Sandhini Agarwal},
  \bibinfo{person}{Katarina Slama}, \bibinfo{person}{Alex Ray},
  {et~al\mbox{.}}} \bibinfo{year}{2022}\natexlab{}.
\newblock \showarticletitle{Training language models to follow instructions
  with human feedback}.
\newblock \bibinfo{journal}{\emph{arXiv preprint arXiv:2203.02155}}
  (\bibinfo{year}{2022}).
\newblock


\bibitem[\protect\citeauthoryear{Park and Youn-kyung}{Park and
  Youn-kyung}{2019}]%
        {park2019design}
\bibfield{author}{\bibinfo{person}{Sunjeong Park} {and} \bibinfo{person}{Lim
  Youn-kyung}.} \bibinfo{year}{2019}\natexlab{}.
\newblock \showarticletitle{Design considerations for explanations made by a
  recommender chatbot}. In \bibinfo{booktitle}{\emph{IASDR Conference 2019}}.
  IASDR.
\newblock


\bibitem[\protect\citeauthoryear{Penha and Hauff}{Penha and Hauff}{2020}]%
        {penha2020does}
\bibfield{author}{\bibinfo{person}{Gustavo Penha} {and}
  \bibinfo{person}{Claudia Hauff}.} \bibinfo{year}{2020}\natexlab{}.
\newblock \showarticletitle{What does bert know about books, movies and music?
  probing bert for conversational recommendation}. In
  \bibinfo{booktitle}{\emph{Fourteenth ACM Conference on Recommender Systems}}.
  \bibinfo{pages}{388--397}.
\newblock


\bibitem[\protect\citeauthoryear{Pradeep, Nogueira, and Lin}{Pradeep
  et~al\mbox{.}}{2021}]%
        {pradeep2021expando}
\bibfield{author}{\bibinfo{person}{Ronak Pradeep}, \bibinfo{person}{Rodrigo
  Nogueira}, {and} \bibinfo{person}{Jimmy Lin}.}
  \bibinfo{year}{2021}\natexlab{}.
\newblock \showarticletitle{The expando-mono-duo design pattern for text
  ranking with pretrained sequence-to-sequence models}.
\newblock \bibinfo{journal}{\emph{arXiv preprint arXiv:2101.05667}}
  (\bibinfo{year}{2021}).
\newblock


\bibitem[\protect\citeauthoryear{Puri, Spring, Patwary, Shoeybi, and
  Catanzaro}{Puri et~al\mbox{.}}{2020}]%
        {puri2020training}
\bibfield{author}{\bibinfo{person}{Raul Puri}, \bibinfo{person}{Ryan Spring},
  \bibinfo{person}{Mostofa Patwary}, \bibinfo{person}{Mohammad Shoeybi}, {and}
  \bibinfo{person}{Bryan Catanzaro}.} \bibinfo{year}{2020}\natexlab{}.
\newblock \showarticletitle{Training question answering models from synthetic
  data}.
\newblock \bibinfo{journal}{\emph{arXiv preprint arXiv:2002.09599}}
  (\bibinfo{year}{2020}).
\newblock


\bibitem[\protect\citeauthoryear{Radlinski, Balog, Diaz, Dixon, and
  Wedin}{Radlinski et~al\mbox{.}}{2022}]%
        {radlinski2022natural}
\bibfield{author}{\bibinfo{person}{Filip Radlinski}, \bibinfo{person}{Krisztian
  Balog}, \bibinfo{person}{Fernando Diaz}, \bibinfo{person}{Lucas Dixon}, {and}
  \bibinfo{person}{Ben Wedin}.} \bibinfo{year}{2022}\natexlab{}.
\newblock \showarticletitle{On Natural Language User Profiles for Transparent
  and Scrutable Recommendation}.
\newblock \bibinfo{journal}{\emph{arXiv preprint arXiv:2205.09403}}
  (\bibinfo{year}{2022}).
\newblock


\bibitem[\protect\citeauthoryear{Raffel, Shazeer, Roberts, Lee, Narang, Matena,
  Zhou, Li, Liu, et~al\mbox{.}}{Raffel et~al\mbox{.}}{2020}]%
        {raffel2020exploring}
\bibfield{author}{\bibinfo{person}{Colin Raffel}, \bibinfo{person}{Noam
  Shazeer}, \bibinfo{person}{Adam Roberts}, \bibinfo{person}{Katherine Lee},
  \bibinfo{person}{Sharan Narang}, \bibinfo{person}{Michael Matena},
  \bibinfo{person}{Yanqi Zhou}, \bibinfo{person}{Wei Li},
  \bibinfo{person}{Peter~J Liu}, {et~al\mbox{.}}}
  \bibinfo{year}{2020}\natexlab{}.
\newblock \showarticletitle{Exploring the limits of transfer learning with a
  unified text-to-text transformer.}
\newblock \bibinfo{journal}{\emph{J. Mach. Learn. Res.}} \bibinfo{volume}{21},
  \bibinfo{number}{140} (\bibinfo{year}{2020}), \bibinfo{pages}{1--67}.
\newblock


\bibitem[\protect\citeauthoryear{Rajani, McCann, Xiong, and Socher}{Rajani
  et~al\mbox{.}}{2019}]%
        {rajani2019explain}
\bibfield{author}{\bibinfo{person}{Nazneen~Fatema Rajani},
  \bibinfo{person}{Bryan McCann}, \bibinfo{person}{Caiming Xiong}, {and}
  \bibinfo{person}{Richard Socher}.} \bibinfo{year}{2019}\natexlab{}.
\newblock \showarticletitle{Explain yourself! leveraging language models for
  commonsense reasoning}.
\newblock \bibinfo{journal}{\emph{arXiv preprint arXiv:1906.02361}}
  (\bibinfo{year}{2019}).
\newblock


\bibitem[\protect\citeauthoryear{Rastogi, Zang, Sunkara, Gupta, and
  Khaitan}{Rastogi et~al\mbox{.}}{2020}]%
        {rastogi2020towards}
\bibfield{author}{\bibinfo{person}{Abhinav Rastogi}, \bibinfo{person}{Xiaoxue
  Zang}, \bibinfo{person}{Srinivas Sunkara}, \bibinfo{person}{Raghav Gupta},
  {and} \bibinfo{person}{Pranav Khaitan}.} \bibinfo{year}{2020}\natexlab{}.
\newblock \showarticletitle{Towards scalable multi-domain conversational
  agents: The schema-guided dialogue dataset}. In
  \bibinfo{booktitle}{\emph{Proceedings of the AAAI Conference on Artificial
  Intelligence}}, Vol.~\bibinfo{volume}{34}. \bibinfo{pages}{8689--8696}.
\newblock


\bibitem[\protect\citeauthoryear{Ren, Yin, Chen, Wang, Huang, and Zheng}{Ren
  et~al\mbox{.}}{2021}]%
        {ren2021learning}
\bibfield{author}{\bibinfo{person}{Xuhui Ren}, \bibinfo{person}{Hongzhi Yin},
  \bibinfo{person}{Tong Chen}, \bibinfo{person}{Hao Wang}, \bibinfo{person}{Zi
  Huang}, {and} \bibinfo{person}{Kai Zheng}.} \bibinfo{year}{2021}\natexlab{}.
\newblock \showarticletitle{Learning to ask appropriate questions in
  conversational recommendation}. In \bibinfo{booktitle}{\emph{Proceedings of
  the 44th International ACM SIGIR Conference on Research and Development in
  Information Retrieval}}. \bibinfo{pages}{808--817}.
\newblock


\bibitem[\protect\citeauthoryear{Reynolds and McDonell}{Reynolds and
  McDonell}{2021}]%
        {reynolds2021prompt}
\bibfield{author}{\bibinfo{person}{Laria Reynolds} {and} \bibinfo{person}{Kyle
  McDonell}.} \bibinfo{year}{2021}\natexlab{}.
\newblock \showarticletitle{Prompt programming for large language models:
  Beyond the few-shot paradigm}. In \bibinfo{booktitle}{\emph{Extended
  Abstracts of the 2021 CHI Conference on Human Factors in Computing Systems}}.
  \bibinfo{pages}{1--7}.
\newblock


\bibitem[\protect\citeauthoryear{Ricci and Nguyen}{Ricci and Nguyen}{2007}]%
        {ricci2007acquiring}
\bibfield{author}{\bibinfo{person}{Francesco Ricci} {and}
  \bibinfo{person}{Quang~Nhat Nguyen}.} \bibinfo{year}{2007}\natexlab{}.
\newblock \showarticletitle{Acquiring and revising preferences in a
  critique-based mobile recommender system}.
\newblock \bibinfo{journal}{\emph{IEEE Intelligent systems}}
  \bibinfo{volume}{22}, \bibinfo{number}{3} (\bibinfo{year}{2007}),
  \bibinfo{pages}{22--29}.
\newblock


\bibitem[\protect\citeauthoryear{Rohde, Bonner, Dunlop, Vasile, and
  Karatzoglou}{Rohde et~al\mbox{.}}{2018}]%
        {rohde2018recogym}
\bibfield{author}{\bibinfo{person}{David Rohde}, \bibinfo{person}{Stephen
  Bonner}, \bibinfo{person}{Travis Dunlop}, \bibinfo{person}{Flavian Vasile},
  {and} \bibinfo{person}{Alexandros Karatzoglou}.}
  \bibinfo{year}{2018}\natexlab{}.
\newblock \showarticletitle{Recogym: A reinforcement learning environment for
  the problem of product recommendation in online advertising}.
\newblock \bibinfo{journal}{\emph{arXiv preprint arXiv:1808.00720}}
  (\bibinfo{year}{2018}).
\newblock


\bibitem[\protect\citeauthoryear{Schnabel, Bennett, Dumais, and
  Joachims}{Schnabel et~al\mbox{.}}{2018}]%
        {schnabel2018short}
\bibfield{author}{\bibinfo{person}{Tobias Schnabel}, \bibinfo{person}{Paul~N
  Bennett}, \bibinfo{person}{Susan~T Dumais}, {and} \bibinfo{person}{Thorsten
  Joachims}.} \bibinfo{year}{2018}\natexlab{}.
\newblock \showarticletitle{Short-term satisfaction and long-term coverage:
  Understanding how users tolerate algorithmic exploration}. In
  \bibinfo{booktitle}{\emph{Proceedings of the Eleventh ACM International
  Conference on Web Search and Data Mining}}. \bibinfo{pages}{513--521}.
\newblock


\bibitem[\protect\citeauthoryear{Shuster, Poff, Chen, Kiela, and
  Weston}{Shuster et~al\mbox{.}}{2021}]%
        {shuster2021retrieval}
\bibfield{author}{\bibinfo{person}{Kurt Shuster}, \bibinfo{person}{Spencer
  Poff}, \bibinfo{person}{Moya Chen}, \bibinfo{person}{Douwe Kiela}, {and}
  \bibinfo{person}{Jason Weston}.} \bibinfo{year}{2021}\natexlab{}.
\newblock \showarticletitle{Retrieval augmentation reduces hallucination in
  conversation}.
\newblock \bibinfo{journal}{\emph{arXiv preprint arXiv:2104.07567}}
  (\bibinfo{year}{2021}).
\newblock


\bibitem[\protect\citeauthoryear{Shuster, Xu, Komeili, Ju, Smith, Roller, Ung,
  Chen, Arora, Lane, et~al\mbox{.}}{Shuster et~al\mbox{.}}{2022}]%
        {shuster2022blenderbot}
\bibfield{author}{\bibinfo{person}{Kurt Shuster}, \bibinfo{person}{Jing Xu},
  \bibinfo{person}{Mojtaba Komeili}, \bibinfo{person}{Da Ju},
  \bibinfo{person}{Eric~Michael Smith}, \bibinfo{person}{Stephen Roller},
  \bibinfo{person}{Megan Ung}, \bibinfo{person}{Moya Chen},
  \bibinfo{person}{Kushal Arora}, \bibinfo{person}{Joshua Lane},
  {et~al\mbox{.}}} \bibinfo{year}{2022}\natexlab{}.
\newblock \showarticletitle{BlenderBot 3: a deployed conversational agent that
  continually learns to responsibly engage}.
\newblock \bibinfo{journal}{\emph{arXiv preprint arXiv:2208.03188}}
  (\bibinfo{year}{2022}).
\newblock


\bibitem[\protect\citeauthoryear{Snell, Yang, Fu, Su, and Levine}{Snell
  et~al\mbox{.}}{2022}]%
        {snell2022context}
\bibfield{author}{\bibinfo{person}{Charlie Snell}, \bibinfo{person}{Sherry
  Yang}, \bibinfo{person}{Justin Fu}, \bibinfo{person}{Yi Su}, {and}
  \bibinfo{person}{Sergey Levine}.} \bibinfo{year}{2022}\natexlab{}.
\newblock \showarticletitle{Context-Aware Language Modeling for Goal-Oriented
  Dialogue Systems}.
\newblock \bibinfo{journal}{\emph{arXiv preprint arXiv:2204.10198}}
  (\bibinfo{year}{2022}).
\newblock


\bibitem[\protect\citeauthoryear{Stiennon, Ouyang, Wu, Ziegler, Lowe, Voss,
  Radford, Amodei, and Christiano}{Stiennon et~al\mbox{.}}{2020}]%
        {stiennon2020learning}
\bibfield{author}{\bibinfo{person}{Nisan Stiennon}, \bibinfo{person}{Long
  Ouyang}, \bibinfo{person}{Jeffrey Wu}, \bibinfo{person}{Daniel Ziegler},
  \bibinfo{person}{Ryan Lowe}, \bibinfo{person}{Chelsea Voss},
  \bibinfo{person}{Alec Radford}, \bibinfo{person}{Dario Amodei}, {and}
  \bibinfo{person}{Paul~F Christiano}.} \bibinfo{year}{2020}\natexlab{}.
\newblock \showarticletitle{Learning to summarize with human feedback}.
\newblock \bibinfo{journal}{\emph{Advances in Neural Information Processing
  Systems}}  \bibinfo{volume}{33} (\bibinfo{year}{2020}),
  \bibinfo{pages}{3008--3021}.
\newblock


\bibitem[\protect\citeauthoryear{Sun and Zhang}{Sun and Zhang}{2018}]%
        {sun2018conversational}
\bibfield{author}{\bibinfo{person}{Yueming Sun} {and} \bibinfo{person}{Yi
  Zhang}.} \bibinfo{year}{2018}\natexlab{}.
\newblock \showarticletitle{Conversational recommender system}. In
  \bibinfo{booktitle}{\emph{The 41st international acm sigir conference on
  research \& development in information retrieval}}.
  \bibinfo{pages}{235--244}.
\newblock


\bibitem[\protect\citeauthoryear{Tay, Tran, Dehghani, Ni, Bahri, Mehta, Qin,
  Hui, Zhao, Gupta, et~al\mbox{.}}{Tay et~al\mbox{.}}{2022}]%
        {tay2022transformer}
\bibfield{author}{\bibinfo{person}{Yi Tay}, \bibinfo{person}{Vinh~Q Tran},
  \bibinfo{person}{Mostafa Dehghani}, \bibinfo{person}{Jianmo Ni},
  \bibinfo{person}{Dara Bahri}, \bibinfo{person}{Harsh Mehta},
  \bibinfo{person}{Zhen Qin}, \bibinfo{person}{Kai Hui}, \bibinfo{person}{Zhe
  Zhao}, \bibinfo{person}{Jai Gupta}, {et~al\mbox{.}}}
  \bibinfo{year}{2022}\natexlab{}.
\newblock \showarticletitle{Transformer memory as a differentiable search
  index}.
\newblock \bibinfo{journal}{\emph{arXiv preprint arXiv:2202.06991}}
  (\bibinfo{year}{2022}).
\newblock


\bibitem[\protect\citeauthoryear{Thompson, Goker, and Langley}{Thompson
  et~al\mbox{.}}{2004}]%
        {thompson2004personalized}
\bibfield{author}{\bibinfo{person}{Cynthia~A Thompson},
  \bibinfo{person}{Mehmet~H Goker}, {and} \bibinfo{person}{Pat Langley}.}
  \bibinfo{year}{2004}\natexlab{}.
\newblock \showarticletitle{A personalized system for conversational
  recommendations}.
\newblock \bibinfo{journal}{\emph{Journal of Artificial Intelligence Research}}
   \bibinfo{volume}{21} (\bibinfo{year}{2004}), \bibinfo{pages}{393--428}.
\newblock


\bibitem[\protect\citeauthoryear{Thoppilan, De~Freitas, Hall, Shazeer,
  Kulshreshtha, Cheng, Jin, Bos, Baker, Du, et~al\mbox{.}}{Thoppilan
  et~al\mbox{.}}{2022}]%
        {thoppilan2022lamda}
\bibfield{author}{\bibinfo{person}{Romal Thoppilan}, \bibinfo{person}{Daniel
  De~Freitas}, \bibinfo{person}{Jamie Hall}, \bibinfo{person}{Noam Shazeer},
  \bibinfo{person}{Apoorv Kulshreshtha}, \bibinfo{person}{Heng-Tze Cheng},
  \bibinfo{person}{Alicia Jin}, \bibinfo{person}{Taylor Bos},
  \bibinfo{person}{Leslie Baker}, \bibinfo{person}{Yu Du}, {et~al\mbox{.}}}
  \bibinfo{year}{2022}\natexlab{}.
\newblock \showarticletitle{Lamda: Language models for dialog applications}.
\newblock \bibinfo{journal}{\emph{arXiv preprint arXiv:2201.08239}}
  (\bibinfo{year}{2022}).
\newblock


\bibitem[\protect\citeauthoryear{Torbati, Yates, and Weikum}{Torbati
  et~al\mbox{.}}{2021}]%
        {torbati2021you}
\bibfield{author}{\bibinfo{person}{Ghazaleh~H Torbati}, \bibinfo{person}{Andrew
  Yates}, {and} \bibinfo{person}{Gerhard Weikum}.}
  \bibinfo{year}{2021}\natexlab{}.
\newblock \showarticletitle{You get what you chat: Using conversations to
  personalize search-based recommendations}. In
  \bibinfo{booktitle}{\emph{European Conference on Information Retrieval}}.
  Springer, \bibinfo{pages}{207--223}.
\newblock


\bibitem[\protect\citeauthoryear{Vaswani, Shazeer, Parmar, Uszkoreit, Jones,
  Gomez, Kaiser, and Polosukhin}{Vaswani et~al\mbox{.}}{2017}]%
        {vaswani2017attention}
\bibfield{author}{\bibinfo{person}{Ashish Vaswani}, \bibinfo{person}{Noam
  Shazeer}, \bibinfo{person}{Niki Parmar}, \bibinfo{person}{Jakob Uszkoreit},
  \bibinfo{person}{Llion Jones}, \bibinfo{person}{Aidan~N Gomez},
  \bibinfo{person}{{\L}ukasz Kaiser}, {and} \bibinfo{person}{Illia
  Polosukhin}.} \bibinfo{year}{2017}\natexlab{}.
\newblock \showarticletitle{Attention is all you need}.
\newblock \bibinfo{journal}{\emph{Advances in neural information processing
  systems}}  \bibinfo{volume}{30} (\bibinfo{year}{2017}).
\newblock


\bibitem[\protect\citeauthoryear{Vu, Lester, Constant, Al-Rfou, and Cer}{Vu
  et~al\mbox{.}}{2021}]%
        {vu2021spot}
\bibfield{author}{\bibinfo{person}{Tu Vu}, \bibinfo{person}{Brian Lester},
  \bibinfo{person}{Noah Constant}, \bibinfo{person}{Rami Al-Rfou}, {and}
  \bibinfo{person}{Daniel Cer}.} \bibinfo{year}{2021}\natexlab{}.
\newblock \showarticletitle{Spot: Better frozen model adaptation through soft
  prompt transfer}.
\newblock \bibinfo{journal}{\emph{arXiv preprint arXiv:2110.07904}}
  (\bibinfo{year}{2021}).
\newblock


\bibitem[\protect\citeauthoryear{Wang, Hu, Sha, Xu, Wong, and Jiang}{Wang
  et~al\mbox{.}}{2021a}]%
        {recindial}
\bibfield{author}{\bibinfo{person}{Lingzhi Wang}, \bibinfo{person}{Huang Hu},
  \bibinfo{person}{Lei Sha}, \bibinfo{person}{Can Xu},
  \bibinfo{person}{Kam{-}Fai Wong}, {and} \bibinfo{person}{Daxin Jiang}.}
  \bibinfo{year}{2021}\natexlab{a}.
\newblock \showarticletitle{Finetuning Large-Scale Pre-trained Language Models
  for Conversational Recommendation with Knowledge Graph}.
\newblock \bibinfo{journal}{\emph{CoRR}}  \bibinfo{volume}{abs/2110.07477}
  (\bibinfo{year}{2021}).
\newblock
\showeprint[arXiv]{2110.07477}
\urldef\tempurl%
\url{https://arxiv.org/abs/2110.07477}
\showURL{%
\tempurl}


\bibitem[\protect\citeauthoryear{Wang, Chen, Yang, Wu, Wu, and Xie}{Wang
  et~al\mbox{.}}{2018}]%
        {wang2018reinforcement}
\bibfield{author}{\bibinfo{person}{Xiting Wang}, \bibinfo{person}{Yiru Chen},
  \bibinfo{person}{Jie Yang}, \bibinfo{person}{Le Wu},
  \bibinfo{person}{Zhengtao Wu}, {and} \bibinfo{person}{Xing Xie}.}
  \bibinfo{year}{2018}\natexlab{}.
\newblock \showarticletitle{A reinforcement learning framework for explainable
  recommendation}. In \bibinfo{booktitle}{\emph{2018 IEEE international
  conference on data mining (ICDM)}}. IEEE, \bibinfo{pages}{587--596}.
\newblock


\bibitem[\protect\citeauthoryear{Wang, Wei, Schuurmans, Le, Chi, and Zhou}{Wang
  et~al\mbox{.}}{2022}]%
        {wang2022self}
\bibfield{author}{\bibinfo{person}{Xuezhi Wang}, \bibinfo{person}{Jason Wei},
  \bibinfo{person}{Dale Schuurmans}, \bibinfo{person}{Quoc Le},
  \bibinfo{person}{Ed Chi}, {and} \bibinfo{person}{Denny Zhou}.}
  \bibinfo{year}{2022}\natexlab{}.
\newblock \showarticletitle{Self-consistency improves chain of thought
  reasoning in language models}.
\newblock \bibinfo{journal}{\emph{arXiv preprint arXiv:2203.11171}}
  (\bibinfo{year}{2022}).
\newblock


\bibitem[\protect\citeauthoryear{Wang, Yu, Firat, and Cao}{Wang
  et~al\mbox{.}}{2021b}]%
        {wang2021towards}
\bibfield{author}{\bibinfo{person}{Zirui Wang}, \bibinfo{person}{Adams~Wei Yu},
  \bibinfo{person}{Orhan Firat}, {and} \bibinfo{person}{Yuan Cao}.}
  \bibinfo{year}{2021}\natexlab{b}.
\newblock \showarticletitle{Towards zero-label language learning}.
\newblock \bibinfo{journal}{\emph{arXiv preprint arXiv:2109.09193}}
  (\bibinfo{year}{2021}).
\newblock


\bibitem[\protect\citeauthoryear{Wei, Tay, Bommasani, Raffel, Zoph, Borgeaud,
  Yogatama, Bosma, Zhou, Metzler, et~al\mbox{.}}{Wei et~al\mbox{.}}{2022a}]%
        {wei2022emergent}
\bibfield{author}{\bibinfo{person}{Jason Wei}, \bibinfo{person}{Yi Tay},
  \bibinfo{person}{Rishi Bommasani}, \bibinfo{person}{Colin Raffel},
  \bibinfo{person}{Barret Zoph}, \bibinfo{person}{Sebastian Borgeaud},
  \bibinfo{person}{Dani Yogatama}, \bibinfo{person}{Maarten Bosma},
  \bibinfo{person}{Denny Zhou}, \bibinfo{person}{Donald Metzler},
  {et~al\mbox{.}}} \bibinfo{year}{2022}\natexlab{a}.
\newblock \showarticletitle{Emergent abilities of large language models}.
\newblock \bibinfo{journal}{\emph{arXiv preprint arXiv:2206.07682}}
  (\bibinfo{year}{2022}).
\newblock


\bibitem[\protect\citeauthoryear{Wei, Wang, Schuurmans, Bosma, Chi, Le, and
  Zhou}{Wei et~al\mbox{.}}{2022b}]%
        {wei2022chain}
\bibfield{author}{\bibinfo{person}{Jason Wei}, \bibinfo{person}{Xuezhi Wang},
  \bibinfo{person}{Dale Schuurmans}, \bibinfo{person}{Maarten Bosma},
  \bibinfo{person}{Ed Chi}, \bibinfo{person}{Quoc Le}, {and}
  \bibinfo{person}{Denny Zhou}.} \bibinfo{year}{2022}\natexlab{b}.
\newblock \showarticletitle{Chain of thought prompting elicits reasoning in
  large language models}.
\newblock \bibinfo{journal}{\emph{arXiv preprint arXiv:2201.11903}}
  (\bibinfo{year}{2022}).
\newblock


\bibitem[\protect\citeauthoryear{Weidinger, Mellor, Rauh, Griffin, Uesato,
  Huang, Cheng, Glaese, Balle, Kasirzadeh, et~al\mbox{.}}{Weidinger
  et~al\mbox{.}}{2021}]%
        {weidinger2021ethical}
\bibfield{author}{\bibinfo{person}{Laura Weidinger}, \bibinfo{person}{John
  Mellor}, \bibinfo{person}{Maribeth Rauh}, \bibinfo{person}{Conor Griffin},
  \bibinfo{person}{Jonathan Uesato}, \bibinfo{person}{Po-Sen Huang},
  \bibinfo{person}{Myra Cheng}, \bibinfo{person}{Mia Glaese},
  \bibinfo{person}{Borja Balle}, \bibinfo{person}{Atoosa Kasirzadeh},
  {et~al\mbox{.}}} \bibinfo{year}{2021}\natexlab{}.
\newblock \showarticletitle{Ethical and social risks of harm from language
  models}.
\newblock \bibinfo{journal}{\emph{arXiv preprint arXiv:2112.04359}}
  (\bibinfo{year}{2021}).
\newblock


\bibitem[\protect\citeauthoryear{Weston, Chopra, and Bordes}{Weston
  et~al\mbox{.}}{2014}]%
        {weston2014memory}
\bibfield{author}{\bibinfo{person}{Jason Weston}, \bibinfo{person}{Sumit
  Chopra}, {and} \bibinfo{person}{Antoine Bordes}.}
  \bibinfo{year}{2014}\natexlab{}.
\newblock \showarticletitle{Memory networks}.
\newblock \bibinfo{journal}{\emph{arXiv preprint arXiv:1410.3916}}
  (\bibinfo{year}{2014}).
\newblock


\bibitem[\protect\citeauthoryear{Weston, Dinan, and Miller}{Weston
  et~al\mbox{.}}{2018}]%
        {DBLP:journals/corr/abs-1808-04776}
\bibfield{author}{\bibinfo{person}{Jason Weston}, \bibinfo{person}{Emily
  Dinan}, {and} \bibinfo{person}{Alexander~H. Miller}.}
  \bibinfo{year}{2018}\natexlab{}.
\newblock \showarticletitle{Retrieve and Refine: Improved Sequence Generation
  Models For Dialogue}.
\newblock \bibinfo{journal}{\emph{CoRR}}  \bibinfo{volume}{abs/1808.04776}
  (\bibinfo{year}{2018}).
\newblock
\showeprint[arXiv]{1808.04776}
\urldef\tempurl%
\url{http://arxiv.org/abs/1808.04776}
\showURL{%
\tempurl}


\bibitem[\protect\citeauthoryear{Wu, Rabe, Hutchins, and Szegedy}{Wu
  et~al\mbox{.}}{2022}]%
        {Wu2022MemorizingT}
\bibfield{author}{\bibinfo{person}{Yuhuai Wu}, \bibinfo{person}{Markus~N.
  Rabe}, \bibinfo{person}{DeLesley~S. Hutchins}, {and}
  \bibinfo{person}{Christian Szegedy}.} \bibinfo{year}{2022}\natexlab{}.
\newblock \showarticletitle{Memorizing Transformers}.
\newblock \bibinfo{journal}{\emph{ArXiv}}  \bibinfo{volume}{abs/2203.08913}
  (\bibinfo{year}{2022}).
\newblock


\bibitem[\protect\citeauthoryear{Xiong, Xiong, Li, Tang, Liu, Bennett, Ahmed,
  and Overwijk}{Xiong et~al\mbox{.}}{2020}]%
        {xiong2020approximate}
\bibfield{author}{\bibinfo{person}{Lee Xiong}, \bibinfo{person}{Chenyan Xiong},
  \bibinfo{person}{Ye Li}, \bibinfo{person}{Kwok-Fung Tang},
  \bibinfo{person}{Jialin Liu}, \bibinfo{person}{Paul Bennett},
  \bibinfo{person}{Junaid Ahmed}, {and} \bibinfo{person}{Arnold Overwijk}.}
  \bibinfo{year}{2020}\natexlab{}.
\newblock \showarticletitle{Approximate nearest neighbor negative contrastive
  learning for dense text retrieval}.
\newblock \bibinfo{journal}{\emph{arXiv preprint arXiv:2007.00808}}
  (\bibinfo{year}{2020}).
\newblock


\bibitem[\protect\citeauthoryear{Xu, Szlam, and Weston}{Xu
  et~al\mbox{.}}{2021}]%
        {xu2021beyond}
\bibfield{author}{\bibinfo{person}{Jing Xu}, \bibinfo{person}{Arthur Szlam},
  {and} \bibinfo{person}{Jason Weston}.} \bibinfo{year}{2021}\natexlab{}.
\newblock \showarticletitle{Beyond goldfish memory: Long-term open-domain
  conversation}.
\newblock \bibinfo{journal}{\emph{arXiv preprint arXiv:2107.07567}}
  (\bibinfo{year}{2021}).
\newblock


\bibitem[\protect\citeauthoryear{Yi, Yang, Hong, Cheng, Heldt, Kumthekar, Zhao,
  Wei, and Chi}{Yi et~al\mbox{.}}{2019}]%
        {yi2019sampling}
\bibfield{author}{\bibinfo{person}{Xinyang Yi}, \bibinfo{person}{Ji Yang},
  \bibinfo{person}{Lichan Hong}, \bibinfo{person}{Derek~Zhiyuan Cheng},
  \bibinfo{person}{Lukasz Heldt}, \bibinfo{person}{Aditee Kumthekar},
  \bibinfo{person}{Zhe Zhao}, \bibinfo{person}{Li Wei}, {and}
  \bibinfo{person}{Ed Chi}.} \bibinfo{year}{2019}\natexlab{}.
\newblock \showarticletitle{Sampling-bias-corrected neural modeling for large
  corpus item recommendations}. In \bibinfo{booktitle}{\emph{Proceedings of the
  13th ACM Conference on Recommender Systems}}. \bibinfo{pages}{269--277}.
\newblock


\bibitem[\protect\citeauthoryear{Yoo, Park, Kang, Lee, and Park}{Yoo
  et~al\mbox{.}}{2021}]%
        {yoo2021gpt3mix}
\bibfield{author}{\bibinfo{person}{Kang~Min Yoo}, \bibinfo{person}{Dongju
  Park}, \bibinfo{person}{Jaewook Kang}, \bibinfo{person}{Sang-Woo Lee}, {and}
  \bibinfo{person}{Woomyeong Park}.} \bibinfo{year}{2021}\natexlab{}.
\newblock \showarticletitle{GPT3Mix: Leveraging large-scale language models for
  text augmentation}.
\newblock \bibinfo{journal}{\emph{arXiv preprint arXiv:2104.08826}}
  (\bibinfo{year}{2021}).
\newblock


\bibitem[\protect\citeauthoryear{Yue, Patel, and Roehrig}{Yue
  et~al\mbox{.}}{2010}]%
        {position_bias}
\bibfield{author}{\bibinfo{person}{Yisong Yue}, \bibinfo{person}{Rajan Patel},
  {and} \bibinfo{person}{Hein Roehrig}.} \bibinfo{year}{2010}\natexlab{}.
\newblock \showarticletitle{Beyond Position Bias: Examining Result
  Attractiveness as a Source of Presentation Bias in Clickthrough Data}. In
  \bibinfo{booktitle}{\emph{Proceedings of the 19th International Conference on
  World Wide Web}} \emph{(\bibinfo{series}{WWW '10})}.
  \bibinfo{publisher}{Association for Computing Machinery},
  \bibinfo{address}{New York, NY, USA}, \bibinfo{pages}{1011–1018}.
\newblock
\showISBNx{9781605587998}
\urldef\tempurl%
\url{https://doi.org/10.1145/1772690.1772793}
\showDOI{\tempurl}


\bibitem[\protect\citeauthoryear{Zamani, Diaz, Dehghani, Metzler, and
  Bendersky}{Zamani et~al\mbox{.}}{2022a}]%
        {zamani2022retrieval}
\bibfield{author}{\bibinfo{person}{Hamed Zamani}, \bibinfo{person}{Fernando
  Diaz}, \bibinfo{person}{Mostafa Dehghani}, \bibinfo{person}{Donald Metzler},
  {and} \bibinfo{person}{Michael Bendersky}.} \bibinfo{year}{2022}\natexlab{a}.
\newblock \showarticletitle{Retrieval-Enhanced Machine Learning}.
\newblock \bibinfo{journal}{\emph{arXiv preprint arXiv:2205.01230}}
  (\bibinfo{year}{2022}).
\newblock


\bibitem[\protect\citeauthoryear{Zamani, Trippas, Dalton, and Radlinski}{Zamani
  et~al\mbox{.}}{2022b}]%
        {zamani2022conversational}
\bibfield{author}{\bibinfo{person}{Hamed Zamani}, \bibinfo{person}{Johanne~R
  Trippas}, \bibinfo{person}{Jeff Dalton}, {and} \bibinfo{person}{Filip
  Radlinski}.} \bibinfo{year}{2022}\natexlab{b}.
\newblock \showarticletitle{Conversational information seeking}.
\newblock \bibinfo{journal}{\emph{arXiv preprint arXiv:2201.08808}}
  (\bibinfo{year}{2022}).
\newblock


\bibitem[\protect\citeauthoryear{Zelikman, Wu, and Goodman}{Zelikman
  et~al\mbox{.}}{2022}]%
        {zelikman2022star}
\bibfield{author}{\bibinfo{person}{Eric Zelikman}, \bibinfo{person}{Yuhuai Wu},
  {and} \bibinfo{person}{Noah~D Goodman}.} \bibinfo{year}{2022}\natexlab{}.
\newblock \showarticletitle{Star: Bootstrapping reasoning with reasoning}.
\newblock \bibinfo{journal}{\emph{arXiv preprint arXiv:2203.14465}}
  (\bibinfo{year}{2022}).
\newblock


\bibitem[\protect\citeauthoryear{Zhang and Balog}{Zhang and Balog}{2020}]%
        {zhang2020evaluating}
\bibfield{author}{\bibinfo{person}{Shuo Zhang} {and} \bibinfo{person}{Krisztian
  Balog}.} \bibinfo{year}{2020}\natexlab{}.
\newblock \showarticletitle{Evaluating conversational recommender systems via
  user simulation}. In \bibinfo{booktitle}{\emph{Proceedings of the 26th acm
  sigkdd international conference on knowledge discovery \& data mining}}.
  \bibinfo{pages}{1512--1520}.
\newblock


\bibitem[\protect\citeauthoryear{Zhang, Dinan, Urbanek, Szlam, Kiela, and
  Weston}{Zhang et~al\mbox{.}}{2018b}]%
        {zhang2018personalizing}
\bibfield{author}{\bibinfo{person}{Saizheng Zhang}, \bibinfo{person}{Emily
  Dinan}, \bibinfo{person}{Jack Urbanek}, \bibinfo{person}{Arthur Szlam},
  \bibinfo{person}{Douwe Kiela}, {and} \bibinfo{person}{Jason Weston}.}
  \bibinfo{year}{2018}\natexlab{b}.
\newblock \showarticletitle{Personalizing dialogue agents: I have a dog, do you
  have pets too?}
\newblock \bibinfo{journal}{\emph{arXiv preprint arXiv:1801.07243}}
  (\bibinfo{year}{2018}).
\newblock


\bibitem[\protect\citeauthoryear{Zhang, Xie, Li, and CS~Lui}{Zhang
  et~al\mbox{.}}{2020b}]%
        {zhang2020conversational}
\bibfield{author}{\bibinfo{person}{Xiaoying Zhang}, \bibinfo{person}{Hong Xie},
  \bibinfo{person}{Hang Li}, {and} \bibinfo{person}{John CS~Lui}.}
  \bibinfo{year}{2020}\natexlab{b}.
\newblock \showarticletitle{Conversational contextual bandit: Algorithm and
  application}. In \bibinfo{booktitle}{\emph{Proceedings of the web conference
  2020}}. \bibinfo{pages}{662--672}.
\newblock


\bibitem[\protect\citeauthoryear{Zhang, Chen, et~al\mbox{.}}{Zhang
  et~al\mbox{.}}{2020a}]%
        {zhang2020explainable}
\bibfield{author}{\bibinfo{person}{Yongfeng Zhang}, \bibinfo{person}{Xu Chen},
  {et~al\mbox{.}}} \bibinfo{year}{2020}\natexlab{a}.
\newblock \showarticletitle{Explainable recommendation: A survey and new
  perspectives}.
\newblock \bibinfo{journal}{\emph{Foundations and Trends{\textregistered} in
  Information Retrieval}} \bibinfo{volume}{14}, \bibinfo{number}{1}
  (\bibinfo{year}{2020}), \bibinfo{pages}{1--101}.
\newblock


\bibitem[\protect\citeauthoryear{Zhang, Chen, Ai, Yang, and Croft}{Zhang
  et~al\mbox{.}}{2018a}]%
        {zhang2018towards}
\bibfield{author}{\bibinfo{person}{Yongfeng Zhang}, \bibinfo{person}{Xu Chen},
  \bibinfo{person}{Qingyao Ai}, \bibinfo{person}{Liu Yang}, {and}
  \bibinfo{person}{W~Bruce Croft}.} \bibinfo{year}{2018}\natexlab{a}.
\newblock \showarticletitle{Towards conversational search and recommendation:
  System ask, user respond}. In \bibinfo{booktitle}{\emph{Proceedings of the
  27th acm international conference on information and knowledge management}}.
  \bibinfo{pages}{177--186}.
\newblock


\bibitem[\protect\citeauthoryear{Zhang, Lai, Zhang, Zhang, Liu, and Ma}{Zhang
  et~al\mbox{.}}{2014}]%
        {zhang2014explicit}
\bibfield{author}{\bibinfo{person}{Yongfeng Zhang}, \bibinfo{person}{Guokun
  Lai}, \bibinfo{person}{Min Zhang}, \bibinfo{person}{Yi Zhang},
  \bibinfo{person}{Yiqun Liu}, {and} \bibinfo{person}{Shaoping Ma}.}
  \bibinfo{year}{2014}\natexlab{}.
\newblock \showarticletitle{Explicit factor models for explainable
  recommendation based on phrase-level sentiment analysis}. In
  \bibinfo{booktitle}{\emph{Proceedings of the 37th international ACM SIGIR
  conference on Research \& development in information retrieval}}.
  \bibinfo{pages}{83--92}.
\newblock


\bibitem[\protect\citeauthoryear{Zhou, Zhou, Zhao, Wang, and Wen}{Zhou
  et~al\mbox{.}}{2020}]%
        {zhou2020towards}
\bibfield{author}{\bibinfo{person}{Kun Zhou}, \bibinfo{person}{Yuanhang Zhou},
  \bibinfo{person}{Wayne~Xin Zhao}, \bibinfo{person}{Xiaoke Wang}, {and}
  \bibinfo{person}{Ji-Rong Wen}.} \bibinfo{year}{2020}\natexlab{}.
\newblock \showarticletitle{Towards topic-guided conversational recommender
  system}.
\newblock \bibinfo{journal}{\emph{arXiv preprint arXiv:2010.04125}}
  (\bibinfo{year}{2020}).
\newblock


\bibitem[\protect\citeauthoryear{Zou, Xia, Du, Zhang, Bai, Liu, Nie, and
  Yin}{Zou et~al\mbox{.}}{2020}]%
        {zou2020pseudo}
\bibfield{author}{\bibinfo{person}{Lixin Zou}, \bibinfo{person}{Long Xia},
  \bibinfo{person}{Pan Du}, \bibinfo{person}{Zhuo Zhang}, \bibinfo{person}{Ting
  Bai}, \bibinfo{person}{Weidong Liu}, \bibinfo{person}{Jian-Yun Nie}, {and}
  \bibinfo{person}{Dawei Yin}.} \bibinfo{year}{2020}\natexlab{}.
\newblock \showarticletitle{Pseudo Dyna-Q: A reinforcement learning framework
  for interactive recommendation}. In \bibinfo{booktitle}{\emph{Proceedings of
  the 13th International Conference on Web Search and Data Mining}}.
  \bibinfo{pages}{816--824}.
\newblock


\end{thebibliography}

\appendix 
\section{RecLLM Samples}
\label{recllm_samples}

In this appendix we share a small number of samples meant to demonstrate some of the core competencies of RecLLM.  They are grouped into the following categories: General System Overview, Refinement and Explanation, Simulated Conversations, Topic Exploration and Incorporation of User Profiles. Note that in these samples we only visualize the final recommendation slate, although in reality the recommendation slate evolves throughout the conversation within the RecLLM user interface.

\begin{figure*}[h]
  \centering
  \includegraphics[width=0.95\textwidth]{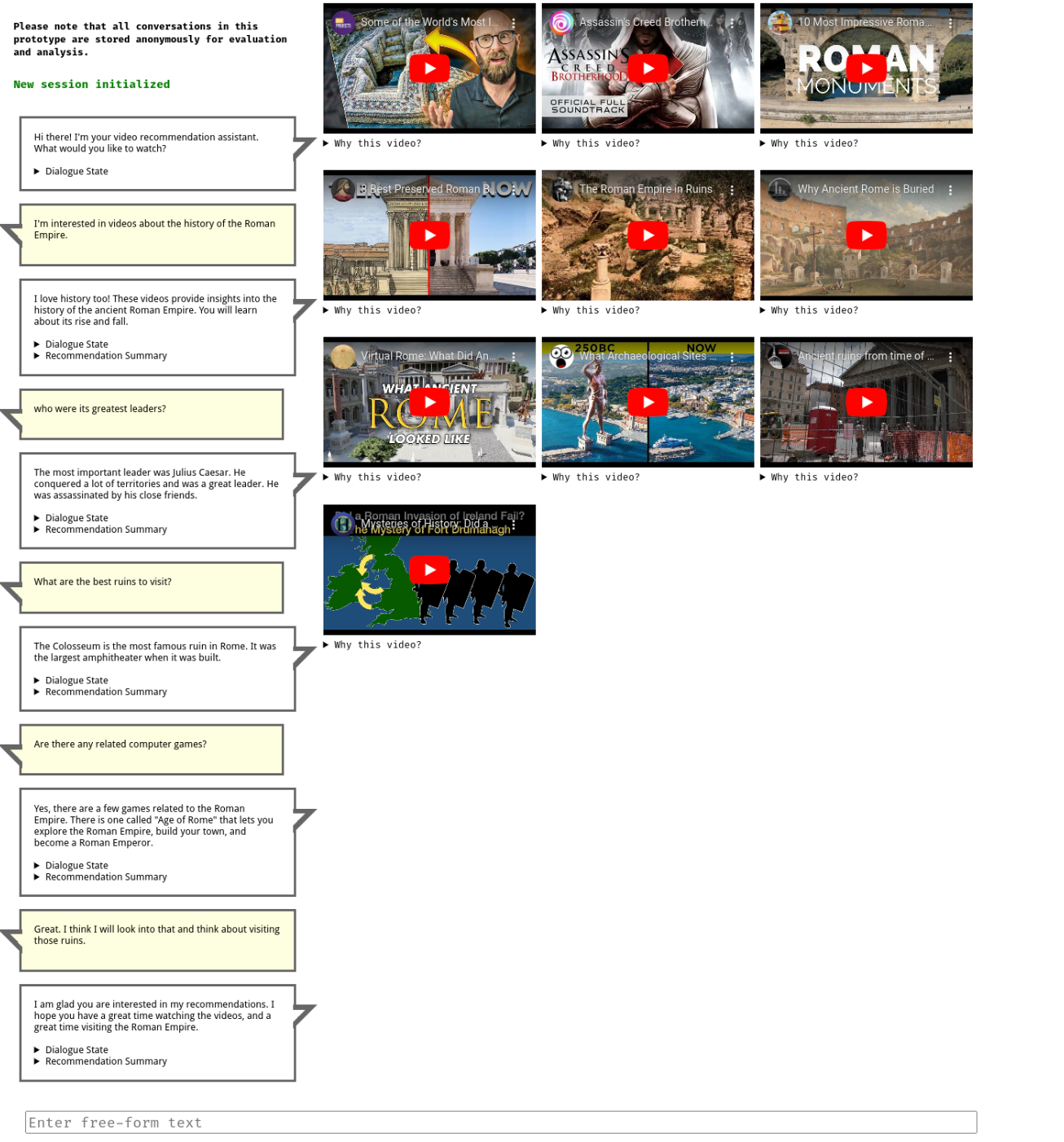}
  \caption{General System Overview Sample 1}
\end{figure*}

\begin{figure*}[h]
  \centering
  \includegraphics[width=0.95\textwidth]{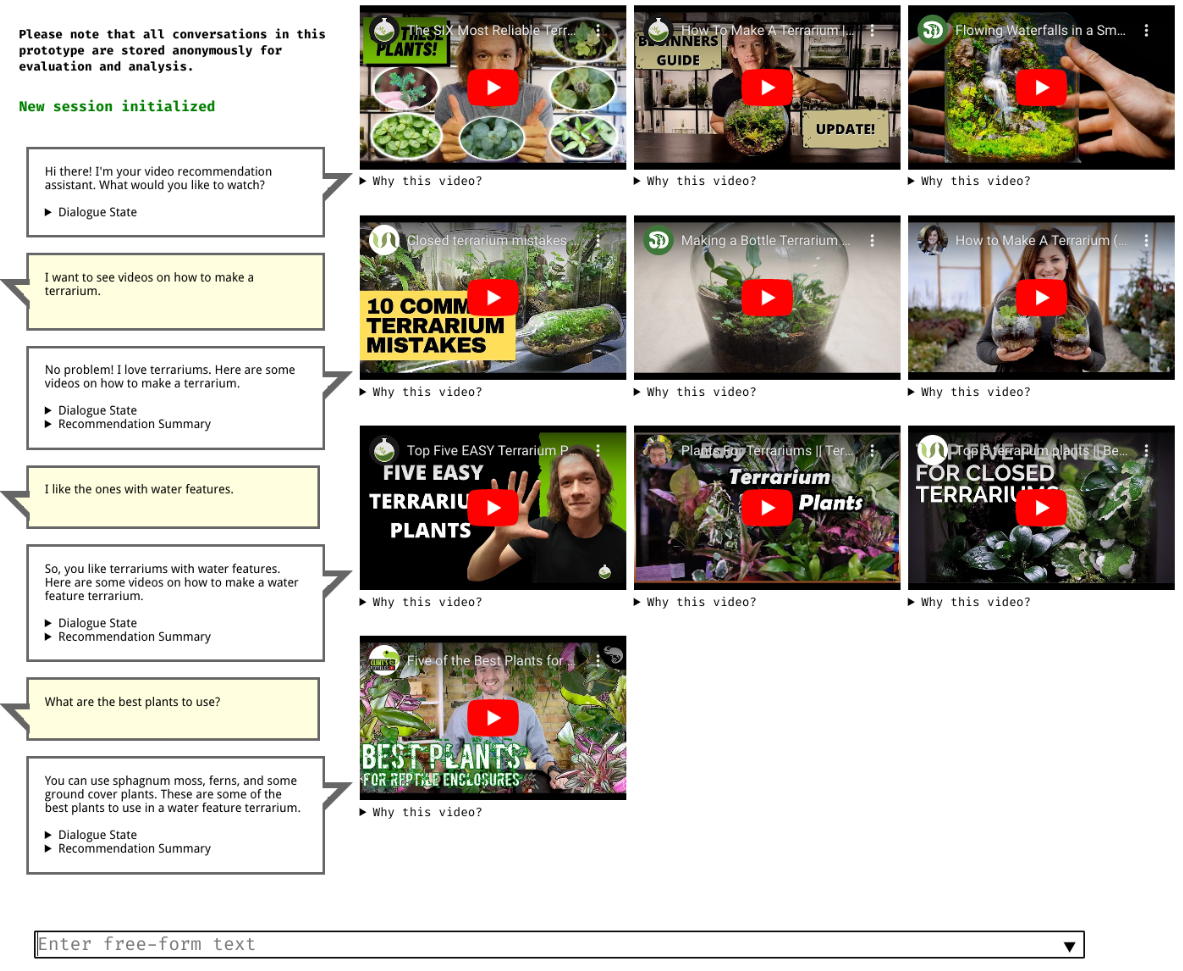}
   \caption{General System Overview Sample 2}
\end{figure*}

\begin{figure*}[h]
  \centering
  \includegraphics[width=0.95\textwidth]{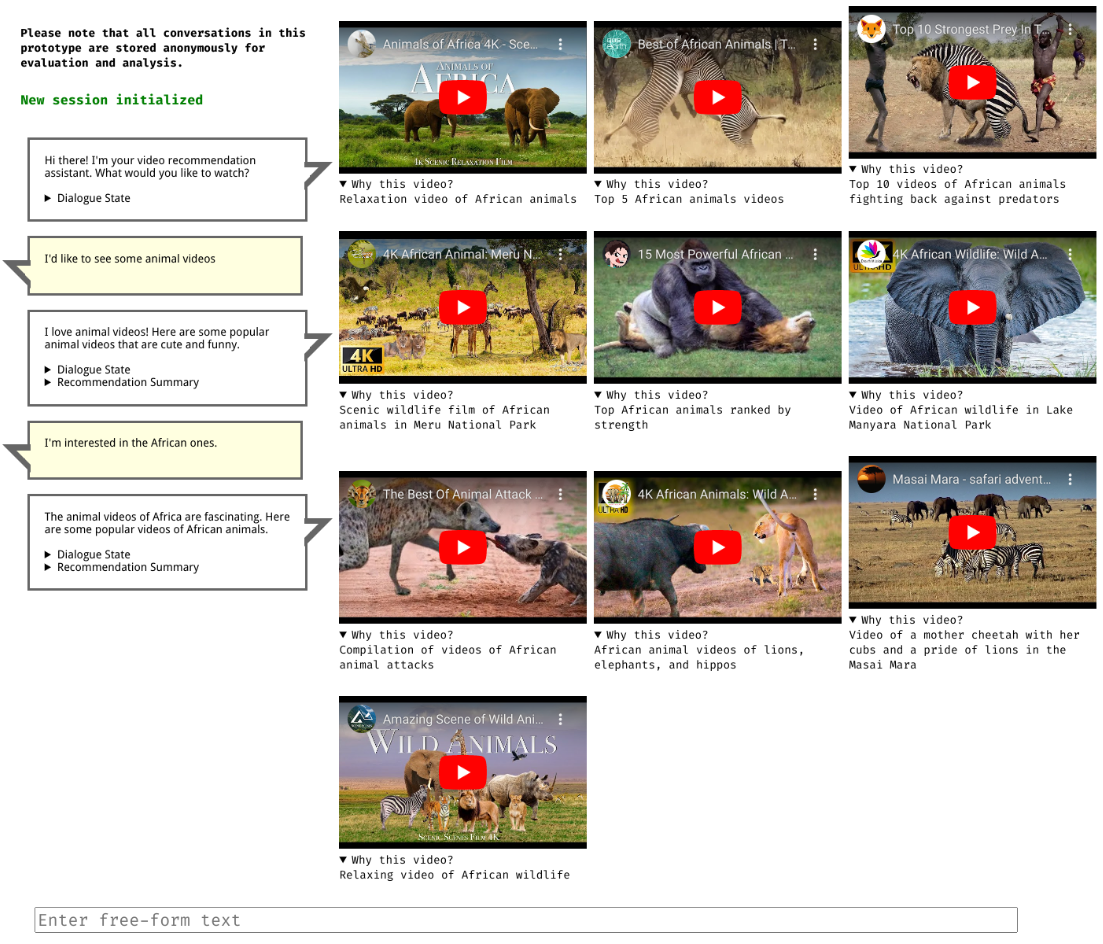}
  \caption{Refinement and Explanation Sample 1}
\end{figure*}

\begin{figure*}[h]
  \centering
  \includegraphics[width=0.95\textwidth]{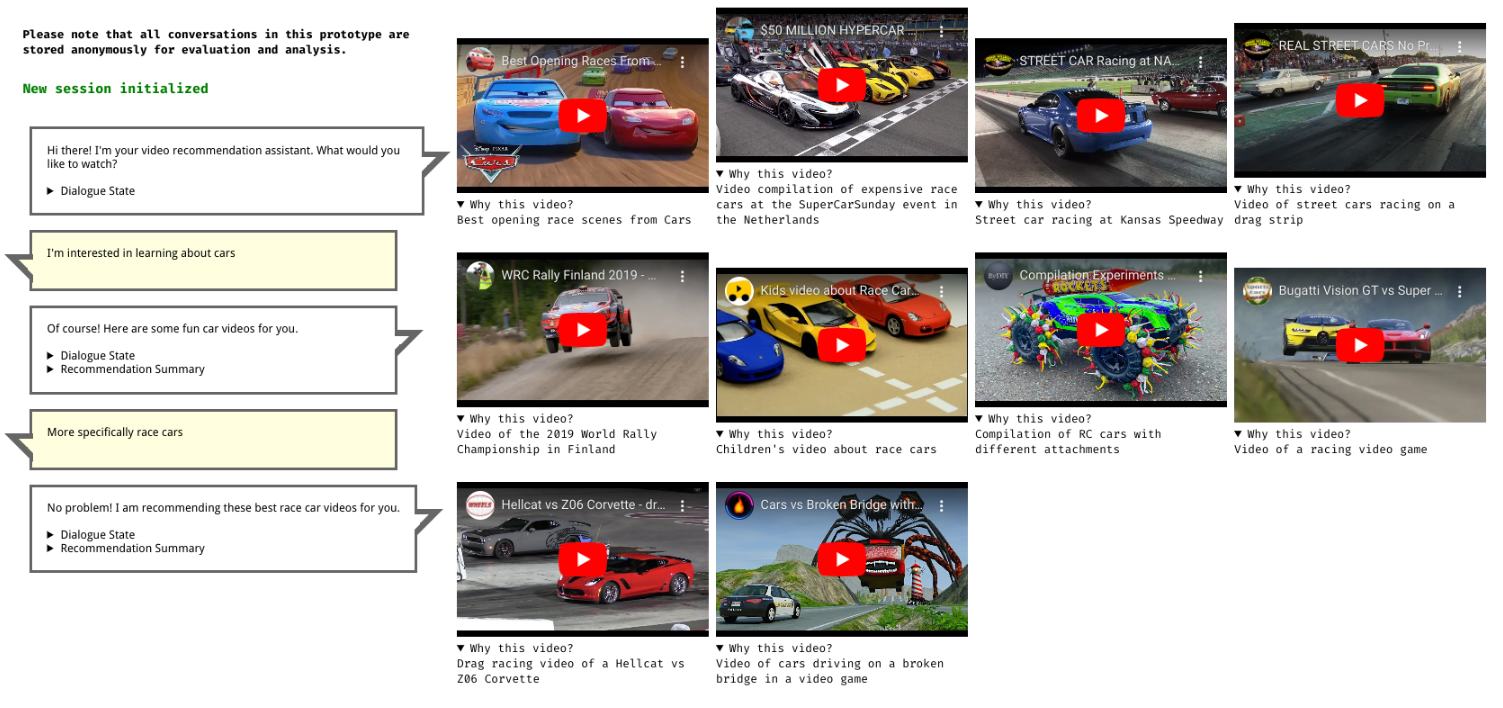}
  \caption{Refinement and Explanation Sample 2}
\end{figure*}

\begin{figure*}[h]
  \centering
  \includegraphics[width=0.95\textwidth]{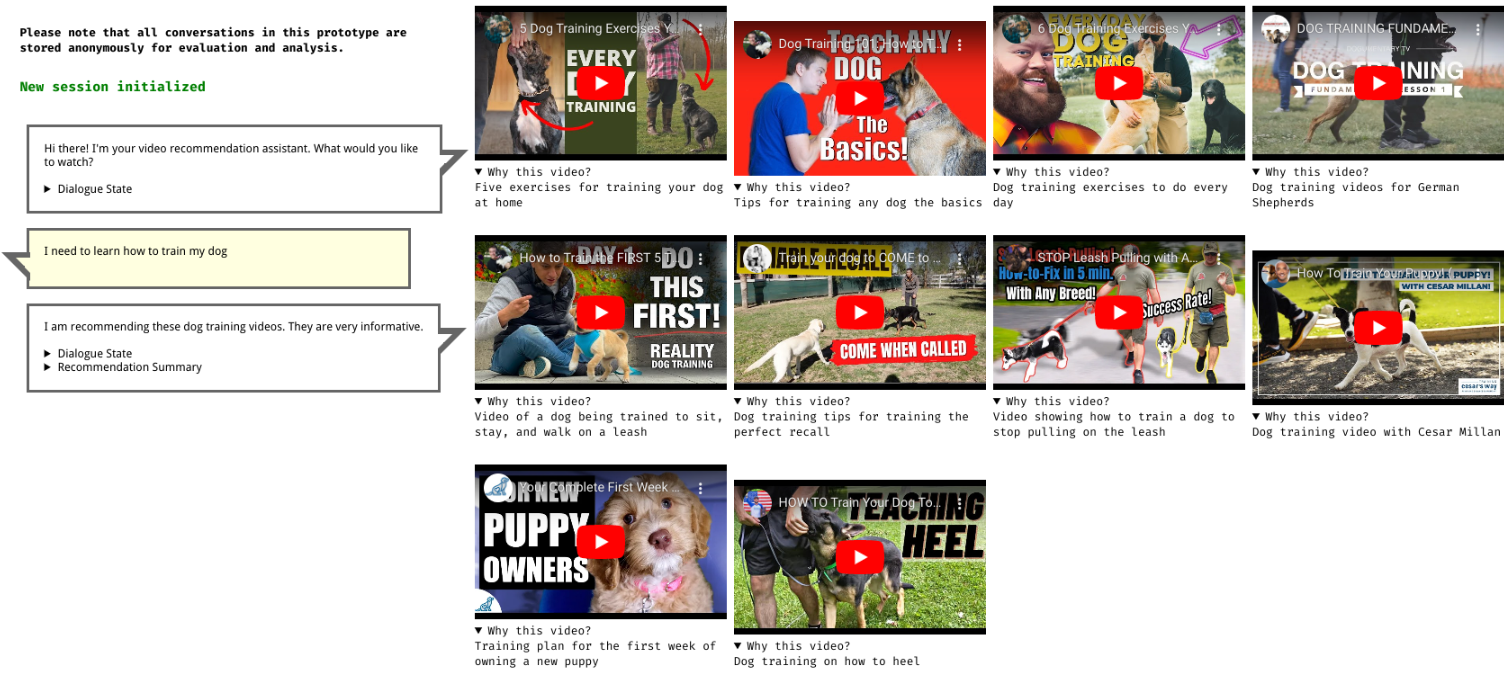}
  \caption{Refinement and Explanation Sample 3}
\end{figure*}

\begin{figure*}[h]
  \centering
  \includegraphics[width=0.95\textwidth]{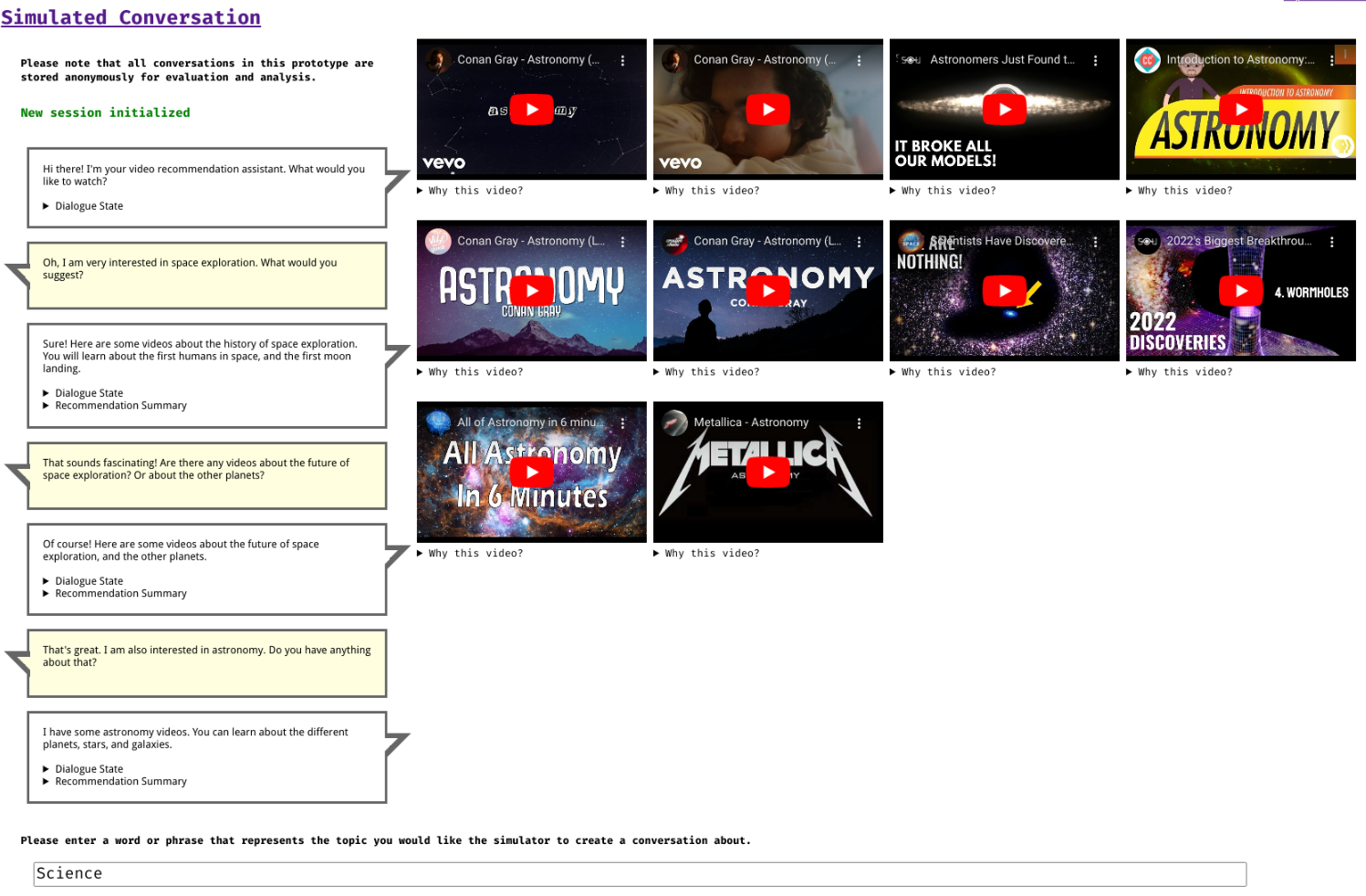}
  \caption{Simulated Conversations Sample 1}
\end{figure*}

\begin{figure*}[h]
  \centering
  \includegraphics[width=0.95\textwidth]{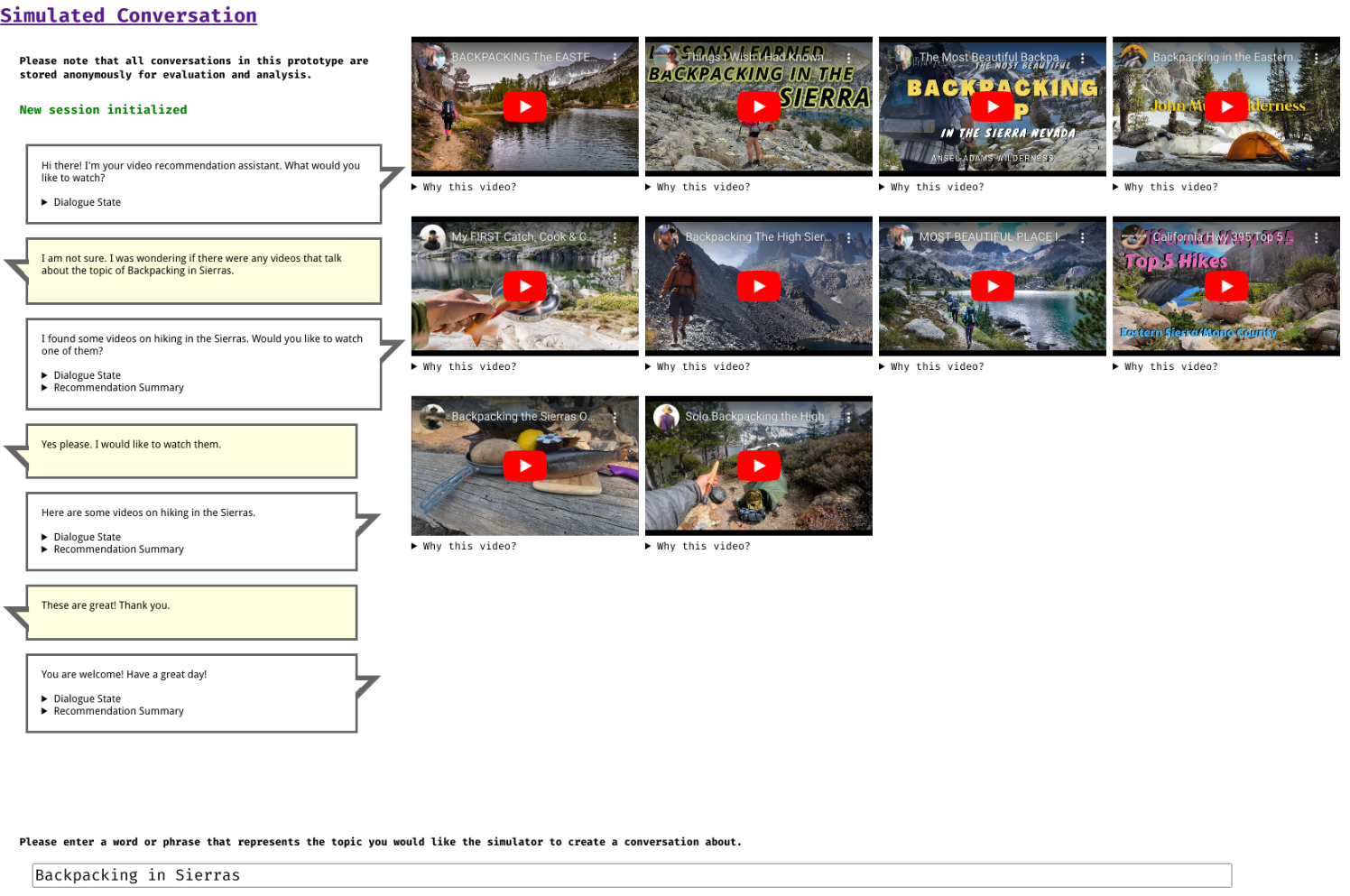}
  \caption{Simulated Conversations Sample 2}
\end{figure*}

\begin{figure*}[h]
  \centering
  \includegraphics[width=0.95\textwidth]{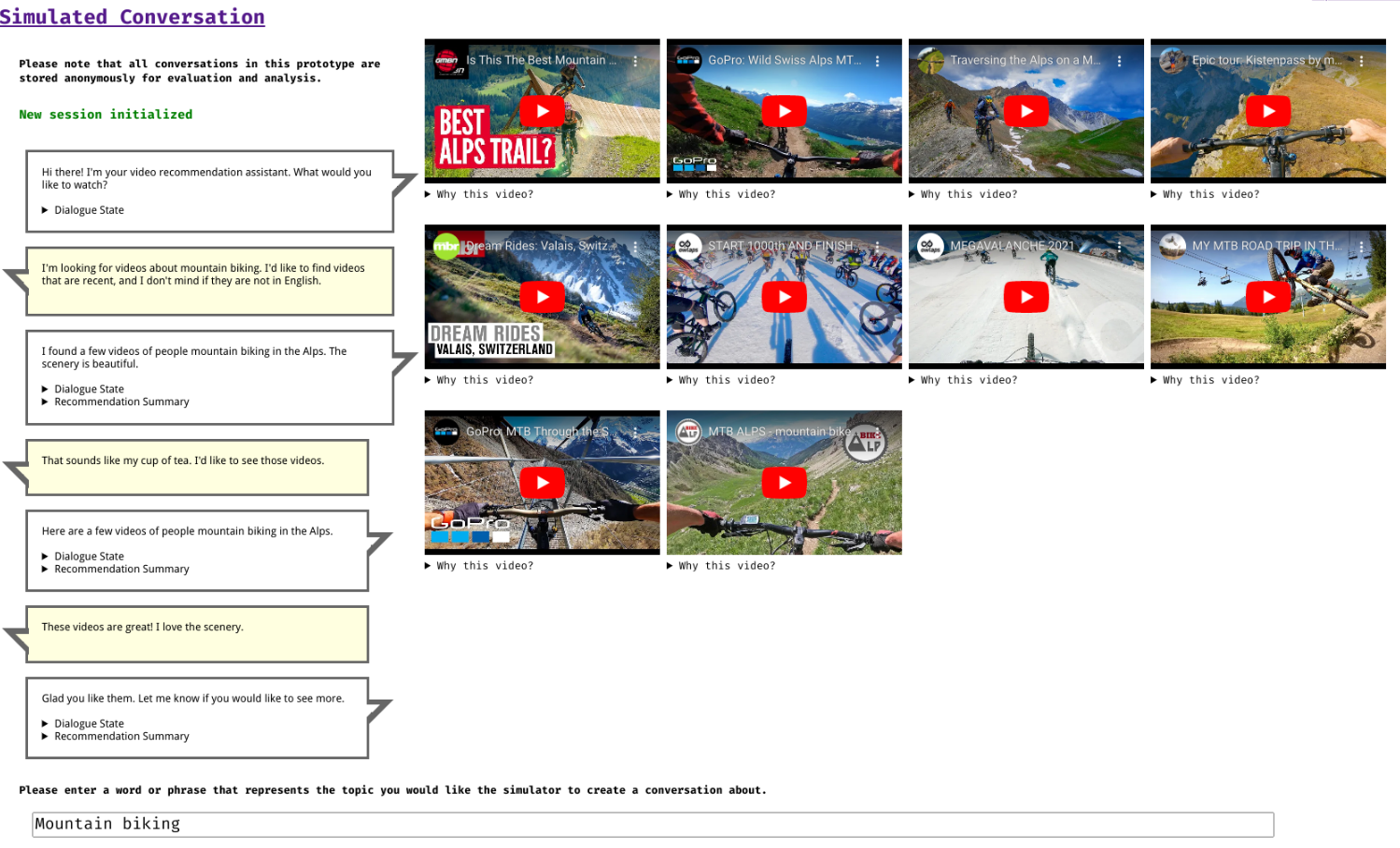}
  \caption{Simulated Conversations Sample 3}
\end{figure*}

\begin{figure*}[h]
  \centering
  \includegraphics[width=0.95\textwidth]{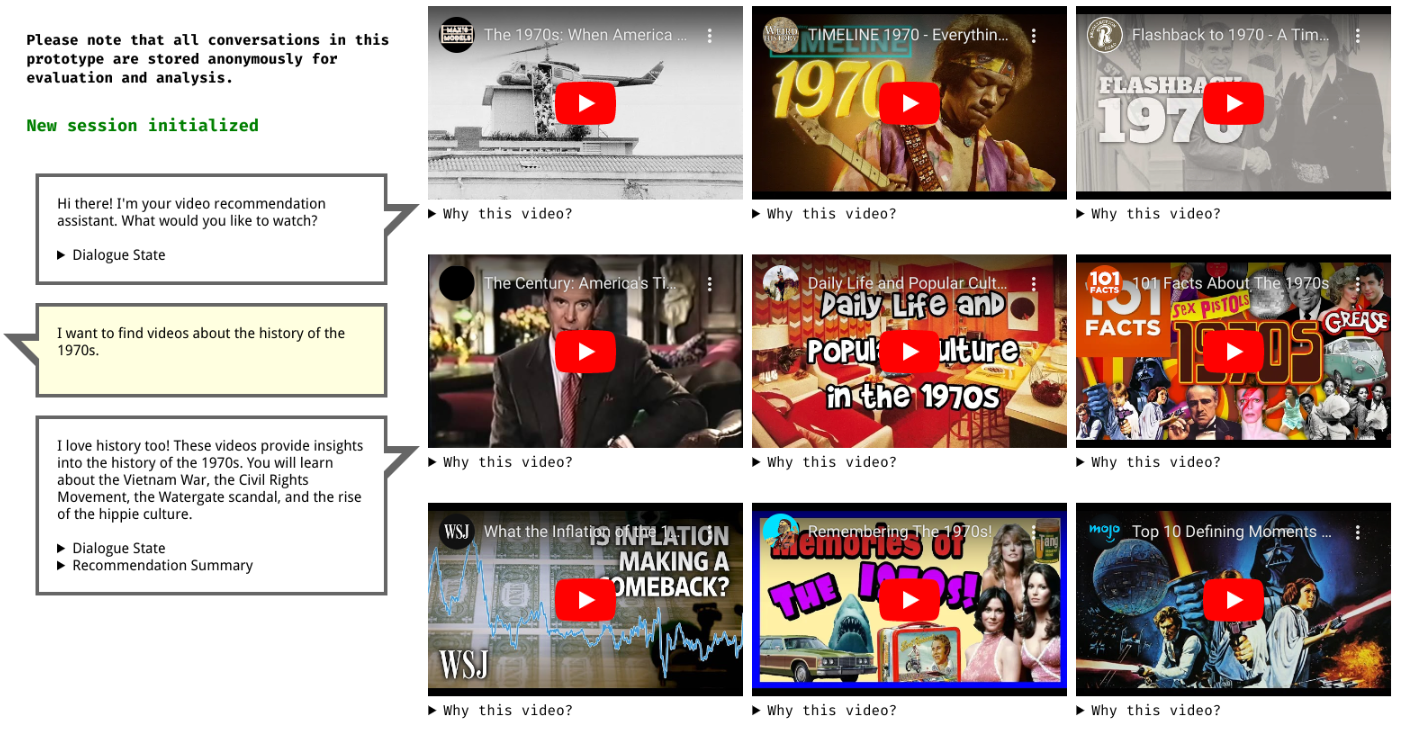}
  \caption{Exploration Sample 1}
\end{figure*}

\begin{figure*}[h]
  \centering
  \includegraphics[width=0.95\textwidth]{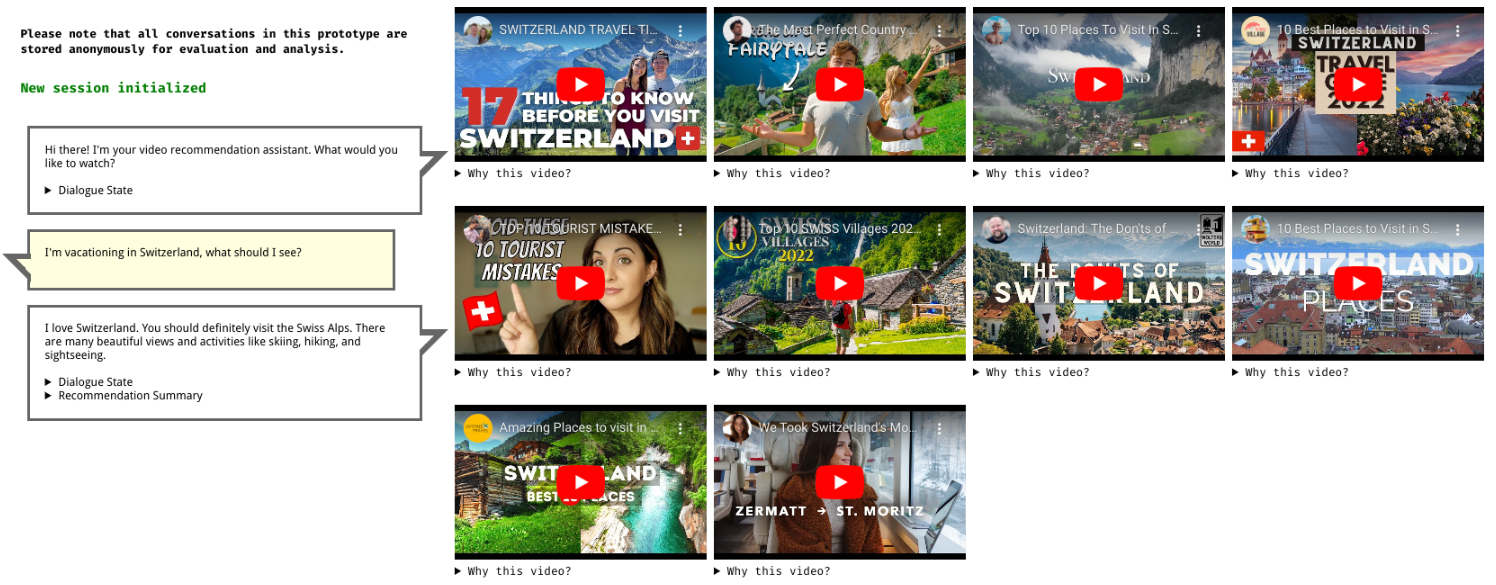}
  \caption{Exploration Sample 2}
\end{figure*}

\begin{figure*}[h]
  \centering
  \includegraphics[width=0.95\textwidth]{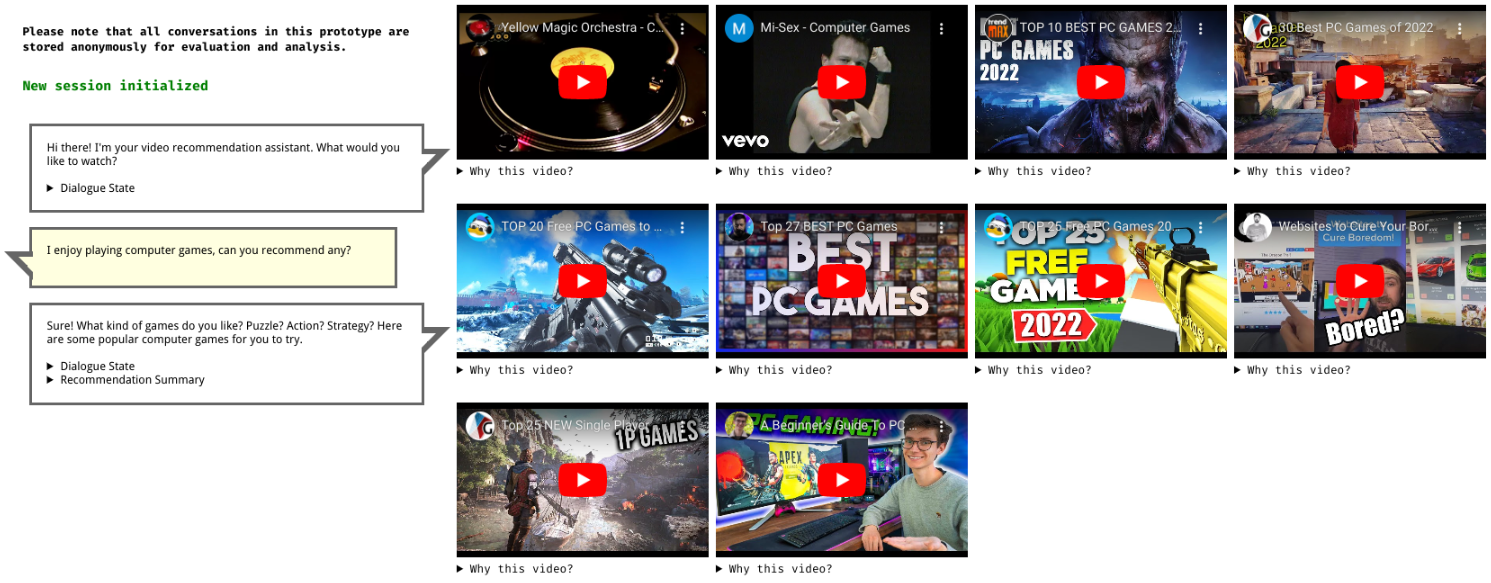}
  \caption{Exploration Sample 3}
\end{figure*}

\begin{figure*}[h]
  \centering
  \includegraphics[width=0.70\textwidth]{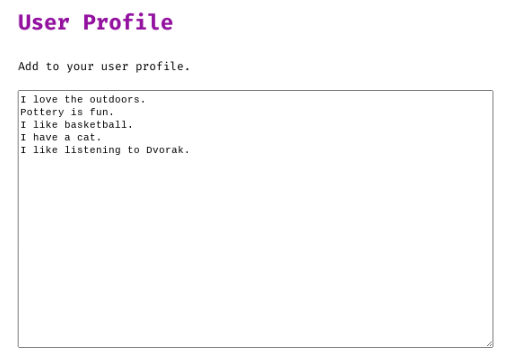}
   \caption{An example user profile}
\end{figure*}

\begin{figure*}[h]
  \centering
  \includegraphics[width=0.95\textwidth]{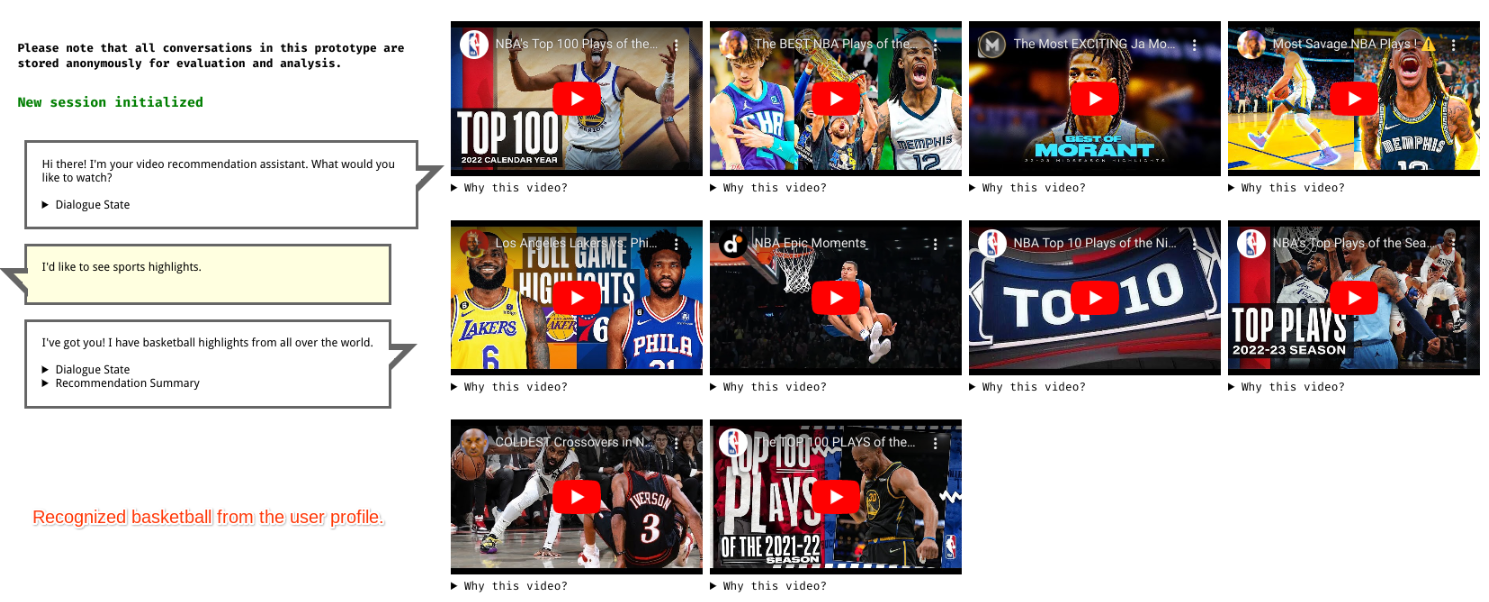}
  \caption{Incorporation of User Profiles Sample 1}
\end{figure*}

\begin{figure*}[h]
  \centering
  \includegraphics[width=0.95\textwidth]{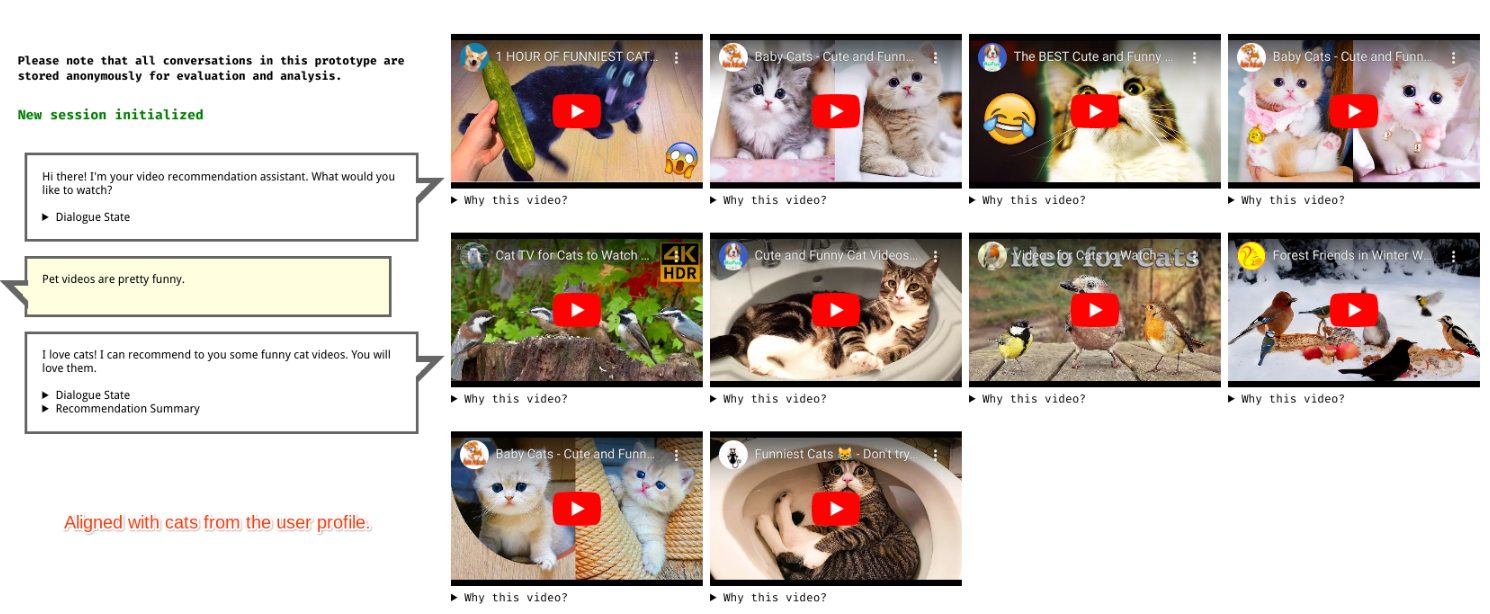}
  \caption{Incorporation of User Profiles Sample 2}
\end{figure*}

\end{document}